\documentclass[epj]{svjour}
\pdfoutput=1 

\usepackage{amsmath}
\usepackage{amssymb} 
\usepackage{graphicx}

\begin{document}

\title{Google matrix of the world network of economic activities} 

\author{V.Kandiah$^{1,2,3}$, H.Escaith$^{2,3}$  \and D.L.Shepelyansky$^{1}$}

\institute{
Laboratoire de Physique Th\'eorique du CNRS, IRSAMC, 
Universit\'e de Toulouse, UPS, F-31062 Toulouse, France
\and
World Trade Organization,
rue de Lausanne 154,
CH-1211 Gen\`eve 21,
Switzerland
\and
Opinions are personal and do not represent WTO's position
}

\titlerunning{Google matrix of the world network of economic activities}
\authorrunning{V.Kandiah, H.Escaith and D.L.Shepelyansky}

\abstract{
Using the new data from the OECD-WTO world network of economic activities
we construct the Google matrix $G$ of 
this directed network and perform its detailed analysis.
The network contains 58 countries and 37 activity sectors
for years 1995 and 2008. The construction of $G$, based on 
Markov chain transitions, treats all countries on equal democratic grounds
while the contribution of activity sectors 
is proportional to their exchange monetary volume. 
The Google matrix analysis allows to obtain reliable ranking of
countries and activity sectors and to determine
the sensitivity of CheiRank-PageRank commercial balance
of countries in respect to price variations and labor cost in various countries.
We demonstrate that the developed approach takes into account 
multiplicity of network links with economy interactions between countries and activity sectors
thus being more efficient compared to the usual export-import
analysis.
The spectrum and eigenstates of $G$ are also analyzed
being related to specific activity communities of countries.
}



\date{Dated:  April 24, 2015}

\maketitle

\section{Introduction}

The recent reports of the Organisation for Economic Co-operation and Development
(OECD) \cite{oecd2014} and of  the World Trade Organization (WTO)
\cite{wto2014} demonstrate all the complexity of global manufactoring
activities, exchange and trade in the modern world.
This complexity is rapidly growing with time
and  now it becomes clear that
traditional statistics are increasingly unable
to provide all the necessary information.
Applying modern mathematical tools and methods
to new data sets can allow to understand
the hidden trends of the world economic activities.
Thus the matrix tools for analysis of Input-Out transactions are broadly
used in economy starting from the fundamental works of Leontief 
\cite{leontief1,leontief2}
with their more recent developments described in \cite{miller2009}.
In the last decade the development of modern society generated 
enormous communication and social networks including the World Wide Web (WWW),
Wikipedia, Twitter and other directed networks (see e.g. \cite{dorogovtsev}). 
It has been found that the concept of Markov chains
provides a very useful and powerful mathematical
approach for analysis of such networks.
Thus  the PageRank algorithm, developed by
Brin and Page in 1998 \cite{brin} for the WWW information retrieval,
became at the mathematical foundation of 
the Google search engine  (see e.g. \cite{meyer}).
This algorithm constructs the Google matrix $G$ of 
Markov chain transitions between network nodes
and allows to rank billions of web pages of the WWW.
The spectral and other properties of the Google matrix are
analyzed in \cite{arxivrmp}. 
The historical overviews of the development of Google matrix methods
and their links with the works of Leontief are given in \cite{franceschet,vignahisto}.

The obtained results 
demonstrate the efficiency of the Google matrix
analysis not only for the WWW but also
for various types of directed networks \cite{arxivrmp}.
One of such  examples is the World Trade Network (WTN)
with multiproduct exchange between the world countries.
The data of trade flows are available at  the United Nations (UN) 
COMTRADE database \cite{comtrade}
for more than $50$ years. The results presented in
\cite{wtngoogle,wtnproducts} for the WTN show that the Google matrix analysis 
is well adapted to the ranking of world countries and trade products
and to determination of the sensitivity of trade 
to price variations of various products.
The new element of such an approach is a democratic treatment of world countries
independently of their richness being different from the usual Import and Export
ranking. At the same time the contributions of various
products are considered being proportional to their
trade volume contribution in the exchange flows.

Here we use the Google matrix analysis developed for the multiproduct WTN
\cite{wtnproducts} showing that it can be directly used for the World Network
of Economic Activities (WNEA) constructed from the OECD-WTO trade in value-added database. 
In a certain sense activities (or sectors)
are correlated to products in the WTN. However, for the WTN 
there is exchange between countries but there is no exchange between
industries and commodities. Thus in \cite{wtnproducts} it was argued that
certain economical features are not captured by the COMTRADE
database since in real economy the traders are industries, not countries;
in particular certain products 
are transferred to each other (e.g. metal
and plastic are used for production of cars).
In contrast to that,  the OECD-WTO WNEA
incorporates the transitions between activity sectors
thus representing the economic reality
of world activities in a more correct manner.

\begin{figure}[!ht]
\begin{center}
\includegraphics[width=0.48\textwidth]{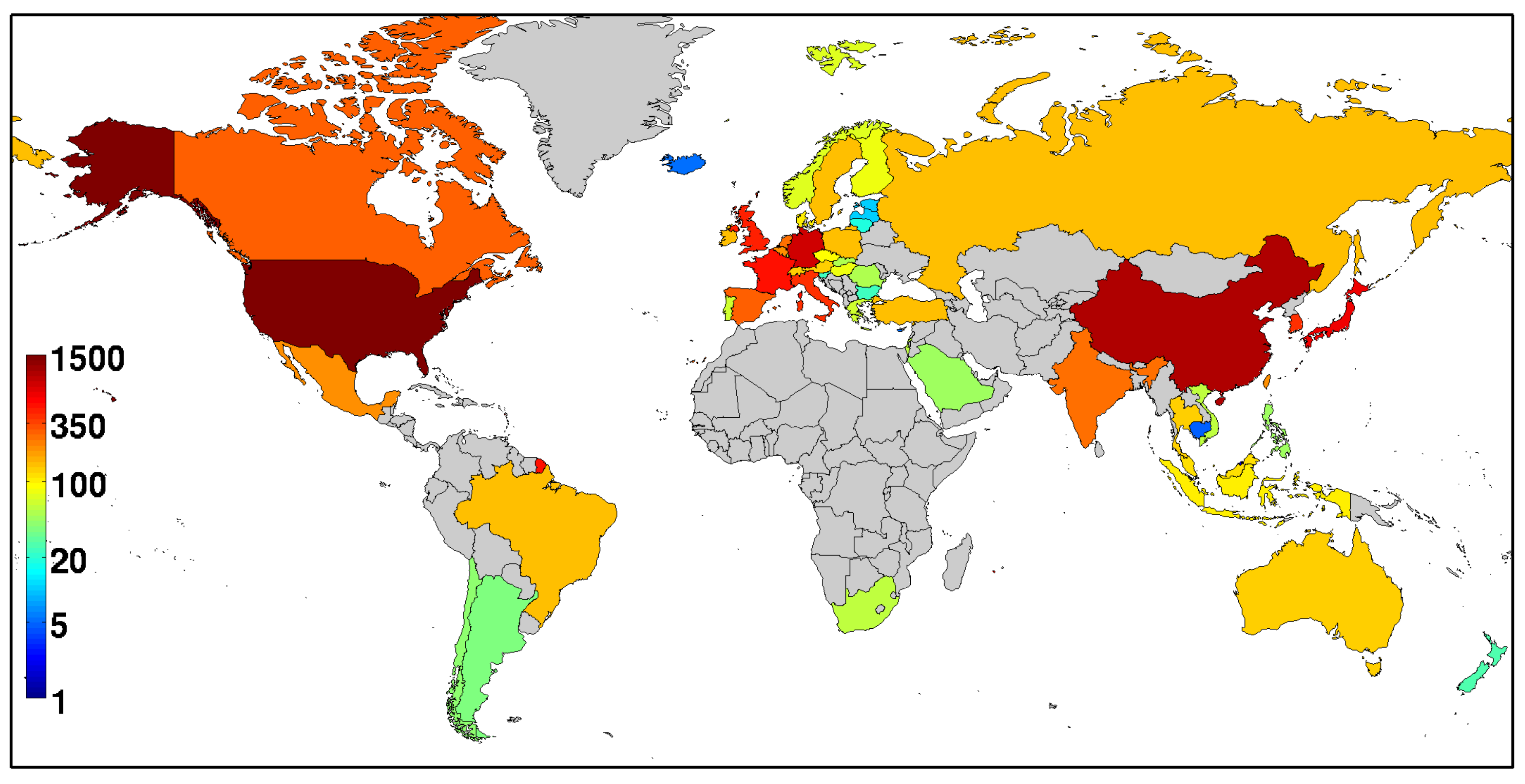}\\
\includegraphics[width=0.48\textwidth]{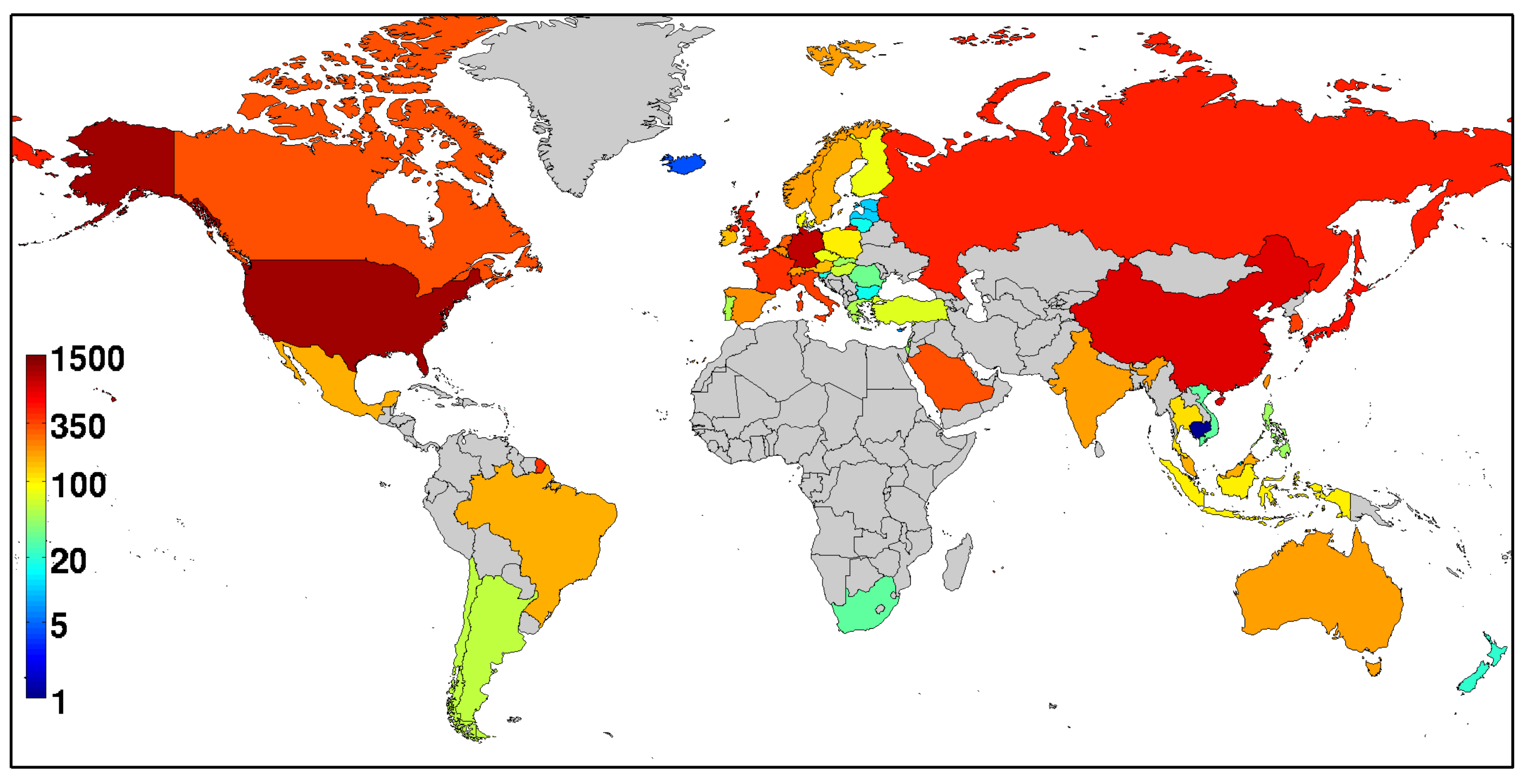}
\end{center}
\vglue -0.1cm
\caption{
World map of countries with color showing country import (top panel) 
and export (bottom panel) 
with economic activity (trade) volume expressed in billions of USD 
and given by numbers at color bars;
the gray color marks countries attributed to the ROW group (rest of the world)
with exchange values $733$ (Import) and $1018$ (Export) in billions of USD.
The data are shown for year 2008 with $N_c=57+1$ countries (with ROW) for the 
economic activities in all $N_s=37$ sectors. 
Country names can be found in Table~\ref{tab1} and 
in the world map of countries \cite{worldmap}.
}
\label{fig1}
\end{figure}

We note that there has been a number of other investigations
of the WTN reported in 
\cite{garlaschelli,hedeem,fagiolo2,garlaschelli2010,benedictis,plosjapan,imfpaper}.
However, in this work we have the new important elements, 
introduced in \cite{wtngoogle,wtnproducts}:
the analysis of PageRank and CheiRank probabilities 
corresponding to direct and inverted network flows and related to
Import and Export; democratic treatment of countries 
combined with the contributions of sectors (or products)
being proportional to their commercial exchange fractions.
We point that the OECD-WTO TiVA database of economic activities
between  world countries and activity sectors
has been created very recently (2013) and thus this work
represents the first Google matrix analysis of these data. 
We stress that the usual Import-Export ranking 
of commercial flows, shown in Fig.~\ref{fig1},
is not able to take into account all the complexity of
chains of links between various countries 
and various activity sectors. In contrast to that
the approach developed here takes all of them into account
due to the powerful method based on the Google matrix.

\section{Methods and data description}

Here we describe the data available for the OECD-WTO TiVA network and the mathematical
methods used for the analysis of this network.
The list of $N_c=58$ countries ($57$ plus $1$ for the Rest Of the World ROW)
is given in Table~\ref{tab1} with their flags.
Following \cite{wtngoogle} we use for countries ISO 3166-1 alpha-3 code
available at Wikipedia.
The list of sectors with their names is given in Table~\ref{tab2} .
The fractions of sectors in the exchange volume are 
given in Table~\ref{tab3} for years $1995$, $2008$.

\subsection{Google matrix construction for the OECD-WTO WNEA}

We use the OECD-WTO TiVA database released in May 2013 which covers years $1995$, $2000$, $2005$, 
$2008$, $2009$ with the main emphasis for years $1995 and 2008$
(2009 data are affected by the global crisis and may not be representative).
The network considers $N_c=58$ world countries  given in Table~\ref{tab1}.
In fact, there are $57$ countries and the rest of the world,
which includes the remaining countries of the world forming
one group called ROW. There are also $N_s=37$ sectors
of economic activities given in Table~\ref{tab2}.
The sectors  are classified according to the International Standard 
Industrial Classification of All Economic Activities
(ISIC) Rev.3 \cite{isic}. 
Here we present results for all $37$ sectors of Table~\ref{tab2},
noting that the sectors $s=1,2,..20$ represent production activities
while $s=21,...,37$ represent service activities.
The transections between service sectors are hard to exctract
and the future improvements of this part of TiVA database
are desirable.

For a given year, the TiVA data extend OECD Input/Out tables of 
economic activity expressed in terms of USD for a given year.
From these data we construct
the matrix $M_{cc^\prime,ss^\prime}$
of money transfer between nodes
expressed in USD:
\begin{equation}
M_{cc^\prime,ss^\prime} = \text{transfer
from country $c^\prime$, sector $s^\prime$ to  $c,s$}
\label{eq1}
\end{equation}
Here the country indexes are $c,c^\prime=1,\ldots,N_c$ and 
activity sector indexes are $s,s^\prime=1,\ldots,N_s$
with $N_c=58$ and $N_s=37$. 
The whole matrix size is $N=N_c \times N_s = 2146$.
Here each node represents a pair of
country and activity sector, a link gives a transfer from
a sector of one country to another sector of another country.
We construct the matrix $M_{cc^\prime,ss^\prime}$
from the TiVA Input/Output tables using the transposed
representation so that the volume of products or sectors flows 
in a column from line to line. In the construction of
 $M_{cc^\prime,ss^\prime}$ we exclude exchanges inside a given country
in order to highlight the trade exchange flows between countries
(elements inside country are zeros).

The ISIC Rev.3 classification of sectors 
have a significant correlation with the UN
Standard International Trade Classification (SITC) Rev. 1
of products used in \cite{wtnproducts}.
There is a clear relationship on the production side 
between ISIC sectors and products of the world 
exports (but not at import level:
if all agricultural exports are produced by the agricultural sector,
agricultural products will be imported 
by manufacturing industries such as food processing of textile and clothing).
There is also another important difference: the transfer matrix from COMTRADE
is diagonal in products \cite{wtnproducts}
(thus there is no transfer from product to product),
while for the TiVA data there are transitions from 
one sector to another sector and thus the  matrix 
of nominal values, in current prices,
(\ref{eq1}) is not diagonal in $s,s^\prime$.

For convenience of future notations we also define the value of imports $V_{cs}$ and exports 
$V^{*}_{cs}$ for a given country $c$ and sector $s$  as
\begin{equation}
V_{cs}=\sum_{c^\prime,s^\prime} M_{cc^\prime,ss^\prime} \, , \,\;
V^{*}_{cs}=\sum_{c^\prime,s^\prime} M_{{{c^\prime}c},{s^\prime}s} .
\label{eq2}
\end{equation}
The import $V_c=\sum_s V_{cs}$ and export $V^*_{c} = \sum_s V^{*}_{cs}$ values 
for countries $c$  are shown on the world map of countries in Fig.~\ref{fig1}
for year 2008. We note that often one uses the notion of volume of export or import
(see. e.g. \cite{wtnproducts}) but from the economic view point it more
correct to speak about value of export or import.

In order to compare later with the PageRank and \\
CheiRank probabilities 
we define exchange value  ranks in
the whole matrix space of dimension $N=N_c\times N_s$. Thus the
ImportRank ($\hat{P}$) and ExportRank ($\hat{P}^*$) probabilities
are given by the normalized import and export values
\begin{equation}
\hat{P}_{i} = {V_{cs}}/{V} \, , \,\;
\hat{P}^*_{i} = {V^{*}_{cs}}/{V} \, ,
\label{eq3}
\end{equation}
where $i=s+(c-1)N_s$, $i=1,\ldots,N$ and the total exchange value is
$V=\sum_{c,c^\prime,s,s^\prime} M_{cc^\prime,ss^\prime}=\sum_{c,s}V_{cs}=\sum_{cs}V^{*}_{cs}$.

The Google matrices $G$ and $G^*$ are defined as $N\times N$ real
 matrices with non-negative elements:
\begin{equation}
G_{ij}= \alpha S_{ij}+(1-\alpha) v_i e_j \, ,\; 
{G^*}_{ij}=\alpha {S^*}_{ij}+(1-\alpha) v^*_i e_j \, ,
\label{eq4}
\end{equation}
where $N=N_c\times N_s$, $\alpha \in (0,1]$ is the damping factor ($0<\alpha<1$), 
$e_j$ is the row 
vector of unit elements ($e_j=1$), and $v_i$ is a 
positive column vector called a \emph{personalization vector} 
with $\sum_i v_i=1$ \cite{meyer,wtnproducts}.
We note that the usual Google matrix corresponds to 
a personalization vector $v_i=e_i/N$ with $e_i=1$. 
In this work, following \cite{wtngoogle,wtnproducts}, we fix $\alpha=0.5$
noting that a variation of $\alpha$ in a range $(0.5,0.9)$
does not significantly affect the probability distributions of PageRank 
and CheiRank vectors \cite{meyer,arxivrmp,wtngoogle}. 
The choice of the personalization vector is specified below.
Following \cite{wtnproducts} we call this approach
the Google Personalized Vector Method (GPVM).

The matrices $S$ and $S^*$  are built from money matrices ${M}_{cc^\prime,ss^\prime}$ as
\begin{eqnarray}
 \nonumber
S_{i,i^\prime}&=&\left\{\begin{array}{cl}   
{M}_{cc^\prime,ss^\prime}/V_{c^\prime s^\prime}& 
\text{    if } V_{c^\prime s^\prime}\ne0\\ 
1/N & \text{    if } V_{c^\prime s^\prime}=0\\ 
\end{array}\right.\\
S^*_{i,i^\prime}&=&\left\{\begin{array}{cl}   
M_{{c^\prime}c,s^\prime s}/V^{*}_{c^\prime s^\prime}& 
\text{    if } V^{*}_{c^\prime s^\prime}\ne0\\ 
1/N & \text{    if } V^{*}_{c^\prime s^\prime}=0\\ 
\end{array}\right.
\label{eq5}
\end{eqnarray}
where $c,c^\prime=1,\ldots,N_c$; $s,s^\prime=1,\ldots,N_s$; 
$i=s+(c-1)N_s$; $i^\prime=s^\prime+(c^\prime-1)N_s$; 
and therefore $i,i^\prime=1,\ldots,N$. 
Here $V_{c's'}=\sum_{cs} M_{cc',ss'}$.
The sum of elements of each column of 
$S$ and $S^*$ is normalized to unity and hence the matrices $G, G^*, S, S^*$
belong to the class of Google matrices and Markov chains.
Thus $S, G$ look at the import perspective
and $S^*, G^*$ at the export side of transactions. 

PageRank and CheiRank ($P$ and $P^*$) are the right eigenvectors of
$G$ and $G^*$ matrices respectively at eigenvalue $\lambda=1$.
The equation for right eigenvectors have the form
\begin{eqnarray}
\sum_j G_{ij} \psi_j= \lambda \psi_i \, , \; 
\sum_j {G^*}_{ij} {\psi^*}_j = \lambda {\psi^*}_j \; .
\label{eq6}
\end{eqnarray}
For the eigenstate at $\lambda=1$ we use the notation
$P_i=\psi_i , P^*={\psi^*}_i$ with the normalization 
$\sum P_i = \sum_i {P^*}_i=1$. For other eigenstates we use
the normalization $\sum_i |\psi_i|^2=\sum_i |\psi^*_i|^2=1$.
The eigenvalues and eigenstates of $G, G^*$ are obtained by a direct numerical
diagonalization using the standard numerical packages.

\subsection{PageRank and CheiRank vectors from GPVM}

The components of $P_i$, ${P^*}_i$ are positive. In the WWW context
they  have a meaning of probabilities
to find a random surfer on a given WWW node 
in the limit of large number of surfer jumps 
over network links \cite{meyer}.
In the WNEA context nodes can be viewed and
markets with a random trader transitions between them.
We will use in the following notation of netwrok nodes.
We define the PageRank $K$ and CheiRank $K^*$ indexes
ordering probabilities $P$ and $P^*$ 
in a decreasing order as
$P(K)\ge P(K+1)$ and $P^*(K)\ge P^*(K^*+1)$ with $K,K^*=1,\ldots,N$. 

We note that the pair of PageRank and CheiRank vectors
is very natural for economy and trade networks
corresponding to Import and Export flows.
For the directed networks the statistical 
properties of the pair of such ranking vectors
have been introduced and studied in \cite{linux,wikizzs,wtngoogle}.

We compute the reduced PageRank and CheiRank probabilities of countries 
tracing probabilities over all sectors and getting 
$P_c=\sum_{s} P_{cs}=\sum_{s}P\left(s+(c-1)N_s\right)$ 
and $P^*_c= \sum_s P^*_{cs}=\sum_{s}P^*\left(s+(c-1)N_s\right)$ 
with the corresponding $K_c$ and $K^*_c$ indexes. 
In a similar way we obtain the reduced PageRank and CheiRank probabilities
for sectors tracing over all countries and getting\\
$P_s=\sum_{c}P\left(s+(c-1)N_s\right) = \sum_{c} P_{cs}$ and \\
$P^*_s=\sum_{c}P^*\left(s+(c-1)N_s\right) = \sum_{c} P^*_{cs} $ 
with their corresponding sector indexes $K_s$ and $K^*_s$.
A similar procedure has been used for 
the multiproduct WTN data \cite{wtnproducts}. 

In summary we have $K_s,K^*_s=1,\ldots,N_s$ and 
$K_{c},K^*_{c}=1,\ldots,N_c$. A similar definition of ranks 
from import and export exchange value can be done 
in a straightforward way via probabilities
$\hat{P}_s,\hat{P}^*_s,\hat{P}_c,\hat{P}^*_c,\hat{P}_{cs},\hat{P}^*_{cs}$ and 
corresponding indexes
$\hat{K}_s,\hat{K}^*_s,\hat{K}_c,\hat{K}^*_c,\hat{K},\hat{K}^*$.

\begin{figure}[!ht]
\begin{center}
\includegraphics[width=0.48\textwidth]{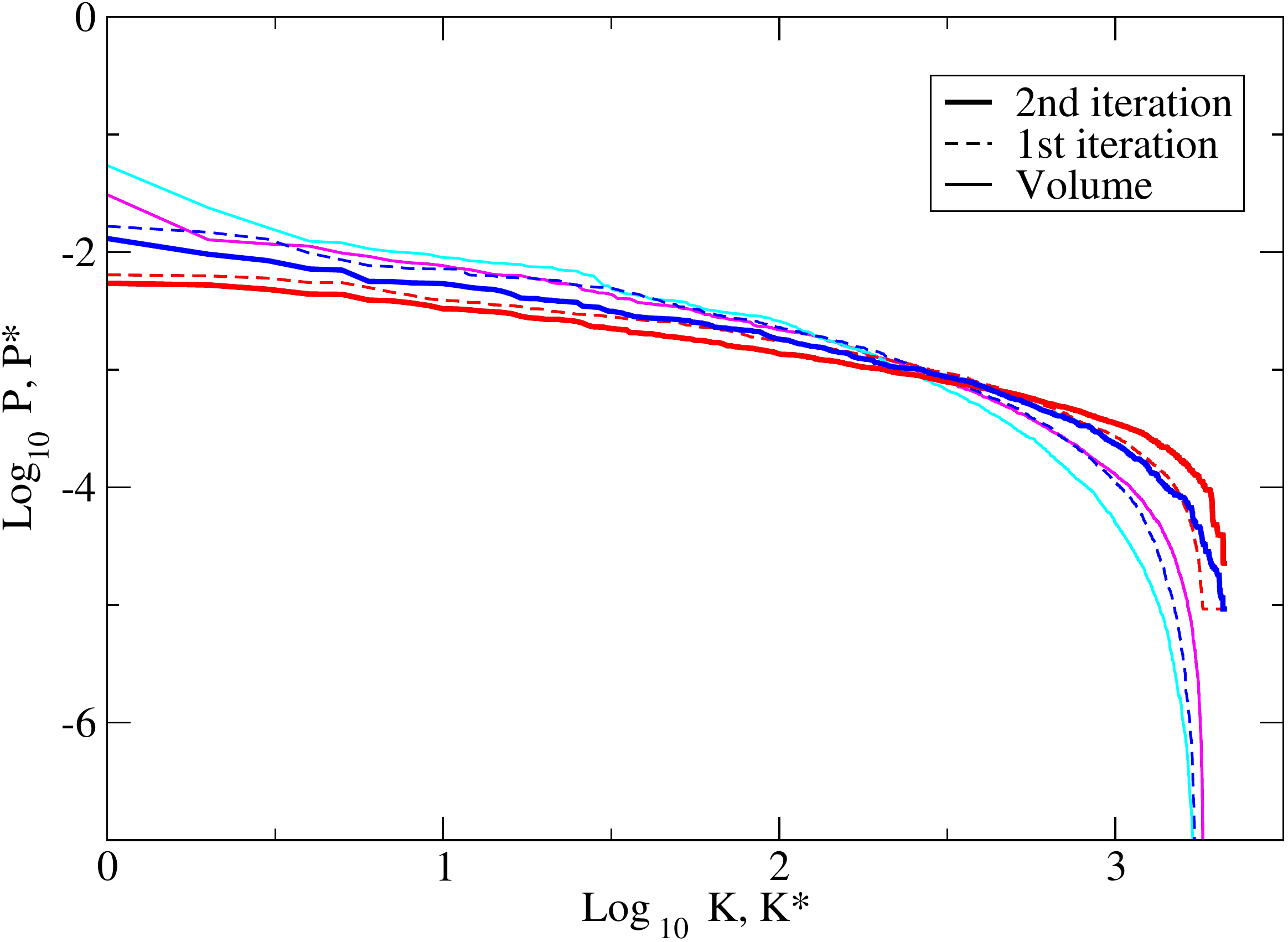}
\end{center}
\vglue -0.1cm
\caption{
Dependence of probabilities of PageRank $P(K)$, CheiRank $P^*(K^*)$, 
ImportRank $\hat{P}(\hat{K})$ and ExportRank $\hat{P}^*(\hat{K}^*)$ 
on their indexes in logarithmic scale for WNEA 
(or OECD-WTO TiVA network) in 2008 
with $\alpha=0.5$, $N_c=58$, $N_s=37$, $N=N_c \times N_s = 2146$. 
Here the results for the GPVM after the first and 
second iterations are shown for 
PageRank (CheiRank) in red (blue) with dashed and 
solid curves respectively. $\;\;\;\;$
Probabilities for ImportRank and ExportRank from exchange value 
are shown by magenta
and cyan thin curves respectively.
}
\label{fig2}
\end{figure}

\begin{figure}[!ht]
\begin{center}
\includegraphics[width=0.48\textwidth]{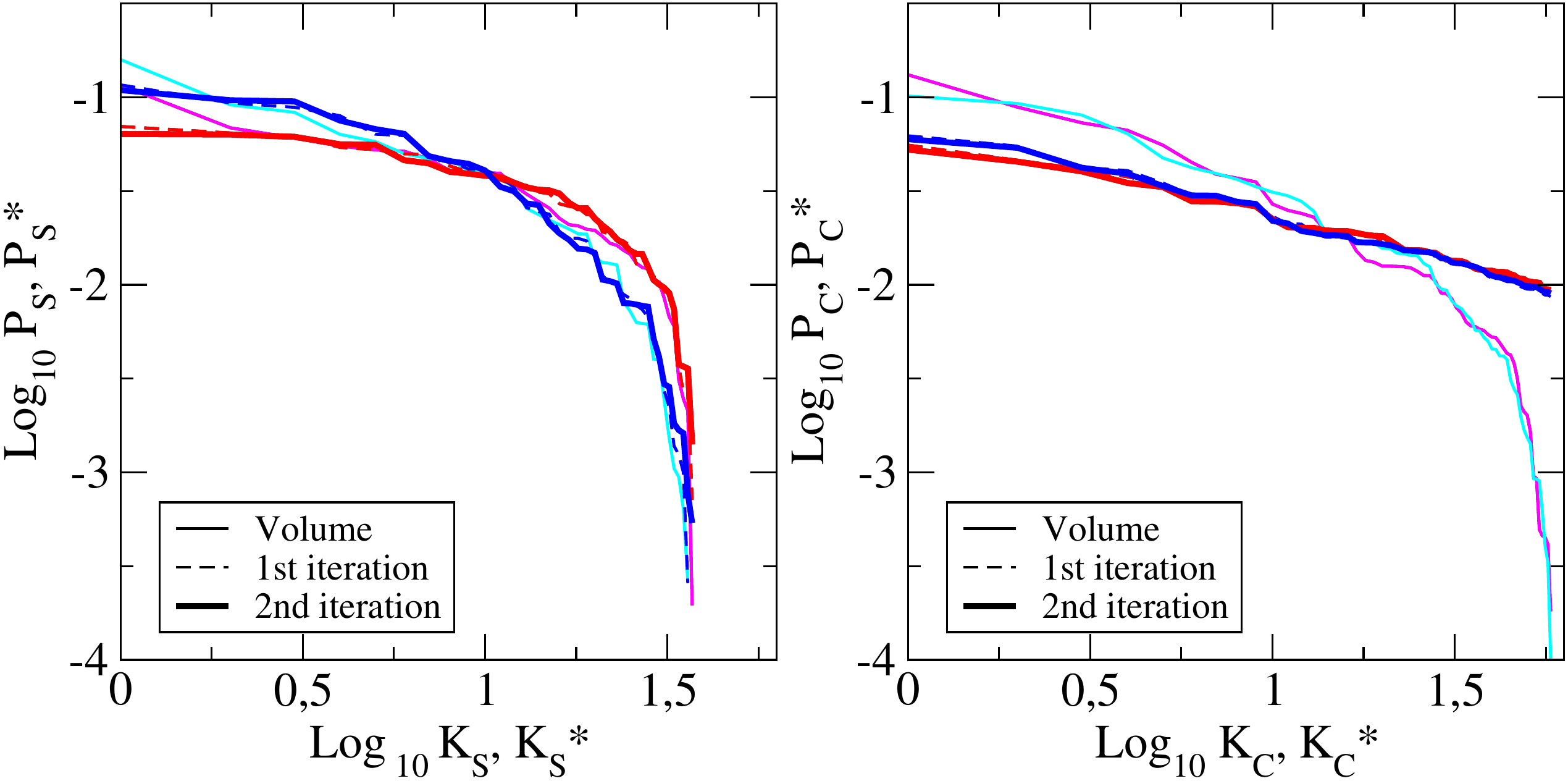}
\end{center}
\vglue -0.1cm
\caption{
Probability distributions of PageRank and CheiRank for sectors $P_s(K_s)$, $P^*_s(K^*_s)$ 
(left panel) and countries $P_c(K_c)$, $P^*_c(K^*_c)$ (right panel) 
in logarithmic scale for WNEA (or OECD-WTO TiVA network) from Fig.\ref{fig2}. 
Here the results for the first and second GPVM iterations are shown 
by red (blue) curves for PageRank (CheiRank) with dashed and solid curves respectively
(with a strong overlap of curves). 
The probabilities from the exchange value ranking are shown 
by thin magenta and cyan lines for ImportRank and ExportRank respectively.
}
\label{fig3}
\end{figure}

To compute the PageRank and CheiRank probabilities
from $G$ and $G^*$, keeping a ``democratic'', or equal, treatment 
of countries (independently of their richness) and at the same time
keeping the 
proportionality of activity sectors to their exchange value,
we use the Google Personalized Vector Method (GPVM) 
developed in \cite{wtnproducts} with
a personalized vector $v_i$ in (\ref{eq4}).
At the first iteration of Google matrix we  take into account 
the relative product value per country using 
the following personalization vectors for $G$ and $G^*$:
\begin{equation}
v_i = \frac{V_{cs}}{N_c \sum_{s^\prime} V_{c s^\prime}} \, , \;
v^*_i = \frac{V^{*}_{cs}}{N_c \sum_{s^\prime} V^{*}_{c s^\prime}} \, ,
\label{eq7}
\end{equation}
using the definitions (\ref{eq2}) and the relation
$i=s+(c-1)N_s$.
This personalized vector depends both on sector and country indexes.
As for the multiproduct WTN in \cite{wtnproducts}
we define the second iteration vector being proportional to the reduced 
PageRank and CheiRank vectors in sectors, obtained from the 
GPVM Google matrix of the first iteration: 
\begin{equation}
v^\prime(i) = \frac{P_s}{N_c} \, , \;
v^{\prime *}(i) = \frac{P^*_s}{N_c} \, .
\label{eq8} 
\end{equation}
In this way we keep democracy in countries but keep contribution of sectors  
proportional to their exchange value.
This second iteration personalized vectors are used in the following
computations and operations with $G$ and $G^*$
giving us the PageRank and CheiRank vectors. 
This procedure with two iterations
forms  our GPVM approach.
The difference between results obtained from
the first and second iterations is not very large 
(see Figs.~\ref{fig2},~\ref{fig3}), but
 the personalized vector for the second iteration
gives a reduction of fluctuations.
In all Figures after Fig.~\ref{fig3}
we show the  GPVM results after the second iteration.

As for the WTN it is convenient to analyze the distribution of nodes on the 
PageRank-CheiRank plane $(K,K^*)$. 
In addition to two ranking indexes $K,K^*$
we use also 2DRank index $K_2$ which describes
the combined contribution of two ranks
as described in \cite{wikizzs}.
The ranking list   $K_2(i)$ is constructed by  
increasing $K \rightarrow K+1$ 
and increasing 2DRank index 
$K_2(i)$ by one if a new entry is present in the list of
first $K^*<K$ entries of CheiRank, then the one unit step is done in
$K^*$ and $K_2$ is increased by one if the new entry is
present in the list of first $K<K^*$ entries of CheiRank.
More formally, 2DRank $K_2(i)$ gives the ordering 
of the sequence of nodes, that $\;$ appear
inside  $\;$ the squares  $\;$
$\left[ 1, 1; \;K = k, K^{\ast} = k; \; \-... \right]$ when one runs
progressively from $k = 1$ to $N$.
Additionally, we analyze the distribution of nodes
for reduced indexes $(K_c,K^*_c)$, $(K_s,K^*_s)$.

The localization properties of eigenstates of $G, G^*$
are characterized 
by the inverse participation ration (IPR)
defined as $\xi = (\sum_i |\psi_i|^2)^2/\sum_i |\psi_i|^4$. This
quantity determines an effective 
number of nodes contributing to a formation
of a given eigenstate (see details in \cite{arxivrmp}). 

\subsection{Correlators of PageRank and CheiRank vectors}

As in previous works \cite{linux,wikizzs,wtngoogle}
we consider 
the correlator of PageRank and CheiRank vectors:
\begin{equation}
\kappa=N \sum_{i=1}^{N} P(i) P^*(i) - 1  \; .
\label{eq9} 
\end{equation}
The typical values of $\kappa$ are given in \cite{arxivrmp}
for various networks.

For the global PageRank and CheiRank probabilities the sector-sector 
correlator matrix is defined as: 
\begin{footnotesize}
\begin{equation}
\kappa_{s s^\prime}=N_c\sum_{c=1}^{N_c}\left[\frac{P(s+(c-1)N_s)P^*(s^\prime+(c-1)N_s)}
{\sum_{c^\prime} P(s+(c^\prime-1)N_s) \sum_{c^{\prime\prime}} 
P^*(s^{\prime}+(c^{\prime\prime}-1)N_s)}\right] -1
\label{eq10} 
\end{equation}
\end{footnotesize}

Then the correlator for a given sector is obtained from (\ref{eq10}) as:
\begin{equation}
\kappa_{s}=\kappa_{s s^\prime} \delta_{s,s^{\prime}} \, , 
  \label{eq11} 
\end{equation}
where $\delta_{s,s^{\prime}}$ is the Kronecker delta.

We also use the correlators obtained from the 
probabilities traced over sectors 
($P_c=\sum_s P_{sc}$) and over countries
($P_s=\sum_c P_{sc}$)  which are defined as
\begin{equation}
\kappa(c)=N_c \sum_{c=1}^{N_c} P_{c} P^*_{c} - 1  \, , \; 
\kappa(s)=N_s \sum_{s=1}^{N_s} P_{s} P^*_{s} - 1  \, .
\label{eq12} 
\end{equation}

In the above equations (\ref{eq9})-(\ref{eq12})
the correlators are computed for PageRank and CheiRank probabilities.
We can also compute the same correlators using probabilities from the exchange
value in ImportRank $\hat{P}$ and ExportRank $\hat{P}^*$ defined 
by (\ref{eq3}).

The obtained results are presented in the next Section 
and at the web site
\cite{ourwebpage}.

\section{Results}

We apply the GPVM approach to the data sets of OECD-WTO TiVA of WNEA
and present the obtained results below.

\subsection{PageRank and CheiRank probabilities}


The dependence of probabilities of PageRank $P(K)$ and CheiRank $P^*(K^*)$
vectors on their indexes $K, K^*$ are shown in Fig.~\ref{fig2}
for a selected year 2008. The results can be approximately described by an
algebraic dependence $P \propto 1/K^\beta$,   $P^* \propto 1/{K^*}^\beta$
with the fit exponent value
$\beta = 0.385 \pm 0.014$ for PageRank and $\beta= 0.486 \pm  0.02$ 
for CheiRank 
for $K,K^* \leq 10^3$. 
In contrast to WWW and Wikipedia networks (see e.g. \cite{arxivrmp})
there is no significant difference of $\beta$ between two ranks
that can be attributed to  an intrinsic property of economy networks
to keep economy balance of commercial exchange.
The probability variation is reduced for the Google ranking compared to
the value ranking. This results from a ``democratic'', or equal grounds ranking 
of countries used in the Google matrix analysis.
The obtained data also show that the variation of probabilities 
for 1st and 2nd GPVM iterations are not very large 
that demonstrates the convergence of this approach.

\begin{figure}[!ht]
\begin{center}
\includegraphics[width=0.48\textwidth]{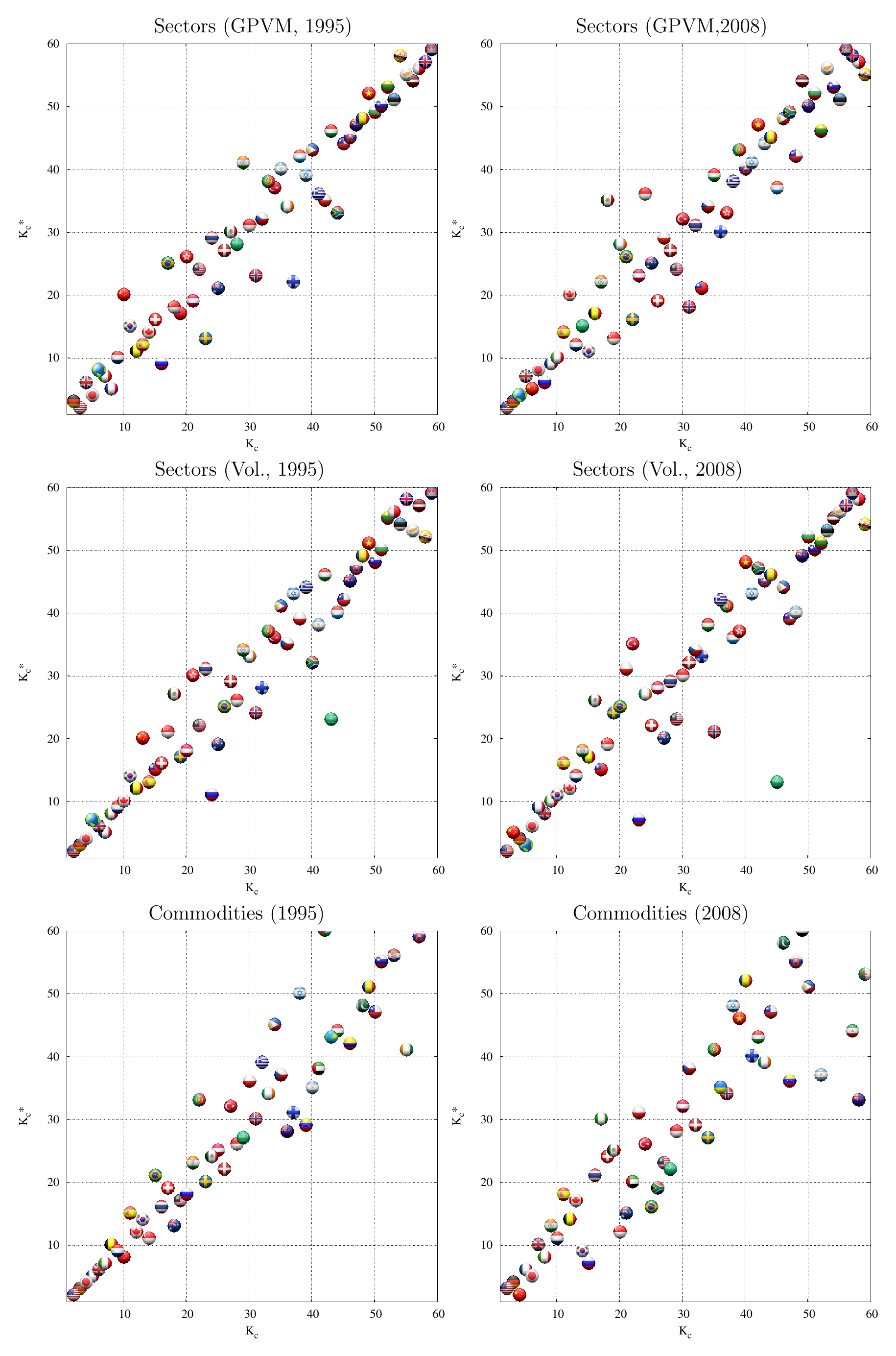}
\end{center}
\vglue -0.1cm
\caption{
Country positions on PageRank-CheiRank plane ($K_c$,$K^*_c$) 
obtained for the  WNEA by 
the GPVM analysis (top panels), ImportRank-ExportRank of 
exchange value (middle panels), and PageRank-CheiRank plane of
 WTN ranking of trade in {\it all commodities} 
from \cite{wtngoogle} (bottom panels)
shown for $K_c , {K^*}_c \leq 60$. 
Left (right) panels show  year 1995 (2008). 
}
\label{fig4}
\end{figure}

\subsection{Ranking of countries and sectors}

After tracing the probabilities $P(K), P^*(K^*)$ 
over sectors we obtain the distribution of world countries on the
PageRank-CheiRank plane $(K_c,K^*_c)$ presented in Fig.~\ref{fig4} 
for WNEA in years 1995, 2008. In the same figure 
we present the rank distributions 
obtained from ImportRank-ExportRank probabilities of exchange value 
and the results obtained in \cite{wtngoogle} 
for the WTN with {\it all commodities}.
For the GPVM data we see the global features already discussed in 
\cite{wtngoogle}: the countries are distributed in a vicinity of diagonal
$K_c=K^*_c$ since for each country the size of imports is correlated
with the size of exports, even if trade is never exactly balanced and some
countries can sustain significant trade surplus or deficit.
The top $20$ list of top $K_2$ countries recover $13$  of 19 countries of
$G20$  major world economies (EU is the number 20) 
thus obtaining 68\% of the whole list.
This is close to the percent obtained in \cite{wtngoogle}
for trade in {\it all commodities}. The Google ranking 
for WNEA and WTN (top and bottom panels in Fig.~\ref{fig4})
gives different  positions for specific
countries (e.g. Russia improves its position
for WNEA with the opposite trend for China) 
but the global features of distributions
of WNEA and WTN remain similar corresponding to the same
economical forces.

\begin{figure}[!ht]
\begin{center}
\includegraphics[width=0.48\textwidth]{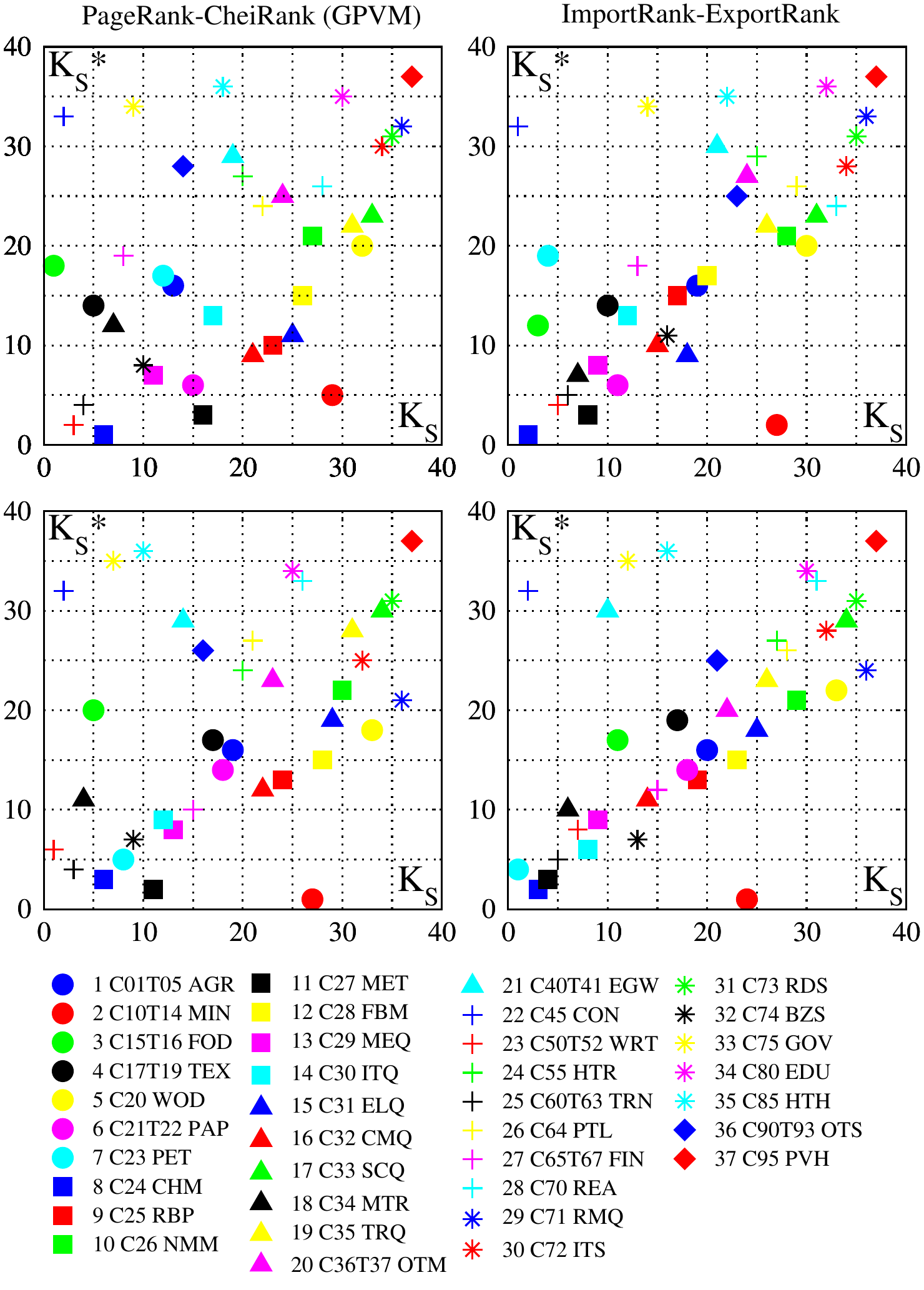}
\end{center}
\vglue -0.1cm
\caption{
Two dimensional ranking of sectors on the ($K_s$,$K^*_s$) plane using 
the GPVM approach for PageRank and CheiRank (left panels) and 
ImportRank-ExportRank (right panels). 
Each sector is represented by its specific combination of color and symbol. 
The list of all $37$ sectors are given in Table \ref{tab2}. 
Top panels show the case for the year 1995 and bottom panels for the year 2008.
}
\label{fig5}
\end{figure}

After tracing over countries we obtain the PageRank-CheiRank
plane of activity sectors shown in Fig.~\ref{fig5}.
We see that some sectors are export oriented
(e.g. $s=2$ {\it C10T14 Mining} at $K_s^*=1$ in 2008)
others are import oriented 
(e.g. $s=23$ {\it C50T52 World Retail and Trade of motors etc.} at $K_s=1$ in 2008).
The ImportRanking gives a rather different
import leader $s=7$ {\it C23 Manufacture of coke, refined petroleum products  etc.}
with $K_s=1$ in 2008. Thus the Google ranking highlights 
highly connected network nodes while Import-Export
gives preference to high value neglecting
existing network relations between various countries and activity sectors.
We can also order sectors by 2DRank index $K_2$ getting for PageRank-CheiRank
top sectors
$s=25, 23, 8$ at $K_2=1, 2, 3$
while Import-Export gives $s=8, 11, 14$
for top $K_2$ values in 2008
(more data are given at \cite{ourwebpage}).
We note that $s=25$ corresponds to Transport
which has many network connections
thus taking the top $K_2$ position.
We note that asymmetry of ranking of products
has been discussed in \cite{wtnproducts}
for COMTRADE data, however, the comparison with these data
is not so simple since the correspondence
between products and activity sectors is not straightforward.
Of course, for the WNEA the asymmetry of sector ranking
exists even for Export-Import ranking,
in a drastic difference from the WTN,
since there are interactions between activity sectors.

The global ranks of top 20 countries and their activities 
are given in Table~\ref{tab4} for 2008. The top 3 places 
of PageRank $K=1, 2, 3$ are 
taken by Germany ({\it Manufacture of motors etc.} $s=18$),
USA ({\it Public administration and defence} $s=33$), ROW (also $s=33$).
Thus imports of arms and weapons play a very important role.
In contrast for ImportRank  $\hat{K}=1, 2, 3$
we find rather different results with
USA (petroleum $s=7$), Japan (also $s=7$), and only then
USA ($s=33$). For CheiRank $K^* =1, 2, 3$
we find ROW, Russia, Saudi Arabia ($s=2$ {\it C10T14 Mining})
while for ExportRank we have
ROW, Saudi Arabia, Russia ($s=2$ {\it C10T14 Mining})
respectively. Thus Russia goes ahead of Saudi Arabia
due to a broad network of activity and trade connections
(a similar effect has been found 
in \cite{wtngoogle,wtnproducts} for trade in petroleum). 
The top 3 positions of 2DRank $K_2=1, 2, 3$
are taken by Germany ($s=8$ {\it Manufacture of chemicals etc.}).
USA ($s=27$ {\it Finance etc.}), Germany ($s=13$ {\it Manufacture of machinery etc.}).

\begin{figure}[!ht] 
\begin{center} 
\includegraphics[width=1\columnwidth,clip=true,trim=0 0 0 0cm]{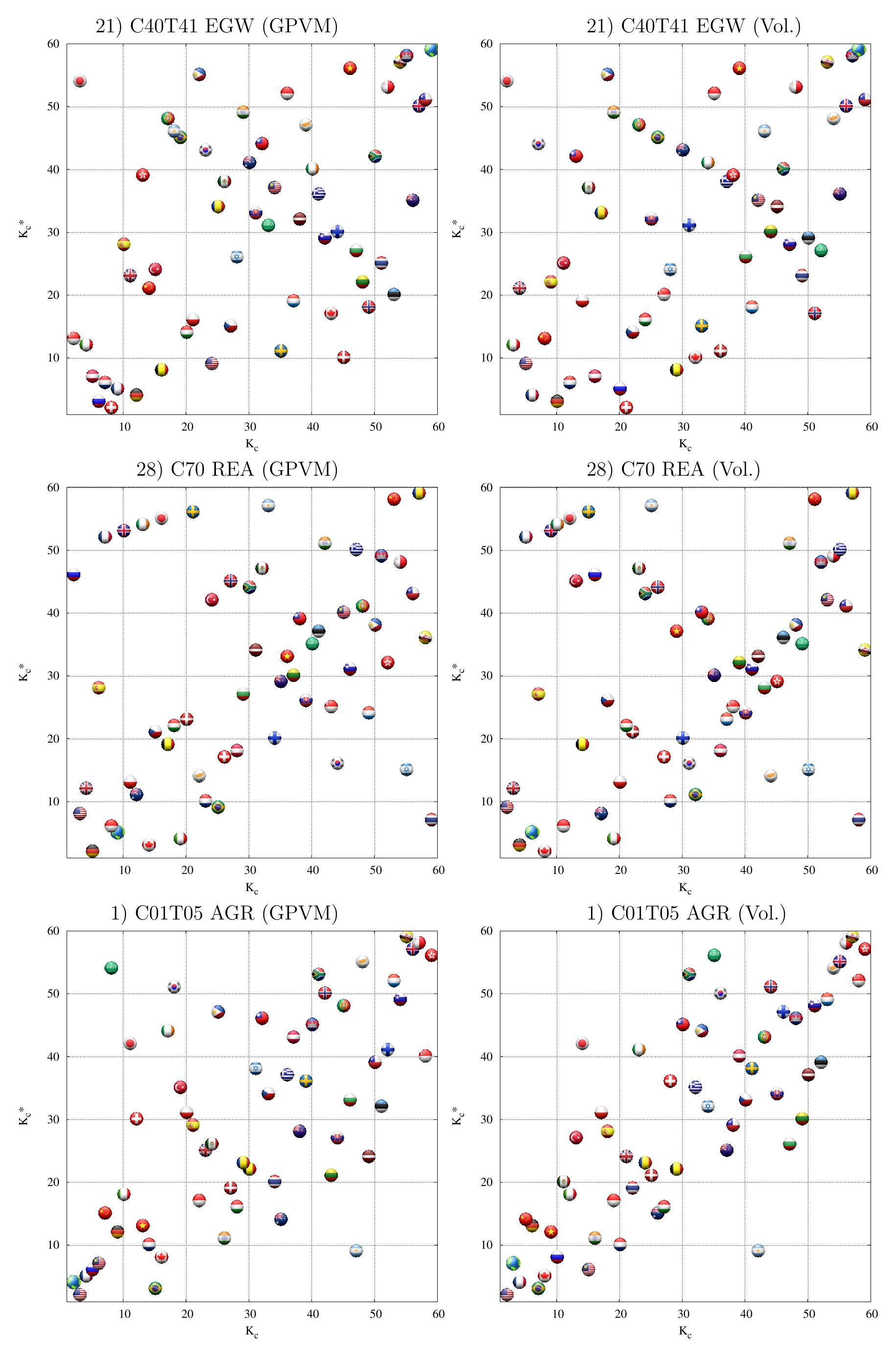}
\vglue -0.1cm
\caption{
Left column panels show results of the GPVM data for country positions on 
PageRank-CheiRank plane of local rank values $K_c, {K_c}^*$ ordered by ($K_{cs}, {K^*}_{cs}$) 
for specific sectors with $s=21$ (top), $s=28$ (center) and $s=1$ (bottom). 
Right column panels show the ImportRank-ExportRank planes respectively for comparison. 
Data are given for year 2008. Each country is shown by its own flag as in Fig \ref{fig4}.}
\label{fig6}
\end{center}
\end{figure}

\begin{figure}[!ht] 
\begin{center} 
\includegraphics[width=1\columnwidth,clip=true,trim=0 0 0 0cm]{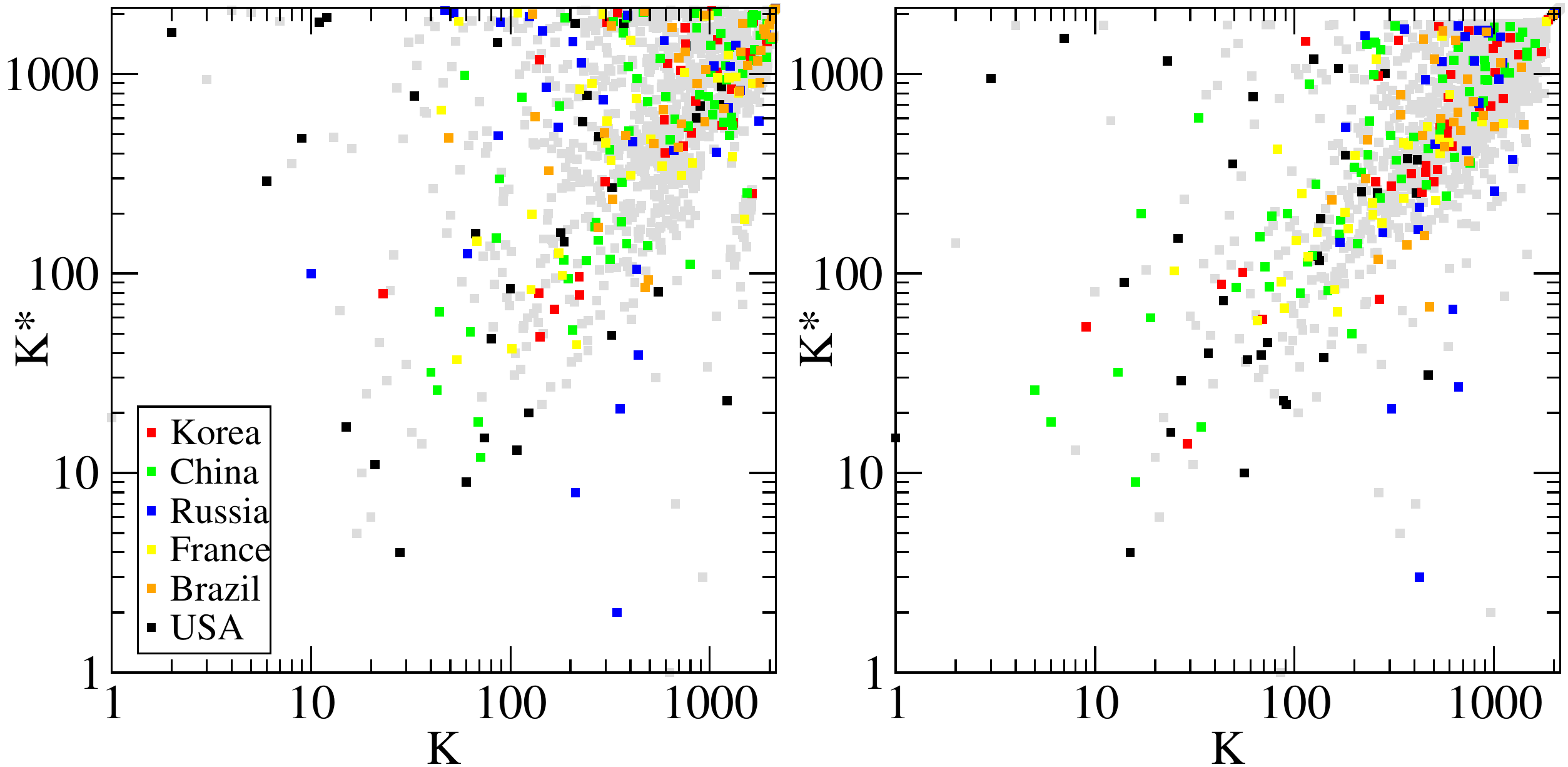}
\vglue -0.1cm
\caption {
Global plane of rank indexes ($K,K^*$) for PageRank-CheiRank (left panel) 
and ImportRank-ExportRank (right panel) for $N=2146$ nodes in year 2008. 
Each country and sector pair is represented by a gray square. 
Some countries are highlighted in colors : USA in black, 
South Korea in red, China (and Taiwan) in green, Russia in blue, 
France in yellow and Brazil in orange.}
\label{fig7}
\end{center}
\end{figure}

We can fix a certain activity sector $s$ 
and then consider local ranking of countries in $(K_c, {K_c}^*)$ plane.
Three examples are shown in Fig.~\ref{fig6}
for $s=21$ ({\it Electricity, gas, water}), 
$28$ ({\it Real estate activity}),
$1$ ({\it Agriculture}). The comparison of Google ranking (left column)
with value Import-Export ranking (right column)
shows  importance of network connections
highlighted by the GPVM, thus Russia moves from ${K_c}^*=4$
on right panel to $K_c^*=2$ on left panel for $s=21$
due to its broad links with Europe and Asia. 
For $s=1$ case in bottom panels of Fig.~\ref{fig6}
we find that the Import-Export ranking distribution is more
clse to diagonal comparing to the PageRank-CheiRank case
that we attribute to effect of indirect links present
in the later case.

The distribution of nodes on the global $(K,K^*)$ plane
is shown in Fig.~\ref{fig7} for Google ranking (left panel)
and Import-Export ranking (right panel) in 2008. 
The majority of countries are shown by gray squares
while 6 selected countries are marked by colors.
The comparison of two panels show that in the Google ranking
the positions of USA are improved (more black symbols at
top $K_2$ positions) while for China the positions (green symbols)
are weakened. We attribute this to a broader 
network connections of USA in important activity sectors
world wide (e.g. military activities and defense).

\subsection{Correlation properties of PageRank and CheiRank}

The directed networks can be characterized by the  correlator
$\kappa$  of PageRank and CheiRank 
vectors.  For various networks 
the properties of $\kappa$ are reported in \cite{linux,arxivrmp}. 
There are directed networks with small or even slightly negative
values of $\kappa$, e.g. Linux Kernel or Physical Review citation networks,
or with $\kappa \sim 4$ for Wikipedia networks
and even larger values $\kappa \approx 116$ for the Twitter network.

\begin{figure}[!ht] 
\begin{center} 
\includegraphics[width=1\columnwidth,clip=true,trim=0 0 0 0cm]{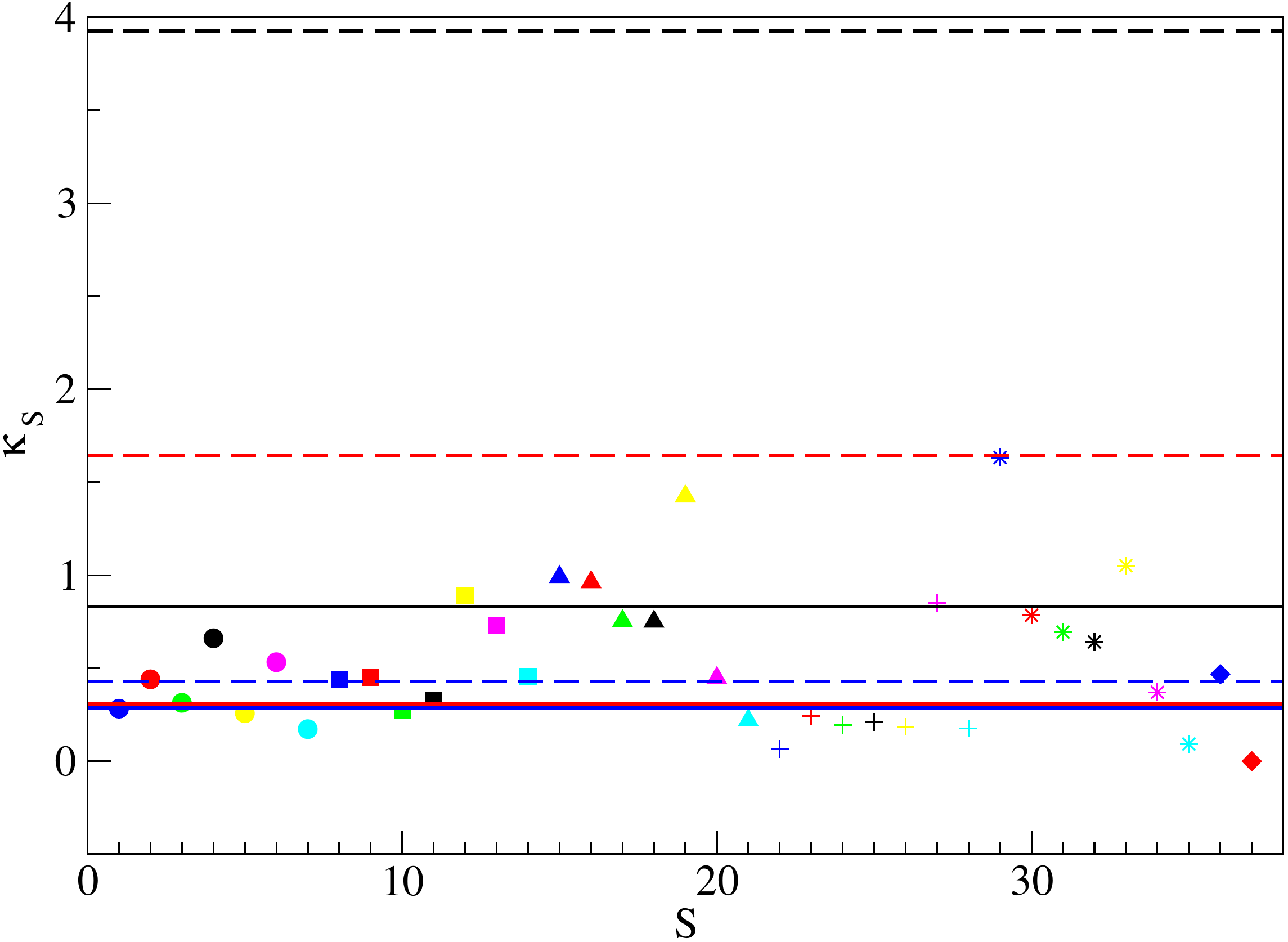}
\vglue -0.1cm
\caption {
PageRank-CheiRank correlators $\kappa_s$ from the GPVM 
(see (\ref{eq10}),  (\ref{eq11}))
are shown as a function of 
the sector index $s$ with the corresponding symbol from Fig.\ref{fig5}. 
PageRank-CheiRank and ImportRank-ExportRank correlators 
are shown by solid and dashed lines respectively, 
where the global correlator $\kappa$ (\ref{eq9}) is shown in black, 
the correlator for countries $\kappa(c)$ (\ref{eq12}) is shown by red lines, 
the correlators for sectors $\kappa(s)$ (\ref{eq12}) is shown by blue lines. 
Here sector index $s$ is counted in order of appearance in 
Table~\ref{tab2}. The data are given for year 2008 
with $N_s=37$, $N_c=58$, $N=2146$.}
\label{fig8}
\end{center}
\end{figure}

The correlators of WNEA for various sectors are shown in Fig.~\ref{fig7}. 
Almost all correlators $\kappa_s$ are positive being distributed in a range
$(0,1)$. A small negative value appears only for $s=37$ 
({\it Private households etc.})
corresponding to anti-correlation between buyers and sellers.
The largest correlator $\kappa_s$ is for $s=29$ 
({\it Renting of machinery etc.})
shows that sales of machinery correlates with their purchases
probably because components are needed to produce machines
produced by firms in the same industrial sectors.

The matrix of correlators between sectors $s,s'$ is shown in Fig.~\ref{fig8}
for years 1995, 2008. It is interesting to see a significant shift
of line of maximal correlators located in 1995 at $s'=28$ 
({\it Real estate activities}) to 
$s=29$ ({\it Renting of machinery etc.}) in 2008.
We also see that there are less correlations between sectors in 2008
compared to 1995. A further more detailed analysis of correlations 
would bring a better understanding of hidden inter-relations between
various sectors of economic activity.

\begin{figure}[!ht] 
\begin{center} 
\includegraphics[width=1\columnwidth,clip=true,trim=0 0 0 0cm]{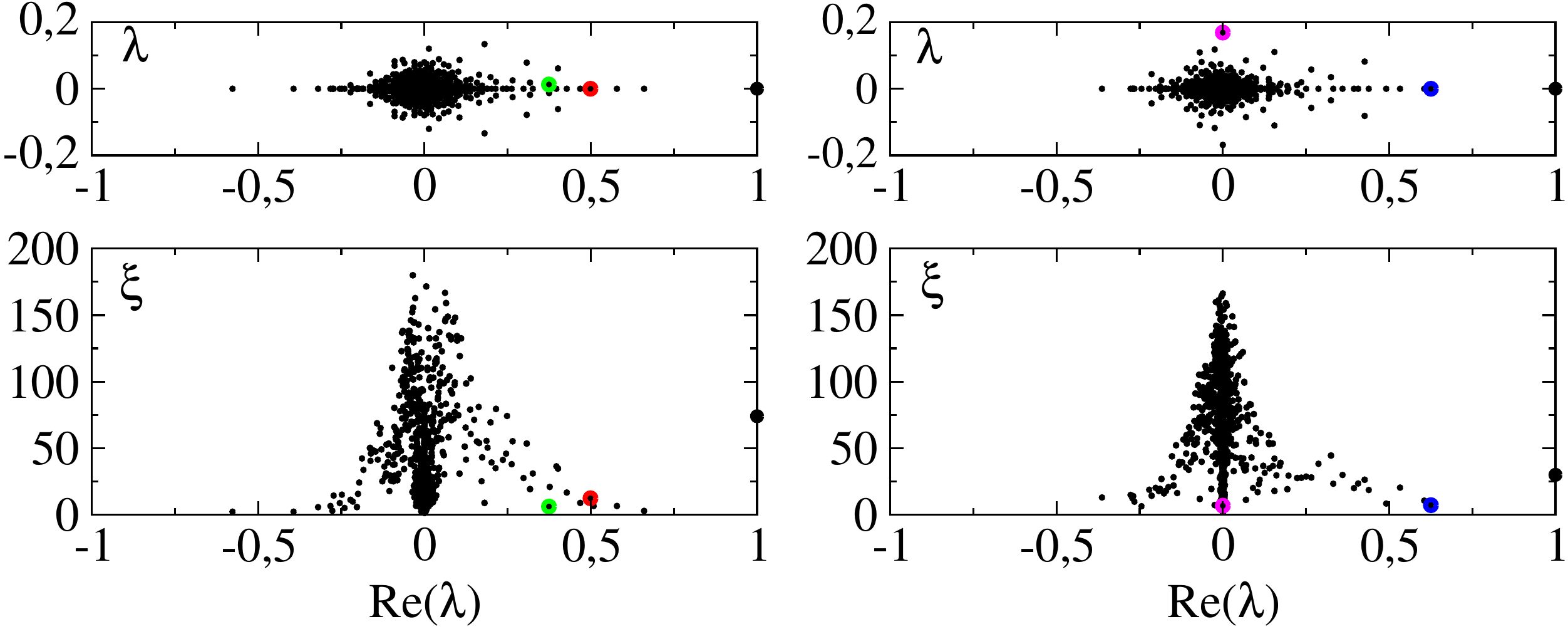}
\vglue -0.1cm
\caption {
\emph{Top panels:} Spectrum of Google matrices $G$ (left) and $G^*$ (right) 
represented in the complex plane of $\lambda$. The data are for year 2008 
with $\alpha=1$, $N=2146$, $N_c=58$, $N_s=37$.
 Four eigenvalues marked by colored circles 
are used for illustration of eigenstates 
in Fig. \ref{fig10} and Table~\ref{tab5}.
\emph{Bottom panels:} Inverse participation ratio (IPR) 
$\xi$ of all eigenstates 
of $G$ (left) and $G^*$ (right) as a function of the real part of 
the corresponding eigenvalue $\lambda$ from the spectrum above.}
\label{fig9}
\end{center}
\end{figure}

\subsection{Spectrum and eigenstates of WNEA Google matrix}

The results obtained for the Wikipedia network \cite{wikispectrum}
and the multiproduct WTN \cite{wtnproducts}
demonstrated that the eigenvectors of $G$ and $G^*$
with large eigenvalue modulus $|\lambda|$
select certain specific communities. Thus it is interesting
to analyze the properties of eigenvalues for the WNEA.
At $\alpha=1$ the gap between $\lambda=1$
and other eigenvalues characterize the rate of 
system relaxation to the equilibrium stationary 
PageRank state (for $G$). The presence of small
gap indicates that the mixing and relaxation 
in the system are developed only after many iterations of $G$ matrix
(see more discussion in \cite{arxivrmp}).

The matrix size of WNEA is relatively small and the whole
spectrum $\lambda$ of $G, G^*$ can be determined 
by direct matrix diagonalization.
The spectrum is shown in top panels of Fig.~\ref{fig9}. 
It is characterized by
a significant gap between $\lambda=1$ and other eigenvalues with 
$|\lambda| < 0.7$ at $\alpha=1$. We attribute this to 
a large number of inter-connected links between
matrix nodes (countries and sectors) which
is usually responsible for appearance of the spectral gap
(see \cite{physrev}, where the gap increases with the increase of number
of random links per node). We also note that the maximal value of
$|\mathit{Im} \lambda|< 0.2$ is relatively small
due to presence of links going in direct and inverse directions
between nodes. These features show that the relaxation
processes to the steady-state PageRank vector
are relatively rapid on the WNEA. Indeed, the relaxation
is governed by the exponent $\exp(-\Delta \lambda t)$
where $\Delta \lambda \approx 0.25$ the gap for for WNEA in Fig.~\ref{fig9}
and $t$ is number of iterations of $G$.

\begin{figure}[!ht] 
\begin{center} 
\includegraphics[width=1\columnwidth,clip=true,trim=0 0 0 0cm]{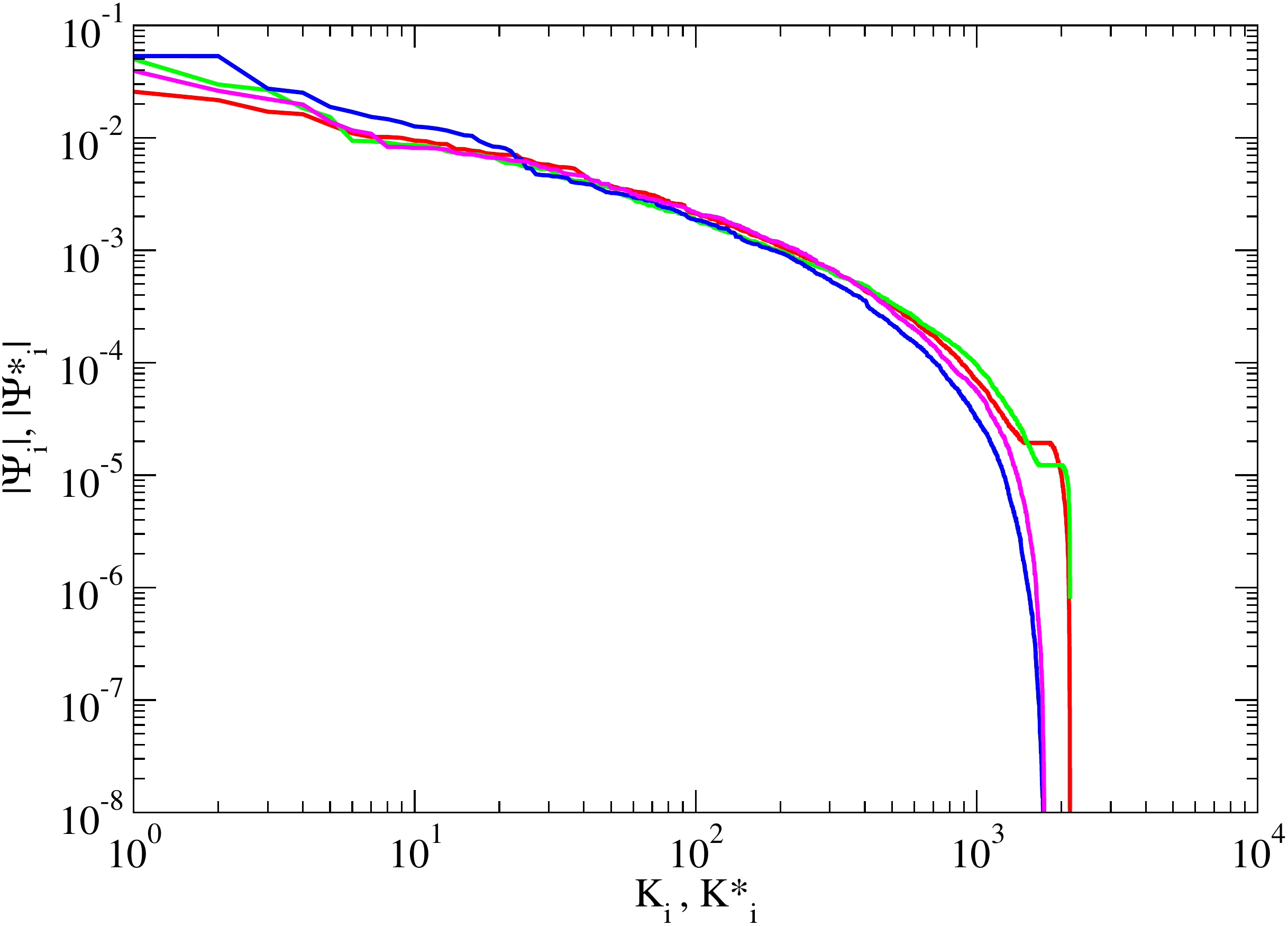}
\vglue -0.1cm
\caption {
Eigenstates amplitudes $|\psi_i|$ ordered by its own decreasing amplitude order 
with local rank index $K_i$ for 4 different eigenvalues of Fig. \ref{fig9} 
(states are normalized as $\sum_i |\psi_i|=1$). The four examples are 
$\lambda=0.4993$ (red), $\lambda=0.3746+0.0126i$ (green), $\lambda=0.6256$ (blue) and 
$\lambda=-0.0001+0.1687i$ (magenta). Node names (country, sector) for top ten largest 
amplitudes of these eigenvectors are shown in Table \ref{tab5}.}
\label{fig10}
\end{center}
\end{figure}

The properties of eigenstates are characterized
by the IPR $\xi$ shown in bottom panels of Fig.~\ref{fig9}.
We find that the main part of states have $\xi \ll N$
so that they occupy only a small fraction of nodes
corresponding to localized states
(see discussion about the Anderson localization
of Google matrix eigenstates in \cite{arxivrmp,zsanderson}). 

The dependence of amplitudes $|\psi_i|$ of a few eigenstates,
ordered by a local rank index $K_i$ corresponding to a monotonic
amplitude decrease, are shown in Fig.~\ref{fig10}. 
The names of top 10 nodes of these eigenstates are given 
in Table~\ref{tab5}. The red curve in Fig.~\ref{fig10} 
selects mainly the sector $s=4$ ({\it Manufacture of textiles etc.})
with close links between China, Italy, USA and ROW;
the green one selects $s=18$ ({\it Manufacture of motor vehicles etc.})
with close links between Argentina, Brasil, Japan and Germany;
the blue state corresponds to $s=16$ 
({\it Manufacture of radio, television and communication equipment and apparatus})
 in the Asian region (China, Korea, Chinese Taipei, Singapore, Malaysia);
the magenta state represents sector $s=2$ ({\it Mining etc.})
with related countries like Russia, Saudi Arabia,
ROW, Norway. These results coincide with the previous observations 
for Wikipedia-type network \cite{wikispectrum}
that the eigenstates of $G$ and $G^*$ select specific communities of the network
nodes. Similar properties of eigenstates of $G$ of the multiproduct
WTN have been found in \cite{wtnproducts}.

\subsection{Sensitivity to price variations}

The ranking of WNEA nodes provides interesting and important information.
In addition,
the established matrix structure of $G, G^*$ of WNEA also
allows to study the sensitivity of
the world economic activities to price variations.
There are certain parallels with the multiproduct WTN
analyzed in \cite{wtnproducts} but there are also new
elements specific to the WNEA.

\begin{figure}[!ht] 
\begin{center} 
\includegraphics[width=1\columnwidth,clip=true,trim=0 0 0 0cm]{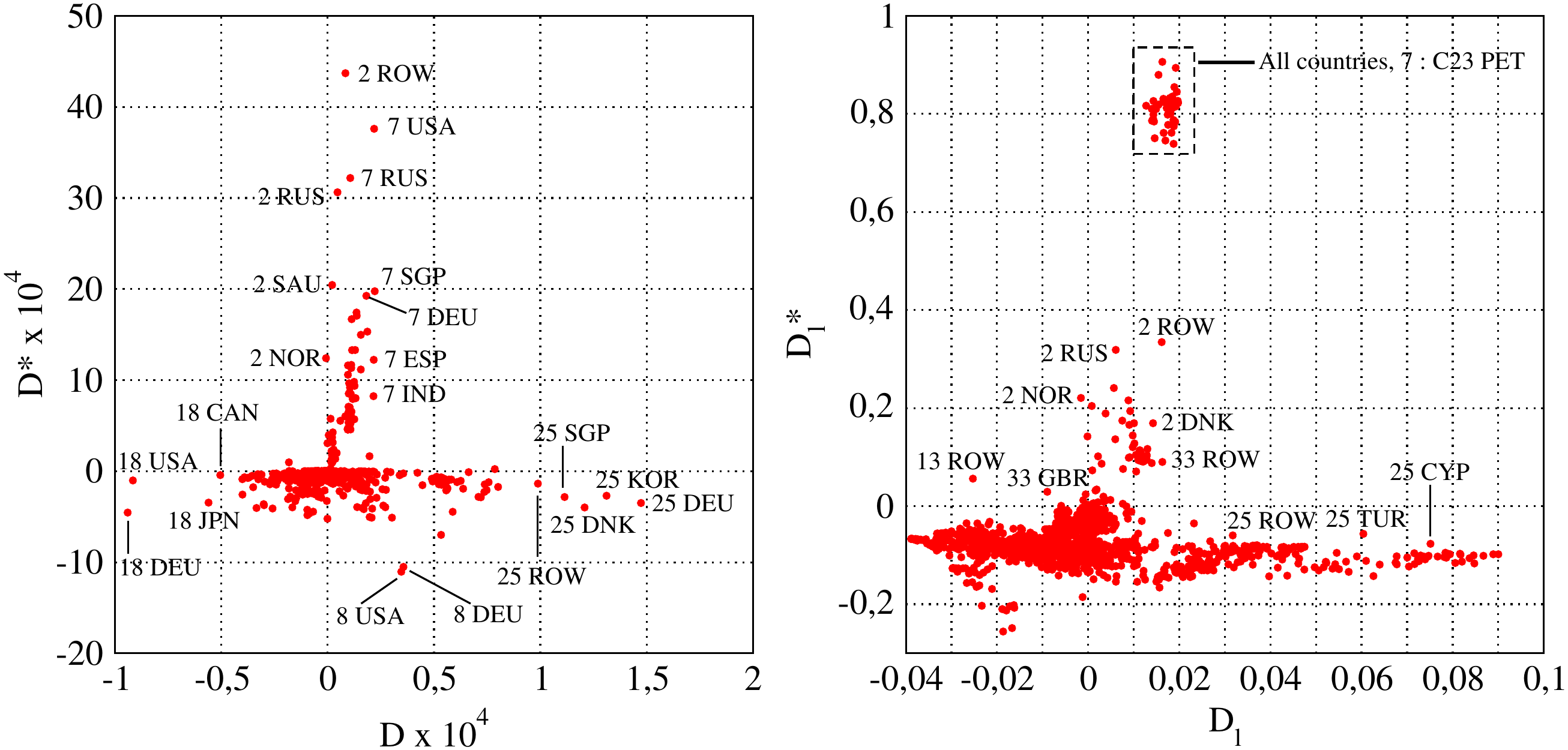}
\vglue -0.1cm
\caption {
\emph{Left panel:} Derivatives $D=dP/d\delta_7$ and $D^*=dP^*/d\delta_7$ 
for a price variation $\delta_7$ of 
\emph{7 C23 PET (Manufacture of coke, refined petroleum products and nuclear fuel)} 
for year 2008. 
\emph{Right panel:} Logarithmic derivatives $D_l=D/P$ and 
$D^*_l=D^*/P^*$ for the same case as left panel. Codes 
in panels give sector number $s=1,...37$ described in Table \ref{tab2},
country codes are from Table~\ref{tab1}.
The group of points, highlighted by the dashed box, represents $58$ nodes of 
the form $(country,s=7)$ where $s=7$ is 
\emph{C23 PET (Manufacture of coke, refined petroleum products and nuclear fuel)}.}
\label{fig11}
\end{center}
\end{figure}

To analyze the sensitivity of price variation in a certain activity sector $s$
we increase from $1$ to $1+\delta_s$  the  money transfer
in the sector $s$ in $M_{cc\, ss'}$ in (\ref{eq1}), where $\delta_s$
is a dimensionless fraction variation of price in this sector.
After that the matrices $G, G^*$ are recomputed in the usual way described above
and their rank probabilities $P, P^*$ are determined.
Then we compute the derivatives of probabilities
of PageRank $D=dP/d\delta_s = \Delta P/\delta_s$ and CheiRank 
$D^*=dP^*/d\delta_s= \Delta P^*/\delta_s$. We do these computations at sufficiently small
$\delta_s$ values checking that the variations of $P, P^*$ are linear in
$\delta_s$. In addition we also compute the logarithmic derivatives
$D_l= d \ln P/d\delta_s$, $D^*_l = d \ln P^*/d\delta_s$
which give us relative changes of $P$, $P^*$.

The sensitivities to price of $s=7$ ({\it Manufacture of
coke, refined petroleum products and nuclear fuel}) are shown in Fig.~\ref{fig11}.
The data for $D,D^*$ in the left panel show a rather complex picture with a significant
derivatives not only for $s=7$ but also
for countries with sectors: $s=18$ 
({\it Manufacture of motor vehicles, trailers and semi-trailers})
at strongly negative $D$ for Germany. USA, Japan;
$s=25$ ({\it Land transport; transport via pipelines etc}) 
at significant positive $D$ for Germany. Korea, Denmark, Singapore;
of course,  for $s=7$ we have positive $D^*$,
but also for $s=2$ related to mining and
negative $D^*$ for $s=8$ ({\it Manufacture of chemicals and chemical products})
for USA and Germany. The logarithmic derivatives
provide strong relative changes and are shown in the right panel of Fig.~\ref{fig11}.

\begin{figure}[!ht] 
\begin{center} 
\includegraphics[width=1\columnwidth,clip=true,trim=0 0 0 0cm]{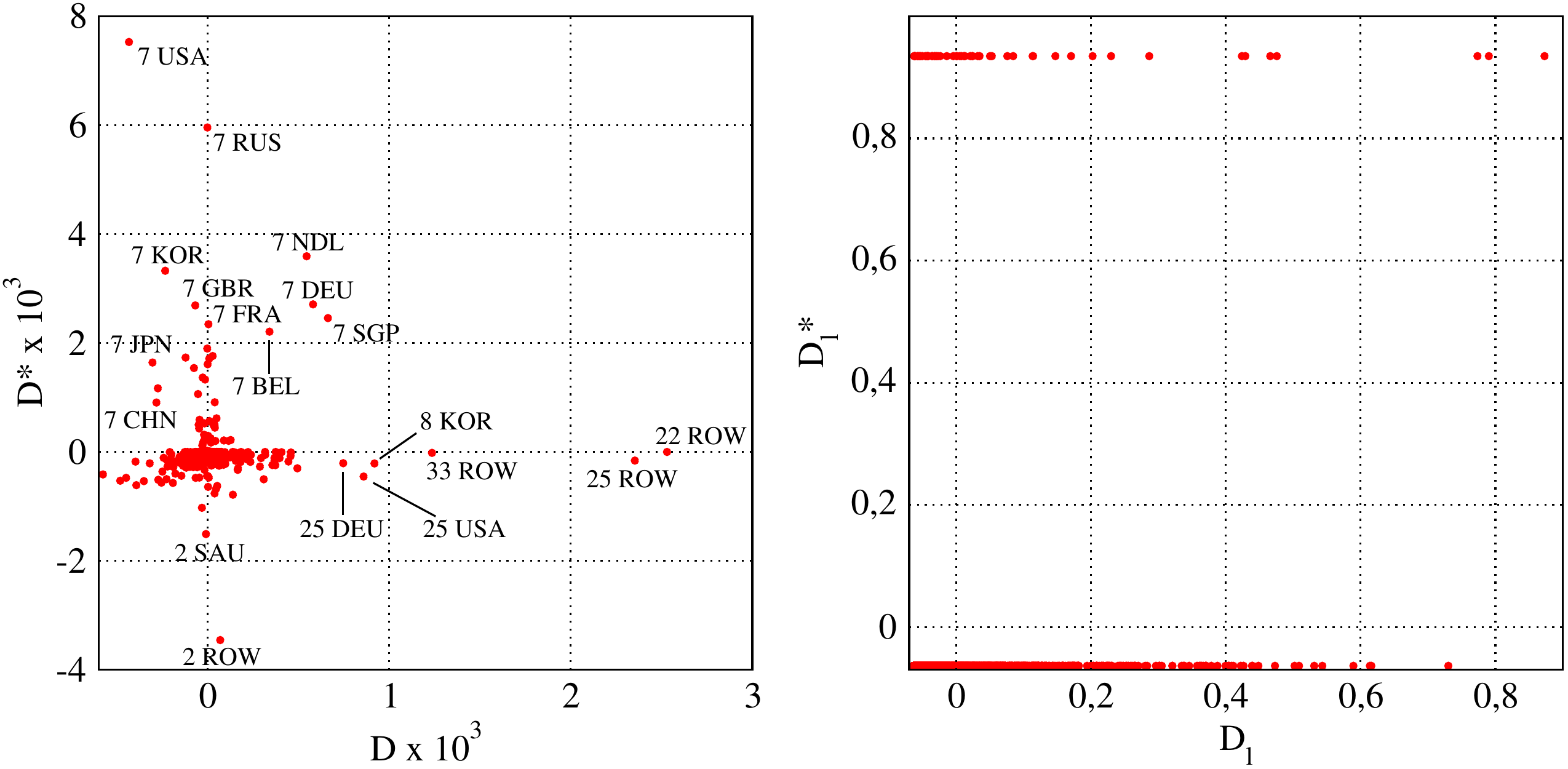}
\vglue -0.1cm
\caption {
Same as the left panel of Fig. \ref{fig11} but using probabilities from the trade value. 
In the right panel, $D^*_l=0.9348$ if $s=s'$ and $D^*_l=-0.0633$ if $s \neq s'$.
}
\label{fig12}
\end{center}
\end{figure}

A similar analysis can be done
using the probabilities $\hat{P},\hat{P}^*$ from the exchange value probabilities
(\ref{eq3}) instead of the above PageRank and CheiRank probabilities.
The results for the value probabilities are presented in Fig.~\ref{fig12}
for the same case as in Fig.~\ref{fig11}. We see that the results are drastically
different especially for the logarithmic derivatives $D_l, D^*_l$.
In fact $D_l, D^*_l$ cannot give correct picture of sensitivity
to price variations since for the monetary exchange the network links between nodes
are not taken into account and there is only
a mechanical re-computation of the value normalization.
A similar situation appears also for the multiproduct WTN \cite{wtnproducts}.
Thus we see from Fig.~\ref{fig11} and Fig.~\ref{fig12}
that the Google matrix approach provides new elements
for the economic activity analysis going significantly beyond the usual
consideration of Import-Export method.

The new element of the WNEA, compared to the multiproduct WTN,
is existence  of transfers between sectors of the same economy. This allows us to
consider the sensitivity not only to sectoral prices but also 
 the sensitivity to labor cost in a given country $c$
(e.g. price shock affecting all industries in the same country).
This can be taken into account by the introduction of
the dimensionless labor cost change in a given country $c$
by replacing the related monetary flows from coefficient $1$
to $1+\sigma_c$ in $M_{c c', s s;}$ (\ref{eq1}) for a selected country $c$. 

Of course, the above derivatives over price of activity sector and labor country cost
give only an approximate consideration of effects of price variations
which is a very complex phenomenon. 
For an economic discussion of the effect of
price shocks on international production networks we address a reader to
the research performed in \cite{escaith}.
We will see below that our
approach gives  results being in a good agreement with economic realities
thus opening complementary possibilities
of economic activity analysis based on the underlying
network relations between countries and activity sectors
which are absent in the usual Import-Export consideration.
We present the results on sensitivity to sector prices and labor cost in next
subsections.

\subsection{Price shocks and trade balance sensitivity}

On the basis of the obtained WNEA Google matrix we can now analyze the trade balance
in various activity sectors for all world countries.
Usually economists consider the export and import of a given country 
as it is shown in Fig.~\ref{fig1}.  Then the trade balance of a given country $c$
can be defined making summation over all sectors:  
\begin{equation}
B_c=\sum_s (P^*_{cs} - P_{cs})/\sum_s (P^*_{cs} + P_{cs}) = (P^*_{c} - P_{c})/(P^*_{c} + P_{c}) .
\label{eq13} 
\end{equation}
In economy, $P_c, P^*_c$ are defined via the probabilities of trade value
$\hat{P}_{cs}, \hat{P}^*_{cs}$  from (\ref{eq3}). In our matrix approach,
we define $P_{cs}, P^*_{cs}$ as PageRank and CheiRank probabilities. 
In contrast to the Import-Export value our approach takes into account the 
multiple network links between nodes. 

\begin{figure}[!ht] 
\begin{center} 
\includegraphics[width=1\columnwidth,clip=true,trim=0 0 0 0cm]{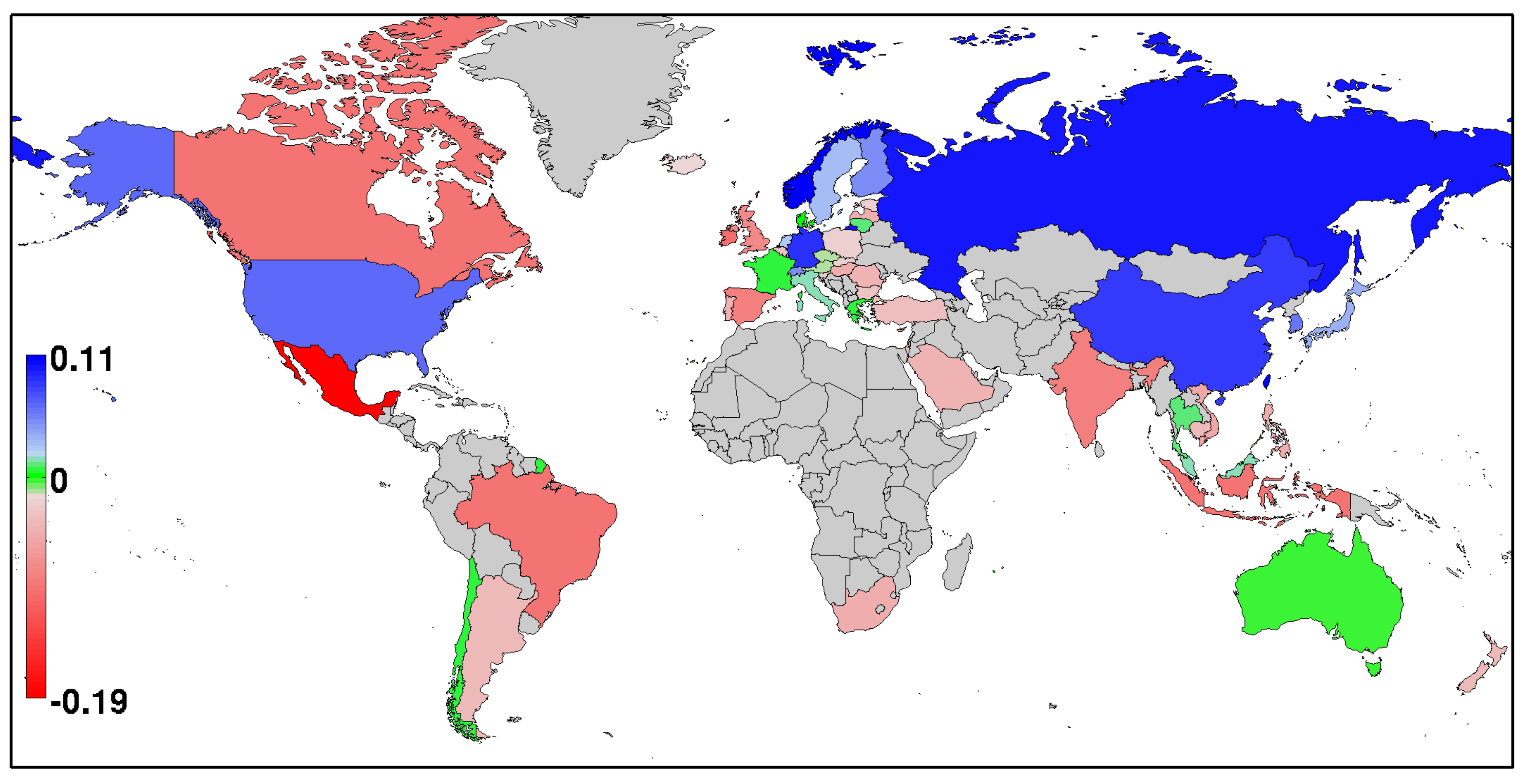} \\
\includegraphics[width=1\columnwidth,clip=true,trim=0 0 0 0cm]{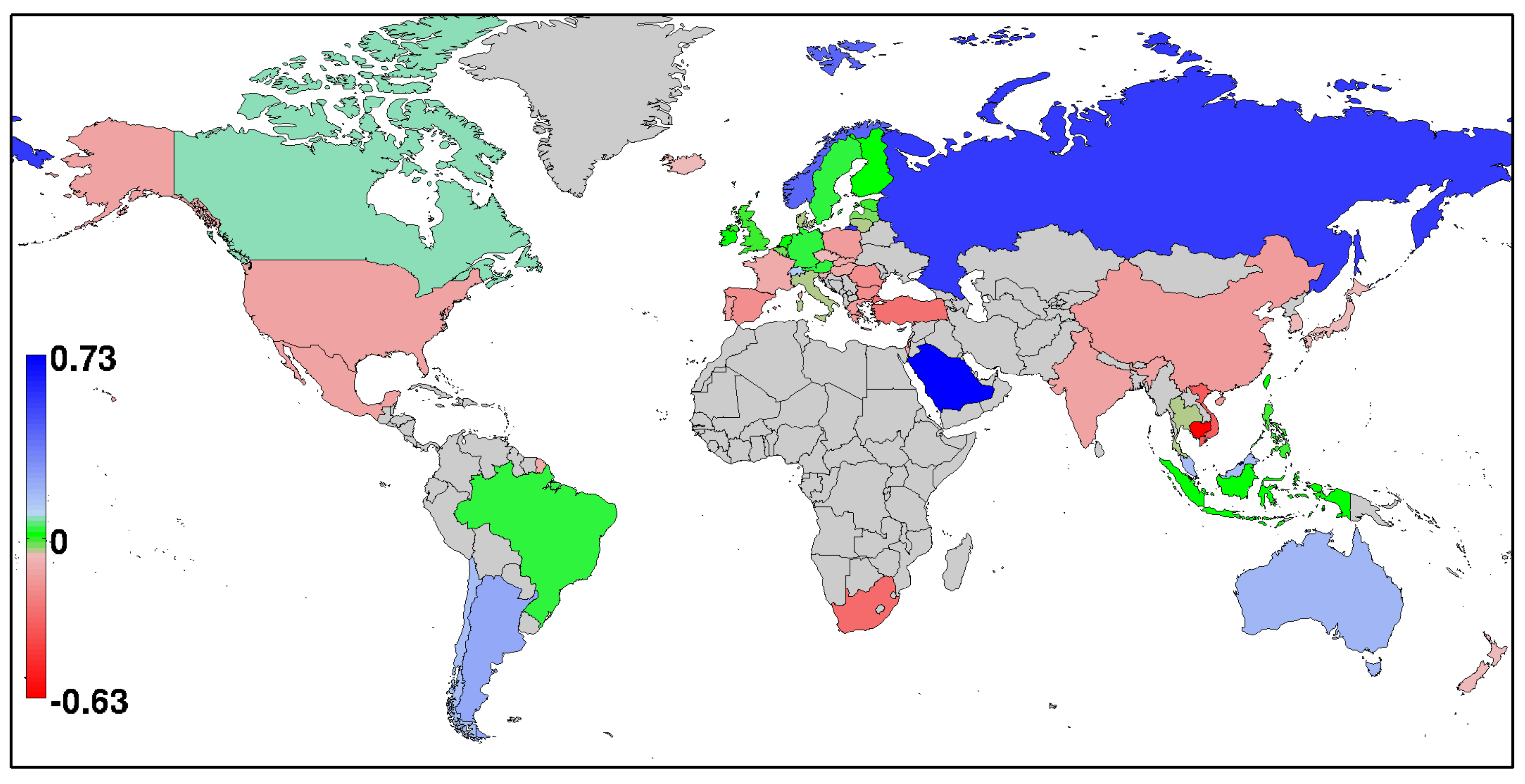}
\vglue -0.1cm
\caption {World map of CheiRank-PageRank balance $B_c=(P^*_c-P_c)/(P^*_c+P_c)$ 
determined for all ${N_c}=58$ countries in year 2008. Top panel shows 
the probabilities $P$ and $P^*$ given by PageRank and CheiRank vectors; 
the value of ROW group is $B_{c=58}=0.023$. Bottom panel shows 
the probabilities $P$ and $P^*$ computed from 
the  Export and Import value; the value of ROW group is $B_{c=58}=0.16$. 
Names of the countries are given in Table \ref{tab1} and 
in the world map of countries \cite{worldmap}.}
\label{fig13}
\end{center}
\end{figure}

The comparison of CheiRank-PageRank balance with
Export-Import balance for the world countries shown in
Fig.~\ref{fig13} for year 2008. Each country is shown by color 
whcih is proportional to the country balance $B_C$ (\ref{eq13})
with the color bar given on the figure.
For Export-Import balance we see the dominance of petroleum
producing countries Saudi Arabia, Russia, Norway
with the largest values.
The $\;\;\;\;$ CheiRank-PageRank balance highlights new features
placing on the top Russia, Norway, 
Germany, China. In fact, USA has now a slightly positive balance 
in top panel of Fig.~\ref{fig13}) while it was negative before
in bottom panel of same figure.
We see that the broad network of economic activity relations and links
makes the economies of the above countries more important in the
world economy while Saudi Arabia, with the largest positive
Export-Import balance, looses its leading position.
Indeed, the trade of this country is mainly oriented to USA
and nearby countries that reduces its importance for
world economy (a similar effect has been observed with 
COMTRADE data \cite{wtngoogle,wtnproducts}). 

\begin{figure}[!ht] 
\begin{center} 
\includegraphics[width=1\columnwidth,clip=true,trim=0 0 0 0cm]{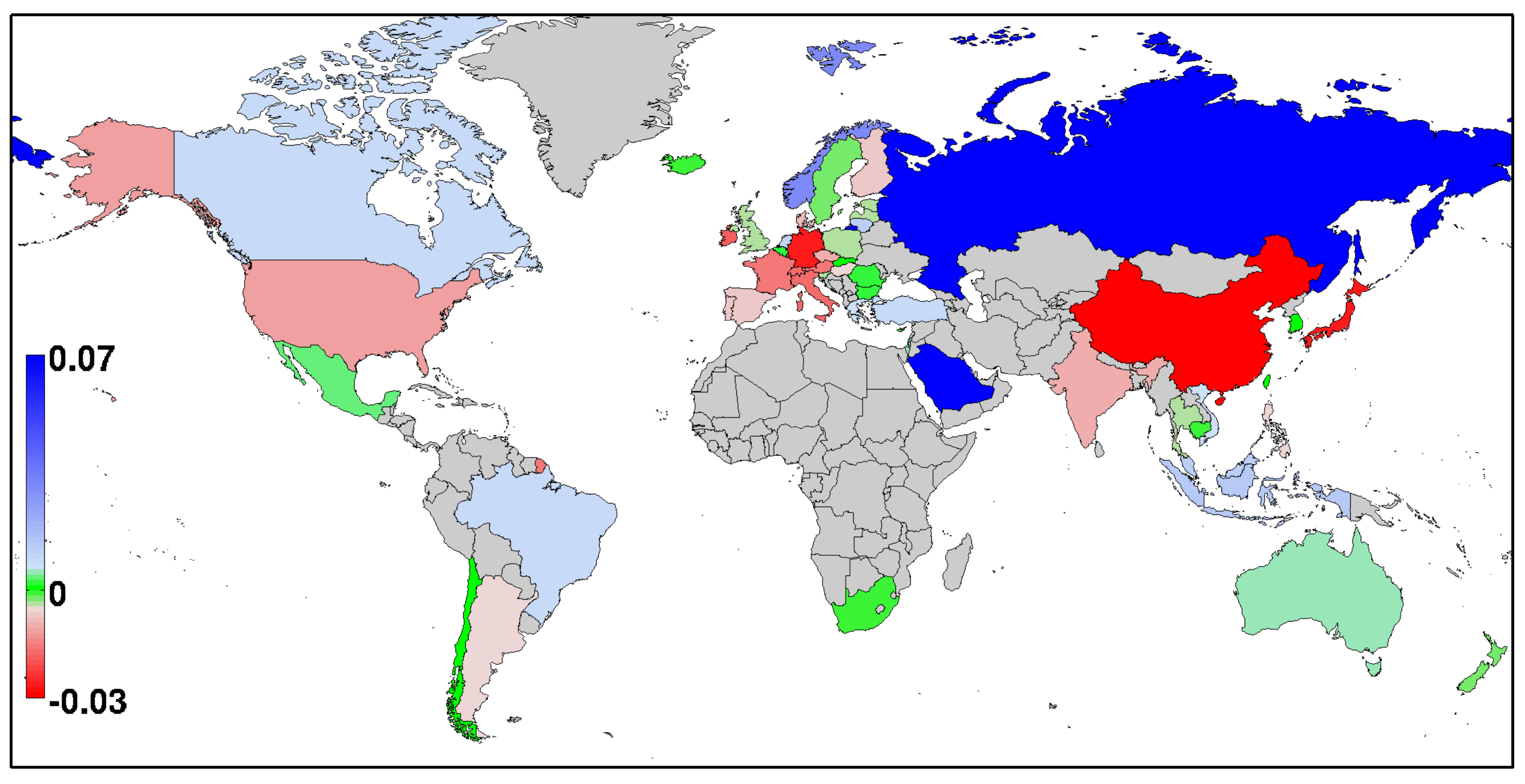} \\
\includegraphics[width=1\columnwidth,clip=true,trim=0 0 0 0cm]{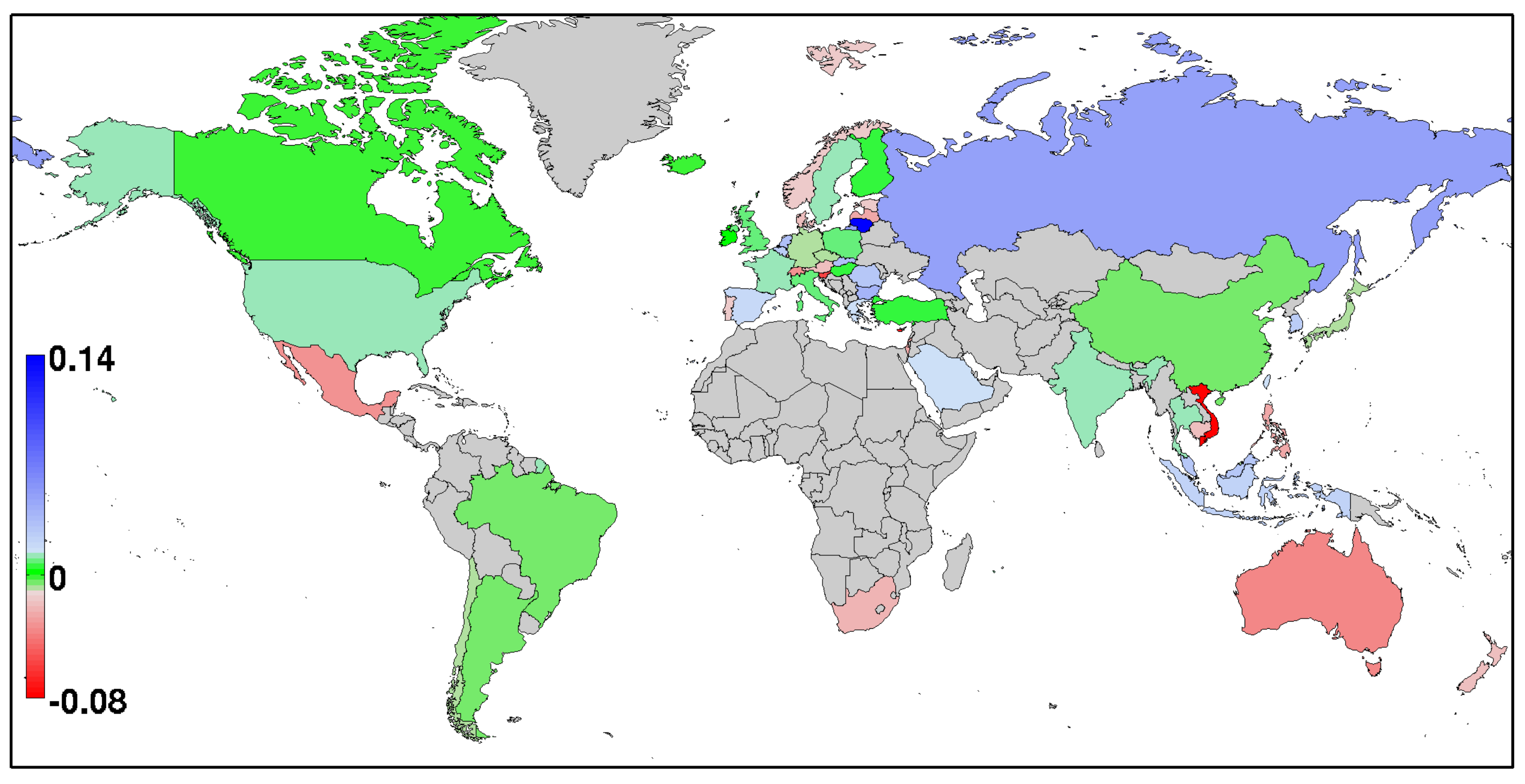}
\vglue -0.1cm
\caption {
Derivative of probabilities balance $dB_c/d\delta_7$ over price of sector
$s=7$ C23PET 
for year 2008. Top panel shows the case when $B_c$ is determined by 
CheiRank and PageRank vectors as in the top panel of Fig.\ref{fig13}; 
the value of ROW group is $dB_{58}/d\delta_7=0.04$. Bottom panel shows the case 
when $B_c$ is computed from the Export-Import value as in the bottom panel of 
Fig.\ref{fig13}; the value of ROW group is $dB_{58}/d\delta_7=-0.07$. 
Names of the countries can be found in Table \ref{tab1} and
in the world map of countries \cite{worldmap}.}
\label{fig14}
\end{center}
\end{figure}

The sensitivity of country balance $d B_c/d \delta_7$
to price variation of sector $s=7$ 
{\it Manufacture of coke, refined petroleum products and nuclear fuel}
is shown in Fig.~\ref{fig14}. For Export-Import in bottom panel
the most sensitive
countries are  Lithuania (positive)
and Vietnam (negative).
Lithuania does not produce petroleum,
but in fact in 2008 there was a  
large oil refinery company there which had a large exportation value
(see e.g. \\
http://en.wikipedia.org/wiki/Economy\_of\_Lithuania).
The Export-Import approach shows that Russia is slightly positive,
even less positive is Saudi Arabia,
China and Germany are close to zero change, USA is only
very slightly positive.
The results of CheiRank-PageRank sensitivity (top panel)
are significantly different
showing strongly positive sensitivity for Saudi Arabia, Russia
and strongly negative sensitivity for China, Germany and Japan;
USA goes from slightly positive side in bottom panel to moderate negative one
in top panel. The CheiRank-PageRank balance 
demonstrates much higher sensitivity of Russia, Saudi Arabia
and China to price variations of $s=7$ sector
comparing to the case of Export-Import value analysis.
The economies of Germany, China and Japan are also very 
sensitive to petroleum prices that is correctly captured by our
analysis.
We consider that the CheiRank-PageRank
approach describes the economic reality 
from a new complementary angle and 
that provides new useful information about
complex trade systems.
 We also note that the highly negative sensitivity
of China to $\;\;\;$ petroleum prices has been also obtained
on the basis of Google matrix analysis of COMTRADE data 
(see Fig.21 in \cite{wtnproducts}).

\begin{figure}[!ht] 
\begin{center} 
\includegraphics[width=1\columnwidth,clip=true,trim=0 0 0 0cm]{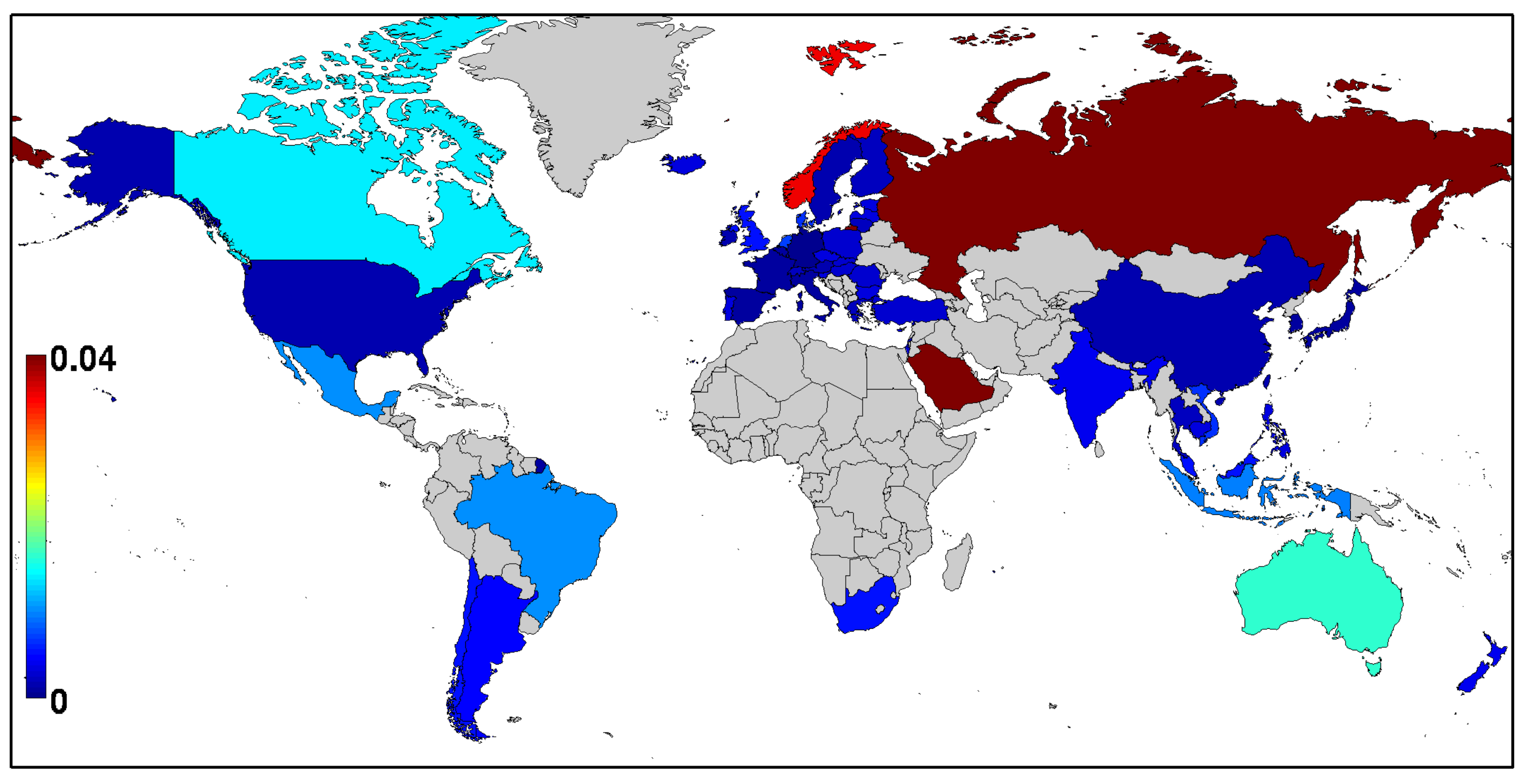} \\
\includegraphics[width=1\columnwidth,clip=true,trim=0 0 0 0cm]{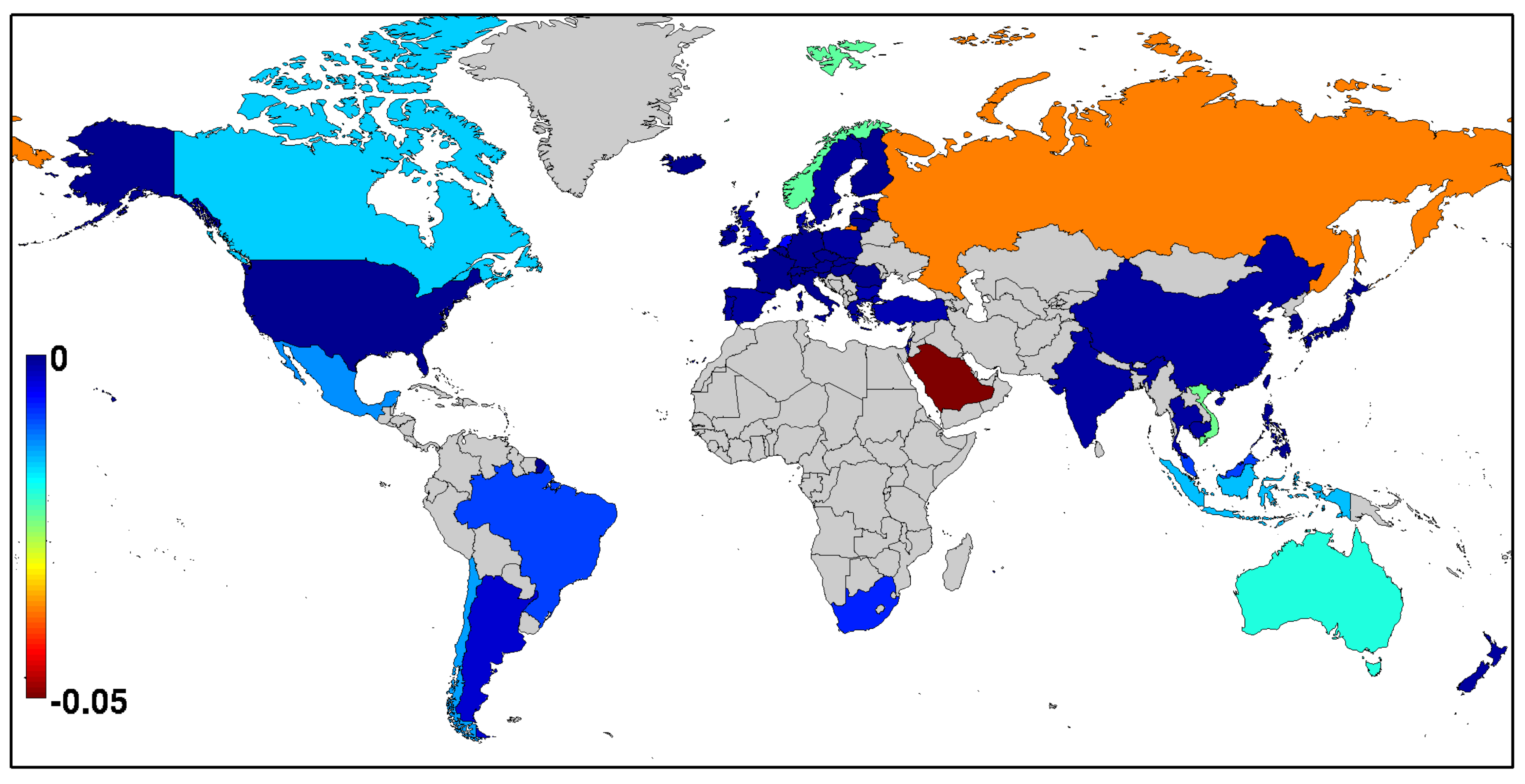}
\vglue -0.1cm
\caption {
Derivative of partial probability balance of sector $s$ defined as 
$dB_{cs}/d\delta_{s'}$ over sector $s'=7$ C23PET price $\delta_7$ for year 2008. 
Here $B_{cs}=(P^*_{cs}-P_{cs})/(P^*_c+P_c)$ and $s=2$ 
\emph{(C10T14MIN, Mining, extraction,...)} from Table \ref{tab2}. 
The sector balance sensitivity of countries $B_{cs}$ is determined from CheiRank and
 PageRank vectors (top panel) and from the exchange value of Export-Import (bottom panel); 
the values of ROW group are $dB_{58,2}/d\delta_7=0.05$ and $dB_{58,2}/d\delta_7=-0.03$ 
respectively. Names of the countries can be found at Table \ref{tab1} and
in the world map of countries \cite{worldmap}.}
\label{fig15}
\end{center}
\end{figure}

It is also possible to determine the cross-sensitivity
of activity sectors to price variation.
For that we determine 
the partial exchange balance for a given sector $s$ defined as
\begin{equation}
B_{cs}=(P^*_{cs} - P_{cs})/\sum_s (P^*_{cs} + P_{cs}) = 
(P^*_{cs} - P_{cs})/(P^*_{c} + P_{c}) ,
\label{eq14} 
\end{equation}
so that the global country balance is $B_c=\sum_s B_{cs}$.
Then the sensitivity of partial balance of a given sector $s$
in respect to a price variation of a sector $s^{\prime}$
is given by the derivative
$d B_{cs}/ d \delta_{s^{\prime}}$. 
The results for $s=2, s'=7$ are shown in Fig.~\ref{fig15}.
We see that two methods give results with even opposite signs.
According to the Google matrix analysis the increase of petroleum prices
stimulates development of mining while for the Export-Import approach the
result is the opposite. In our opinion, the absence of links and
next step relations between countries and sectors in the Export-Import
methods does not allow to take into account all complexity of economy
relations. In contrast
the CheiRank-PageRank approach captures effects of  all links
providing more advanced indications.

\begin{figure}[!ht] 
\begin{center} 
\includegraphics[width=1\columnwidth,clip=true,trim=0 0 0 0cm]{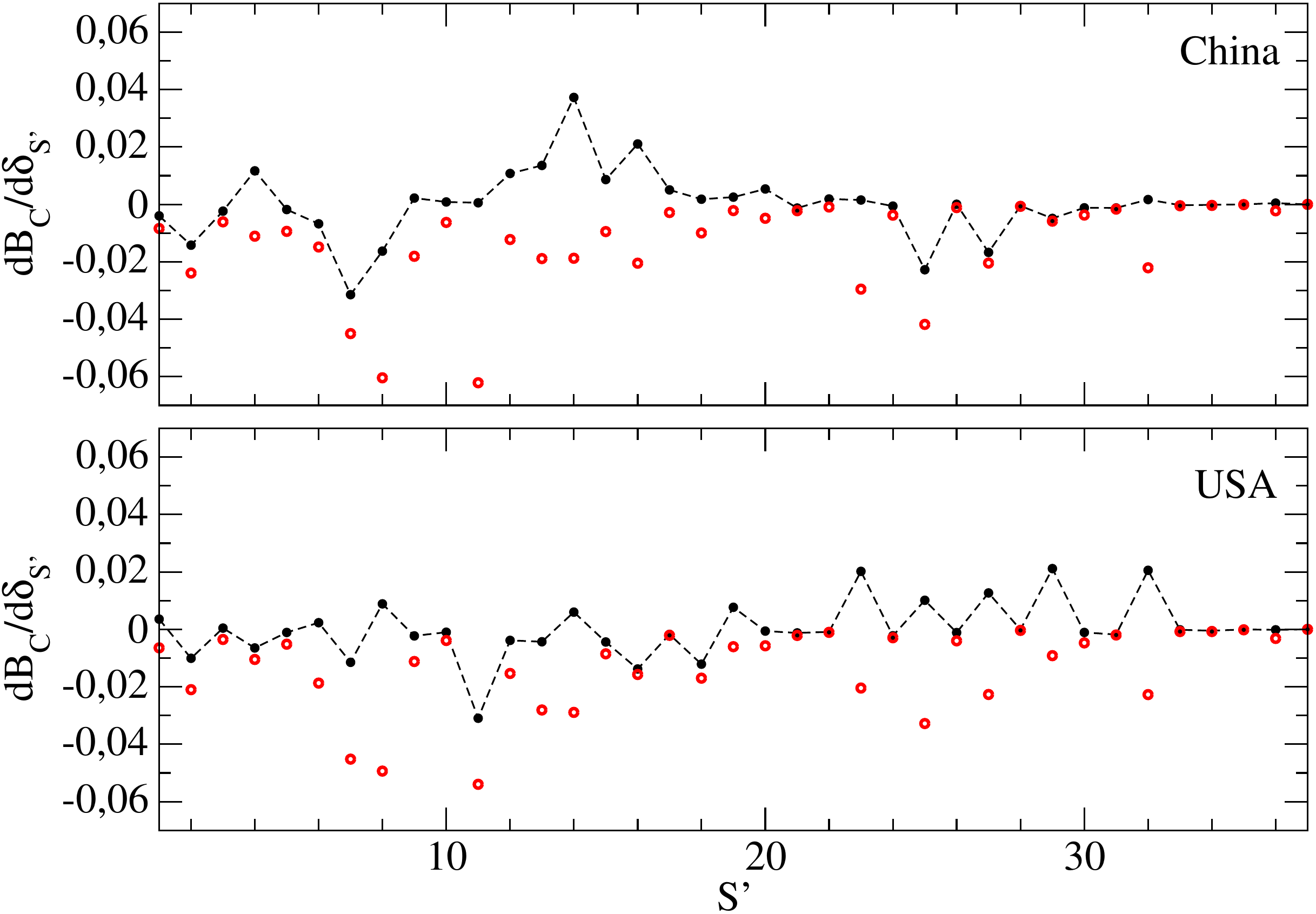}
\vglue -0.1cm
\caption {
Top (China) and bottom (USA) panels show derivative $dB_c/d\delta_{s'}$ 
of country total probability balance $B_c$ over price $\delta_{s'}$ 
of sector $s'$ for year 2008 (black points connected by dashed line); 
derivatives of balance without diagonal term 
($dB_c/d\delta_{s'} - dB_{cs'}/d\delta_{s'}$) are represented by open red circles. 
The sector balance of countries $B_{cs}$ and $B_c$ are determined 
from CheiRank and PageRank vectors. The sectors corresponding 
to sector index $s$ or $s'$ are listed in Table\ref{tab2}.
}
\label{fig16}
\end{center}
\end{figure}

\begin{figure}[!ht] 
\begin{center} 
\includegraphics[width=1\columnwidth,clip=true,trim=0 0 0 0cm]{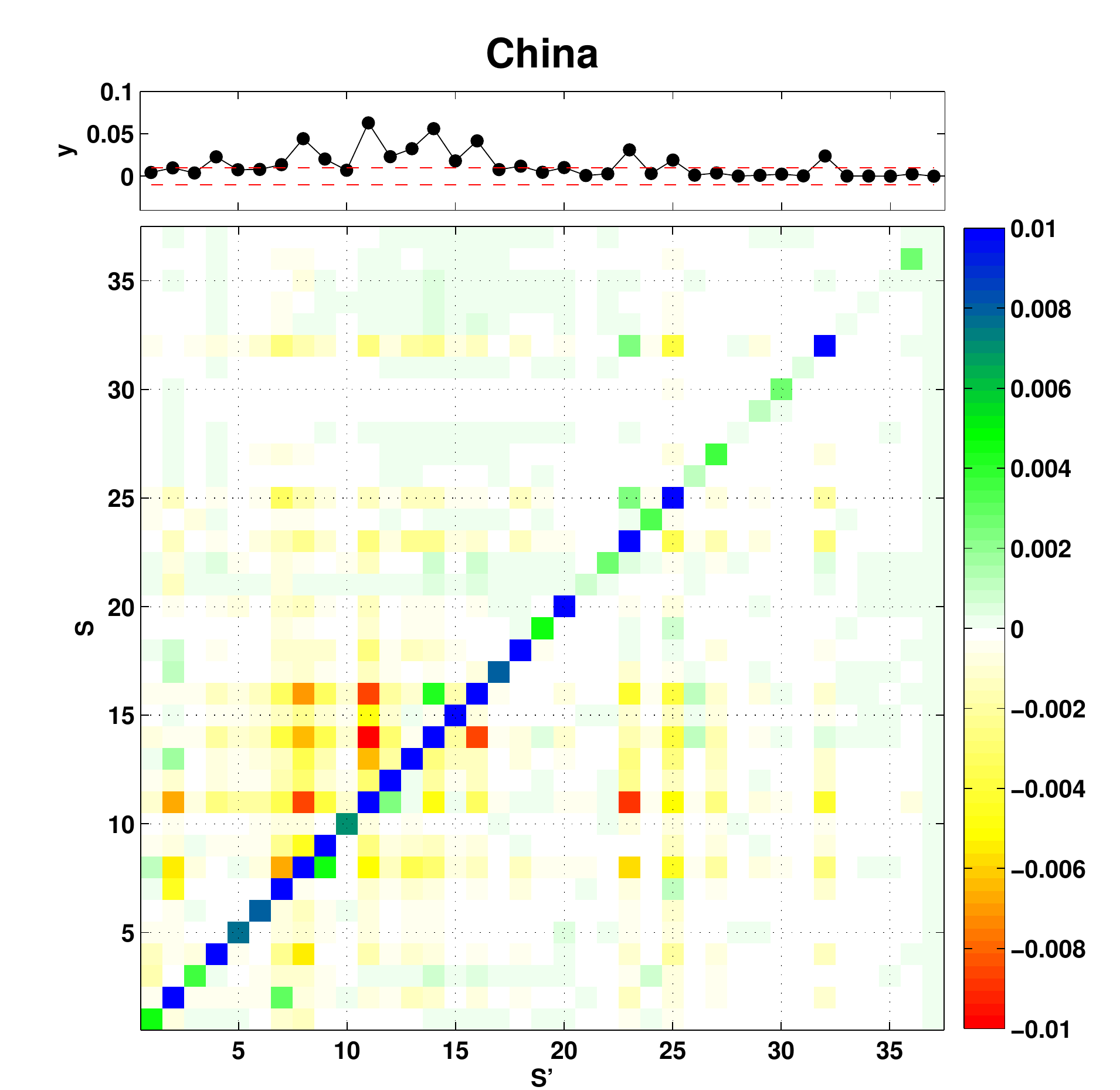} \\
\includegraphics[width=1\columnwidth,clip=true,trim=0 0 0 0cm]{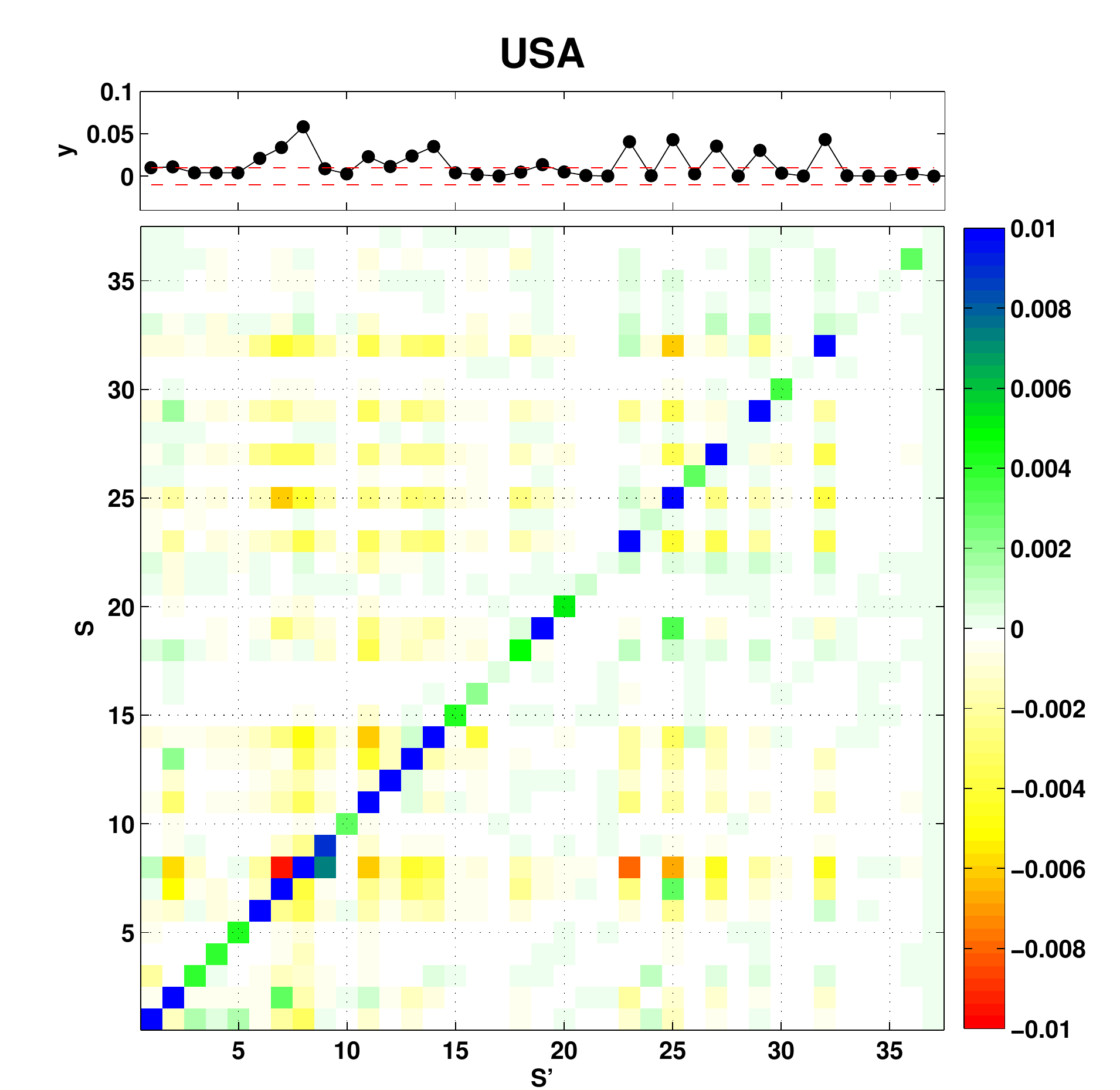}
\vglue -0.1cm
\caption {
China (top) and USA (bottom) examples of derivative $dB_{cs}/d\delta_{s'}$ 
of partial probability balance $B_{cs}$ of sector $s$ over price $\delta_{s'}$ 
of sector $s'$ for year 2008. Diagonal terms, given by $y=dB_{cs}/d\delta_{s}$ for $s=s'$, 
are shown on the top panels of each example. Sectors $s'$ and 
$s$ are shown in $x$-axis and $y$-axis respectively (indexed as in Table\ref{tab2}
from $1$ to $37$), 
while $dB_{cs}/d\delta_{s'}$ is represented by colors with a threshold value given 
by $+\epsilon$ and $-\epsilon$ for negative and positive values respectively, 
also shown in red dashed lines on top panels with diagonal terms. 
Here $\epsilon=0.01$ for USA and China; partial balance $B_{cs}$ 
is defined by CheiRank and PageRank probabilities.}
\label{fig17}
\end{center}
\end{figure}

The sensitivities $dB_c/d \delta_{s'}$ of CheiRank-PageRank balance of China and USA
to price variation of sectors $s'$ are presented in Fig.~\ref{fig16}.
We see two rather different profiles. Thus, for China 
the derivative $dB_c/d \delta_{s'}$
is positive for sectors $s=4, 14, 16$ ({\it Manufacture of textiles; office machinery; radio etc,})
and negative for $s=7, 25, 27$ ({\it Petroleum; Land transport etc.; Financial intermediation etc.}). 
For USA the sensitivity is significantly positive for $s=23, 29, 32$
({\it Sale of motor vehicles etc.; Renting of machinery and equipment etc.; 
Other business activities}) and negative for $s=11$ ({\it Manufacture of basic metals}).
Thus the economic activities of these two countries have very different strong and weak points.
We note that the sensitivity without the diagonal term
($dB_c/d\delta_{s'} - dB_{cs'}/d\delta_{s'}$) 
has negative values for almost all sectors for both countries.

The  matrices of cross-sector sensitivity $d B_{cs}/d \delta s'$
are shown for China and USA in Fig.~\ref{fig17}.
Such matrices provide a detailed information of
interconnections of various activity sectors.
Thus for USA  we see that its $s=8$ ({\it Manufacture of chemicals etc.})
has a significant negative sensitivity to $s'=7, 23, 25$
({\it Petroleum;  Renting of machinery and equipment etc.; Land transport etc.}).
Indeed, chemical production is linked with petroleum, machinery and transport.
For China we find that its sector $s=11$ ({\it Manufacture of basic metals}) has 
a negative sensitivity to $s'= 8, 23$
({\it Manufacture of chemicals etc.;  Renting of machinery and equipment etc.});
also $s=14, 16$ have a negative derivative in respect to $s'=11$).

Of course, the cross sensitivity to price variations in one sector 
and their effects on another sector, based on (\ref{eq14}), 
is a very delicate thing since a price in one sector can affect prices in other
sectors also in other manner since economic systems learn and adapt
while here we considered only linear algebraic relations
without any adaptation features.
However, even being linear, the Google matrix approach provides a detailed information
on hidden interactions and inter-dependencies of various economic
activities for various countries that can provide a useful message
even for nonlinear adapting systems. 

\begin{figure}[!ht] 
\begin{center} 
\includegraphics[width=1\columnwidth,clip=true,trim=0 0 0 0cm]{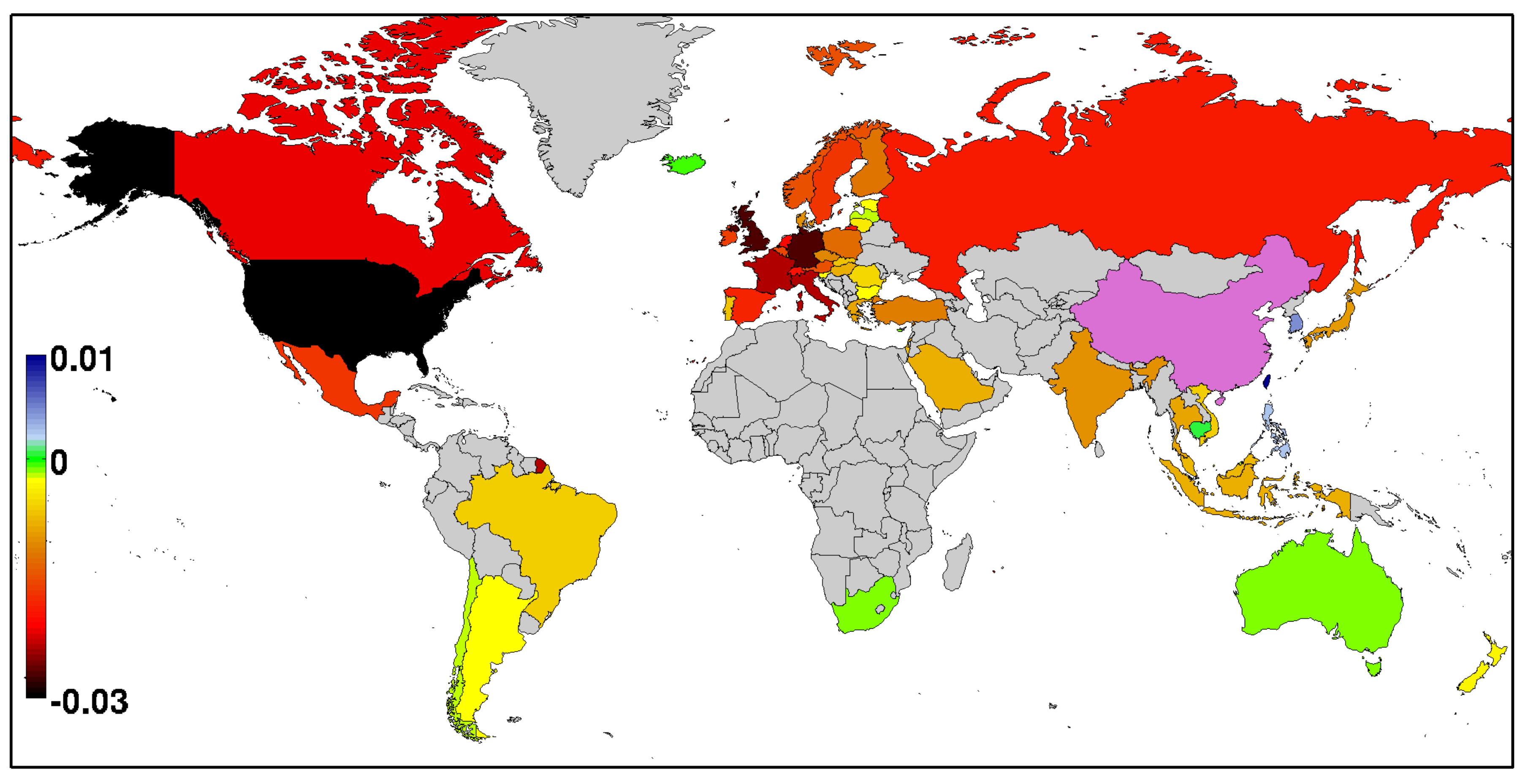} \\
\includegraphics[width=1\columnwidth,clip=true,trim=0 0 0 0cm]{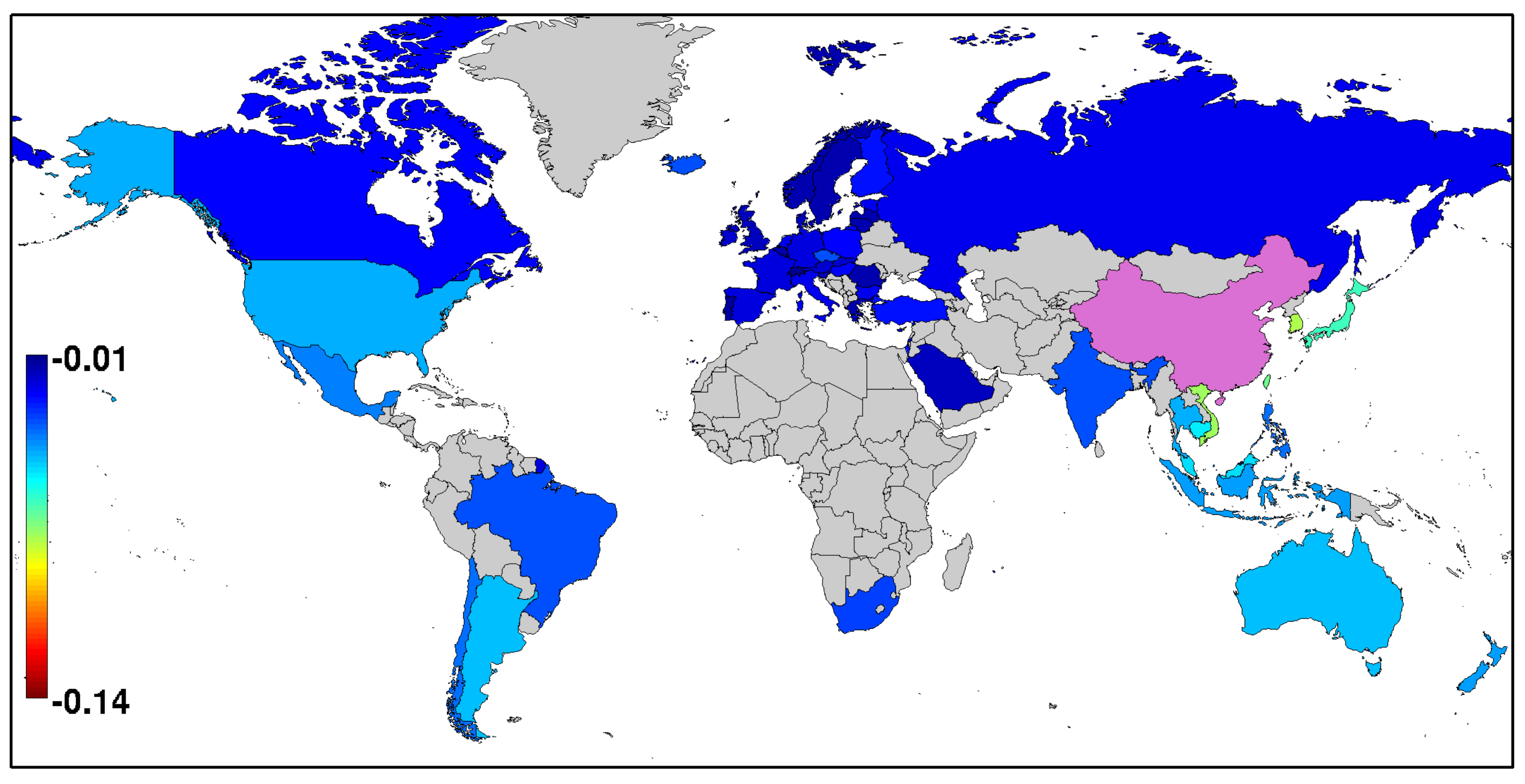}
\vglue -0.1cm
\caption {
Derivative of probabilities balance $dB_c/d\sigma_{c'}$ over labor cost of China $c'=37$ for year 2008. 
Top panel shows the case when $B_c$ is determined by CheiRank and PageRank vectors; 
here the special values are $dB_{58}/d\sigma_{37}=-0.0146$ for ROW group (gray) and 
$dB_{37}/d\sigma_{37}=0.3217$ for China (magenta). 
Bottom panel shows the case when $B_c$ is computed from the  Export-Import value; 
the special values are $dB_{58}/d\sigma_{37}=-0.0352$ for ROW group (gray) and 
$dB_{37}/d\sigma_{37}=0.4810$ for China (magenta). 
Names of the countries can be found in Table \ref{tab1} and
in the world map of countries \cite{worldmap}.}
\label{fig18}
\end{center}
\end{figure}

\subsection{World map of sensitivity to labor cost}

Using the established structure of WNEA we 
can study the sensitivity of
country balance $d B_c/d \sigma_c'$ to the labor cost in different countries.
At the difference of sectoral shocks on one product,
here the price shock affects all industries in a country.
As before, the change in price has to be small enough for the
resulting simulation to remain in a neighbourhood
of the original data. Indeed, larger shocks would trigger
a series of substitution effects diverting
trade to other partners.

The derivative $d B_c/d \sigma_c'$ is computed numerically as described in Sec.~3.5.
The world sensitivity to the labor cost of China is shown in Fig.~\ref{fig18}.
Of course, the largest derivative is found for China itself
($d B_c/d \sigma_c$ at $c=37$ from Table~\ref{tab1}).
The effect on other countries is given by non-diagonal derivatives
at $c \neq c'=37$. From the CheiRank-PageRank balance we find that the most
strong negative effect (minimal negative $d B_c/d \sigma_{c'}$)
is obtained for USA, Germany, UK; a positive derivative is visible  only
for Chinese Taipei ($s=38$) and S.Korea ($s=19$).
For the Export-Import balance the results are rather different: at first all derivatives
at $c \neq c'$ are negative; among the most negative values are such countries as
Hong Kong (most negative with dark red color but hardly visible due to its small size),
Chinese Taipei,  S.Korea, Vietnam. Thus the Google matrix approach
bring a new perspective for analysis of complex of economical relations
between countries and sectors.

\begin{figure}[!ht] 
\begin{center} 
\includegraphics[width=1\columnwidth,clip=true,trim=0 0 0 0cm]{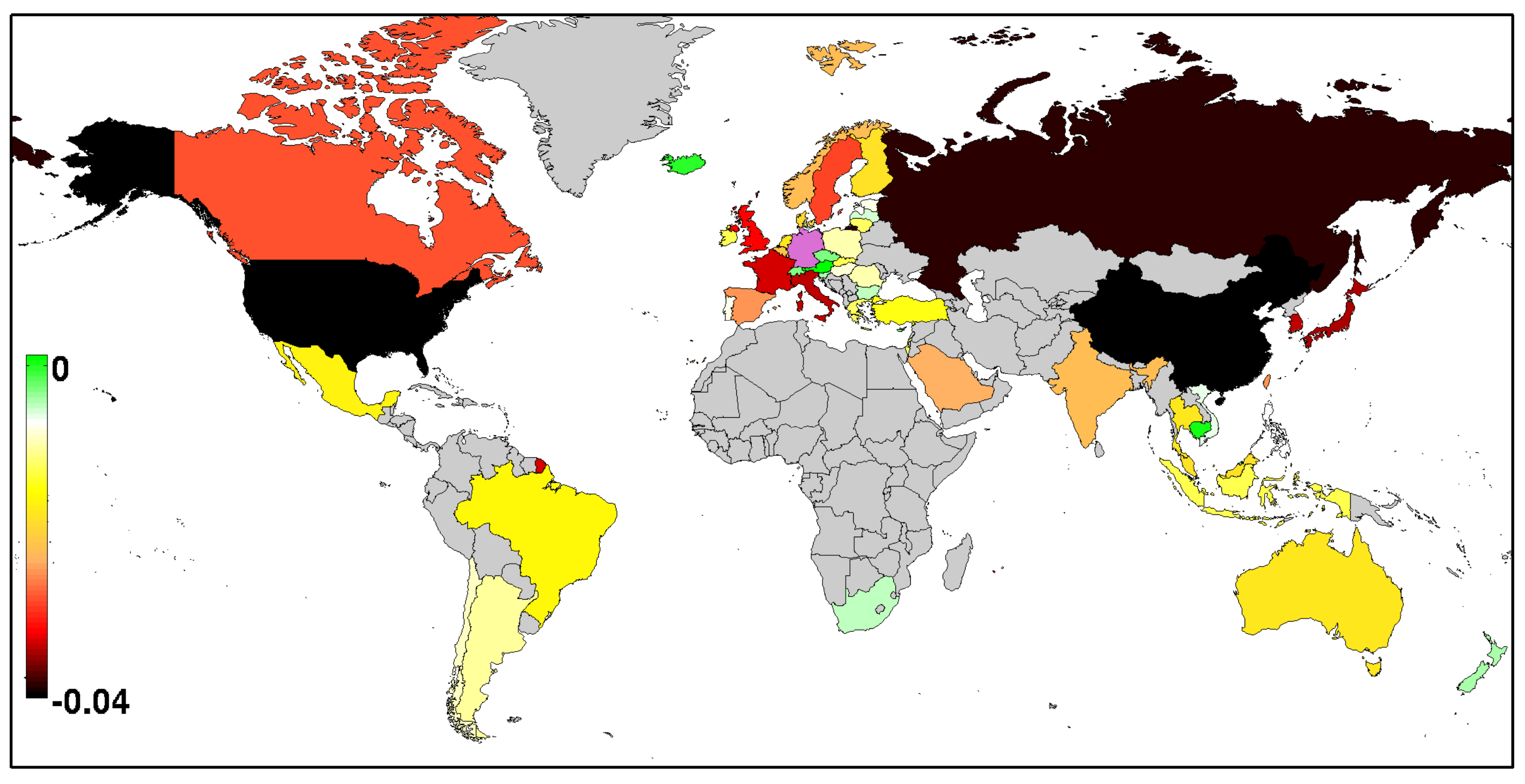} \\
\includegraphics[width=1\columnwidth,clip=true,trim=0 0 0 0cm]{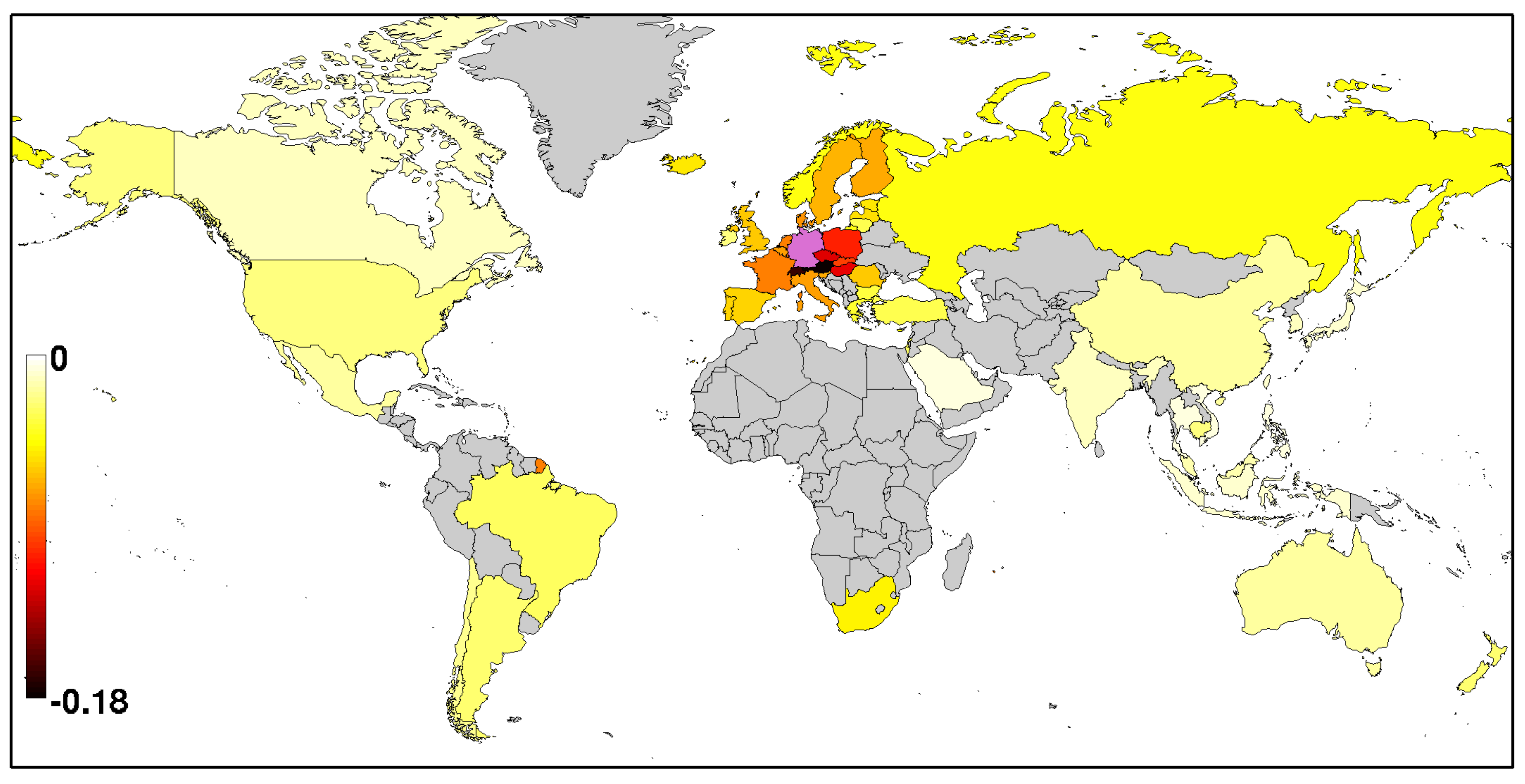}
\vglue -0.1cm
\caption {
Same as in Fig.~\ref{fig18} with  the derivative $dB_c/d\sigma_{c'}$ 
over the labor cost $c'=11$ of Germany for year 2008. 
Top panel shows the case when $B_c$ is determined by CheiRank and PageRank vectors; 
the special values are $dB_{58}/d\sigma_{11}=-0.0367$ for ROW group (gray) and 
$dB_{11}/d\sigma_{11}=0.3248$ for Germany (magenta). 
Bottom panel shows the case when $B_c$ is computed from the Export-Import value; 
the special values are $dB_{58}/d\sigma_{11}=-0.0280$ fro ROW group (gray) and 
$dB_{11}/d\sigma_{11}=0.4911$ for Germany (magenta). 
Names of the countries can be found in Table \ref{tab1} and
in the world map of countries \cite{worldmap}.}
\label{fig19}
\end{center}
\end{figure}

\begin{figure}[!ht] 
\begin{center} 
\includegraphics[width=1\columnwidth,clip=true,trim=0 0 0 0cm]{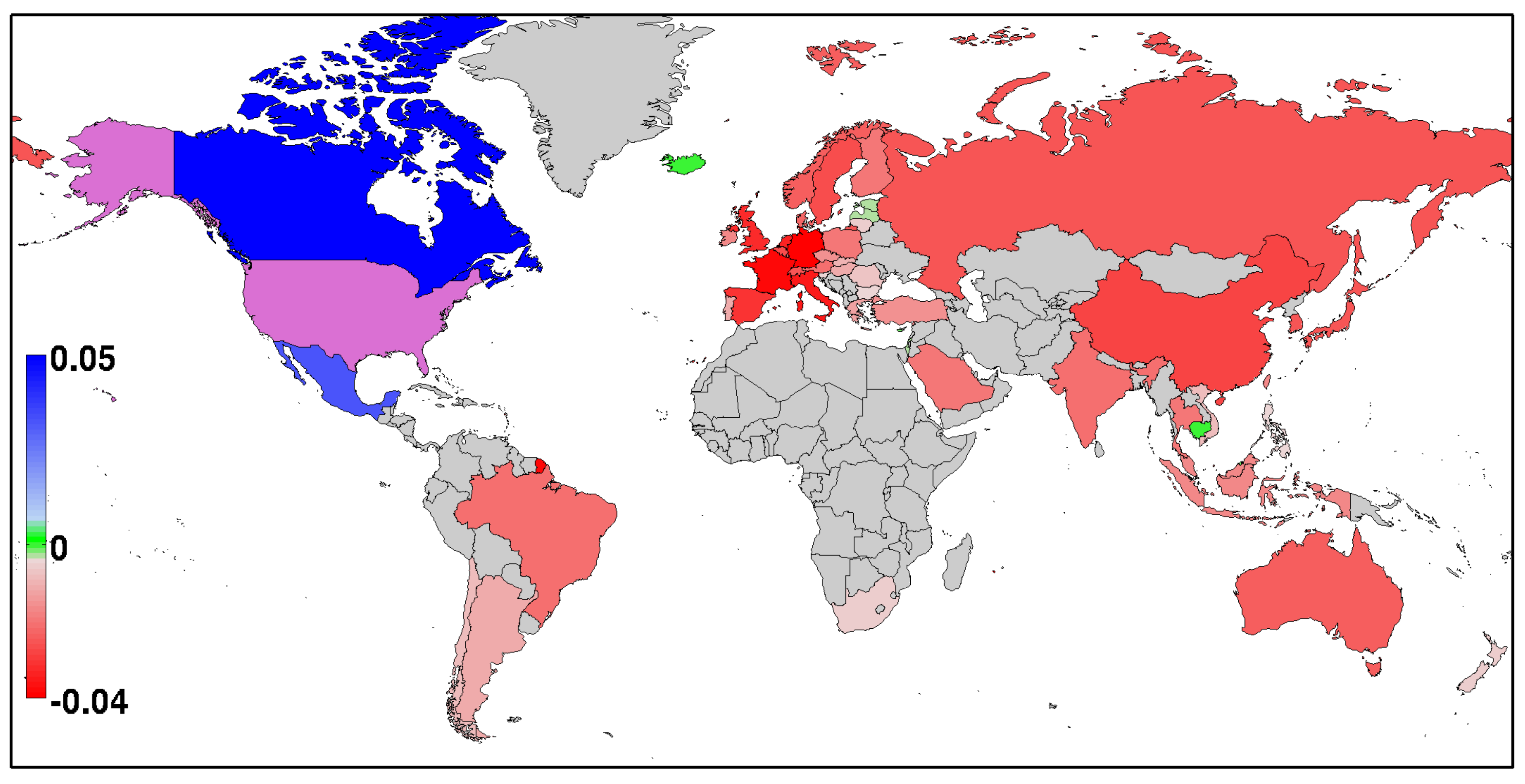} \\
\includegraphics[width=1\columnwidth,clip=true,trim=0 0 0 0cm]{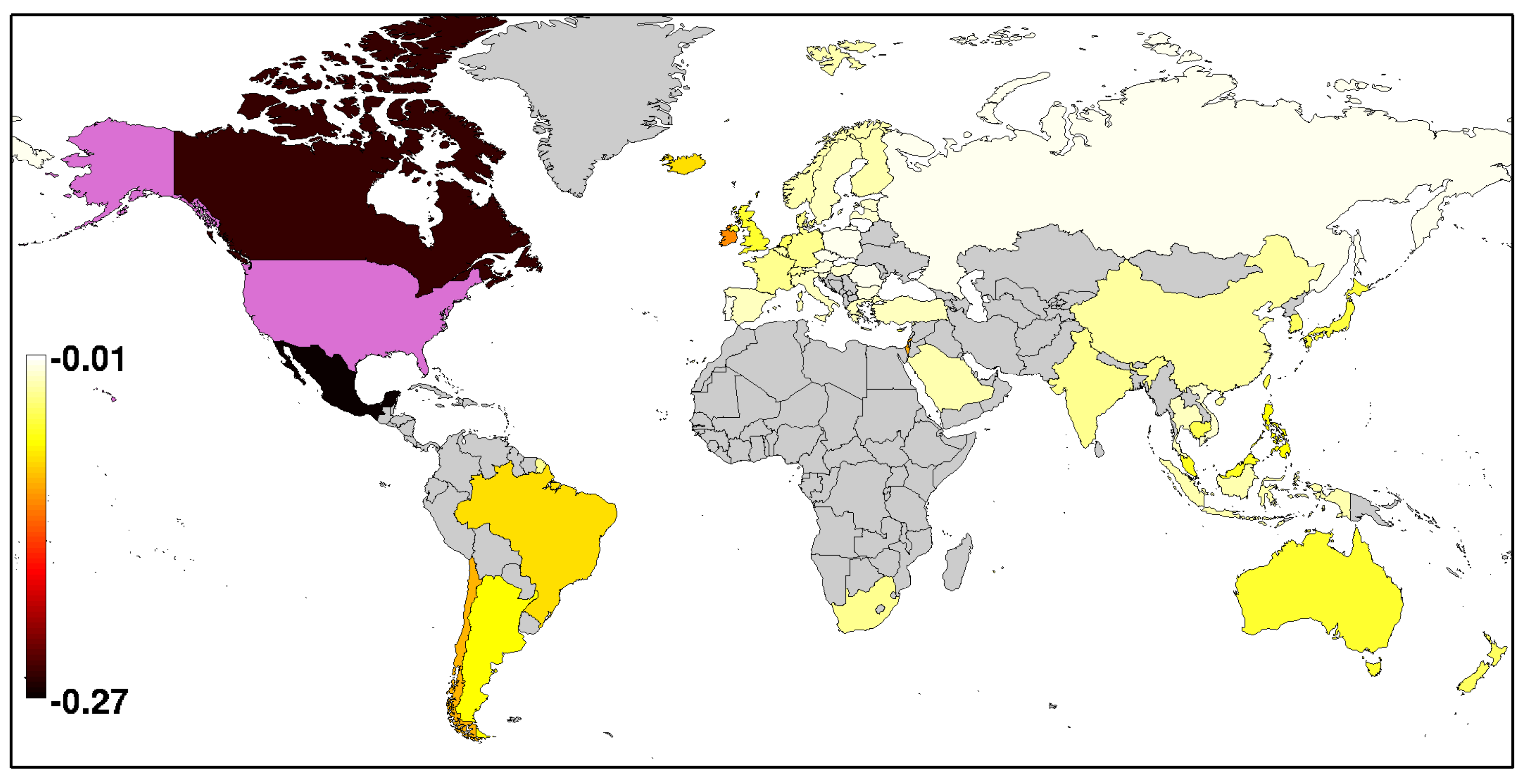}
\vglue -0.1cm
\caption {
Same as in Fig.~\ref{fig18} with  the derivative $dB_c/d\sigma_{c'}$ 
over labor cost $c'=34$ of USA for year 2008. 
Top panel shows the case when $B_c$ is determined by CheiRank and PageRank vectors; 
the special values are $dB_{58}/d\sigma_{34}=-0.0257$ for ROW group (gray) 
and $dB_{34}/d\sigma_{34}=0.3148$ for USA (magenta). 
Bottom panel shows the case when $B_c$ is computed from the Export-Import value; 
the special values are $dB_{58}/d\sigma_{34}=-0.0632$ for ROW group (gray)
and $dB_{34}/d\sigma_{34}=0.4852$ for USA (magenta). 
Names of the countries can be found in Table \ref{tab1} and
in the world map of countries \cite{worldmap}.}
\label{fig20}
\end{center}
\end{figure}

Another results for the effects of labor cost in Germany and 
in USA are shown in Fig.~\ref{fig19} and Fig.~\ref{fig20}.
In the case of Germany the most strong negative sensitivity is
for USA, Russia, China for CheiRank-PageRank balance while for 
Import-Export it is Switzerland and Austria. However,
USA and Russia are relatively weakly affected.
This again stresses the qualitative difference between these two approaches.

The increase of USA labor cost in Fig.~\ref{fig20} 
produces positive derivatives of CheiRank-PageRank balance for Canada and Mexico
that looks reasonable from a view point of economy since these countries will
profit from higher production costs in USA. In opposite, Export-Import
gives most strong negative derivatives for  Canada and Mexico.

\begin{figure}[!ht] 
\begin{center} 
\includegraphics[width=1\columnwidth,clip=true,trim=0 0 0 0cm]{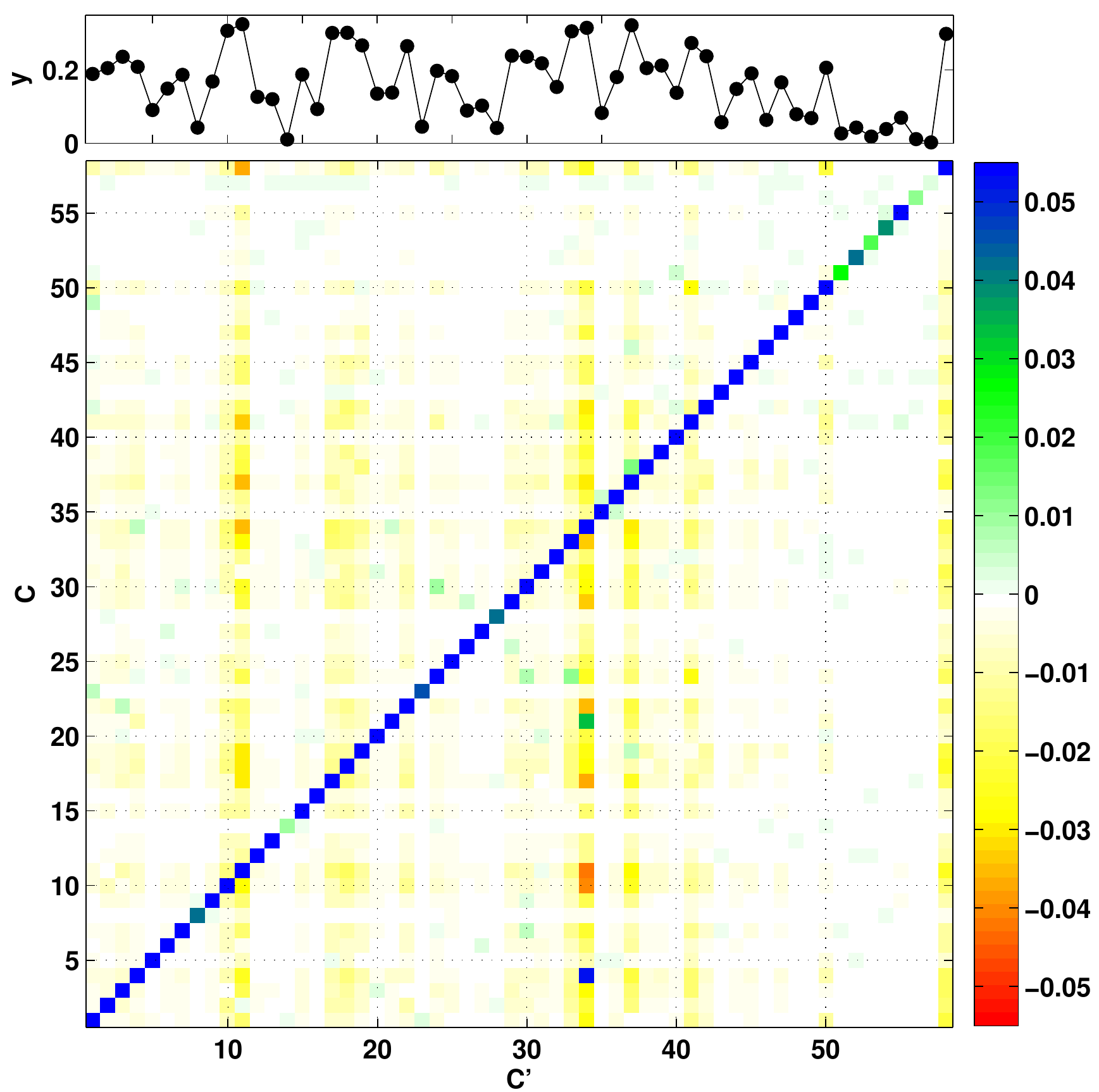}
\vglue -0.1cm
\caption {
Global view of the effect of labor cost variation in country $c'$ on country $c$ in 2008. 
Matrix elements $dB_c/d\sigma_{c'}$ are given in colors shown by the truncated color scale; 
matrix elements above the scale (diagonal terms) are shown in the top inset 
where $y=dB_c/d\sigma_{c'}$. In the matrix of derivatives shown by color,
$x$-axis shows the index $c'$ of country where a labor cost variation $\sigma_{c'}$
takes place and $y$-axis shows 
the country $c$ affected by the change. 
Here $B_c$ is computed from CheiRank and PageRank probabilities.
Country identification numbers $c=1, ..., 58$ are given in Table~\ref{tab1}.}
\label{fig21}
\end{center}
\end{figure}

The whole matrix of labor cost derivatives $d B_c/d \sigma_{c'}$
of the CheiRank-PageRank balance $B_c$ is shown in Fig.~\ref{fig21}
(numerical values of derivatives are given at 
\cite{ourwebpage}). Of course, the diagonal terms 
have the strongest positive derivatives, but off-diagonal terms change signs and
characterize the sensitivity of one country to labor cost in other country.
The vertical lines with high derivative values correspond to
Germany ($c'=11$), Japan ($c'=18$), S.Korea ($c'=19$), USA ($c'=34)$,
China ($c'=37$), Russia ($c'=41$). The rest of the world (ROW) group also have 
a visible effect of other countries ($c'=58$). Thus is it desirable to
obtain individual OECD data for  countries of the ROW group.

\begin{figure}[!ht] 
\begin{center} 
\includegraphics[width=1\columnwidth,clip=true,trim=0 0 0 0cm]{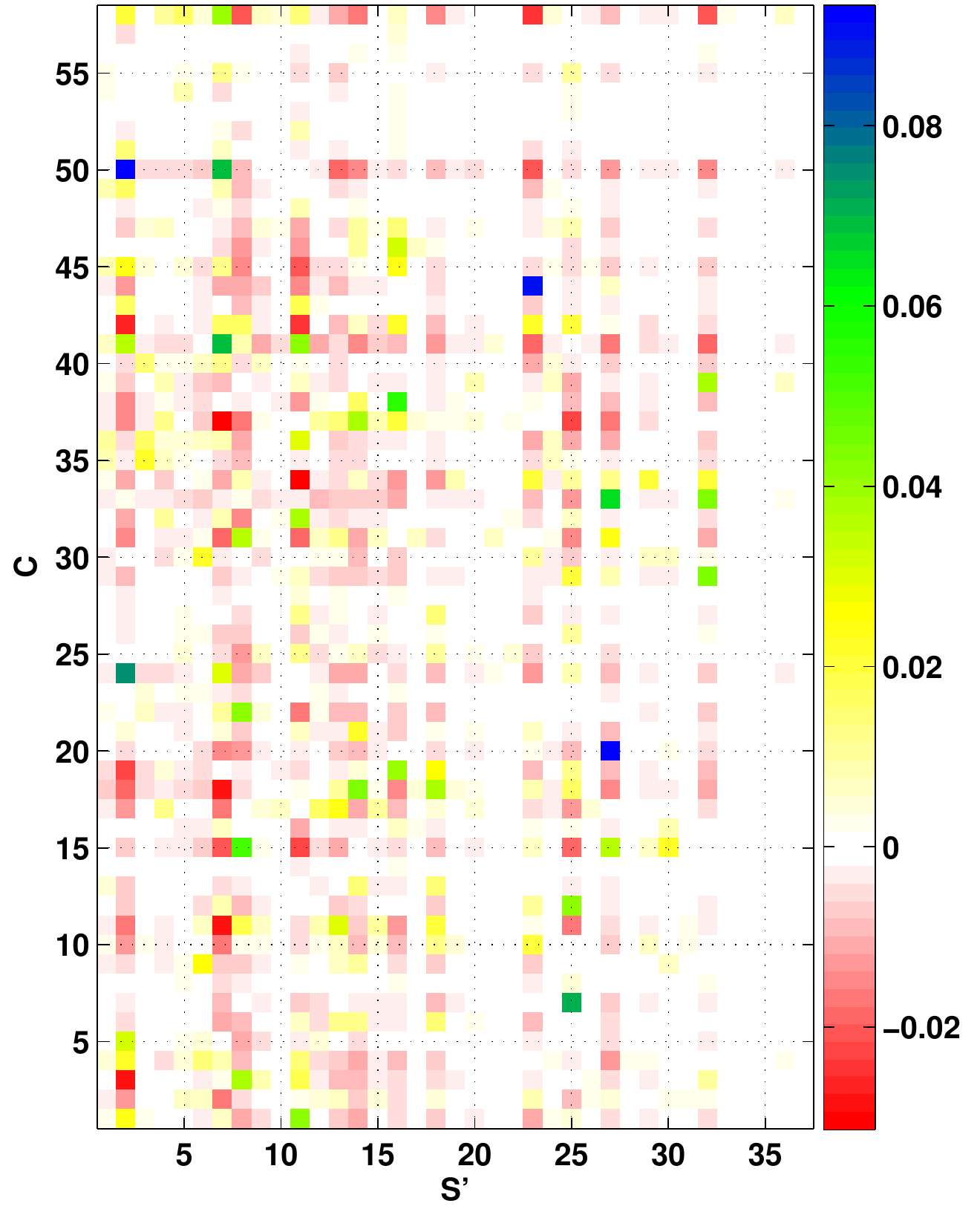}
\vglue -0.1cm
\caption {
Global view of the effect of sector $s'$ price variation on 
balance of country $c$ in 2008. Colors are proportional to matrix elements 
$dB_c/d\delta_{s'}$, $x$-axis shows the sector index $s'$ 
(sectors are given in Table~\ref{tab2}) and $y$-axis gives the country index
$c$ affected by the change (countries are given in Table~\ref{tab1}).
Here  $B_c$ is computed from CheiRank and PageRank probabilities.}
\label{fig22}
\end{center}
\end{figure}

In Fig.~\ref{fig21} we considered the effects of the labor cost in various countries.
We can also see the effect of price variation $\delta_{s'}$ in a given sector $s'$
on the CheiRank-PageRank balance $B_c$ of country $c$. This sensitivity is given by 
the rectangular matrix of derivatives $d B_c/d \delta_{s'}$ shown in Fig.~\ref{fig22}
(numerical data are given at \cite{ourwebpage}).
The strongest positive derivatives (blue squares) are
for $s'=2, c=50$ (mining and Saudi Arabia), $s'=23, c=44$ (motors and Hong Kong),
$s'=27, c=20$ (finance and Luxembourg).  The strongest negative derivatives
(red squares) are for $s'=2, c=3$ (mining and Belgium),  $s'=2, c=42$
(mining and Singapore which economy is very sensitive to
mining products), $s'=7, c=11$ (petroleum and Germany),
 $s'=7, c=18$ (petroleum and Japan), $s'=7, c=37$ (petroleum and China),
$s'=11, c= 34$ (manufacture of basic metals and USA),
$s'=11, c= 42$ (manufacture of basic metals and Singapore).
All these results are in agreement with the economic realities of 
sensitivity of the above countries to given activity sectors.
This shows the strength of the Google matrix approach to
analysis of WNEA. 

\subsection{World transformation matrix of activity sectors}

From the obtained Google matrices $G, G^*$ of WNEA
we can analyze the transformation of the activity sectors by the world economy.
For this analysis we compute the transfer matrix 
\begin{equation}
T = (1-\eta)  (1 - \eta G^*)^{-1} G\;\; ,
\label{eq15} 
\end{equation}
where $\eta$ is a numerical constant. Our study show that as in the case of 
damping factor $\alpha$ the results are robust to variations of $\eta$ in the range
$0.5 < \eta < 0.9$ and thus in the following we present the results
for $\eta=0.7$. We note that a similar construction for ImpactRank has been used
for Wikipedia networks \cite{physrev} and the C.elegans neural network \cite{celegans}.
In a certain sense (\ref{eq15}) can be considered as a scattering matrix of
particles entering in a system by $G$ term and then going out
by the expansion term $1+\eta G^*+(\eta G^*)^2 .... = 1/(1-\eta G^*)$.
In this approach $\eta$ describes a relaxation rate in the system.
We note that $T$ belongs to the Google matrix class.

\begin{figure}[!ht] 
\begin{center} 
\includegraphics[width=1\columnwidth,clip=true,trim=0 0 0 0cm]{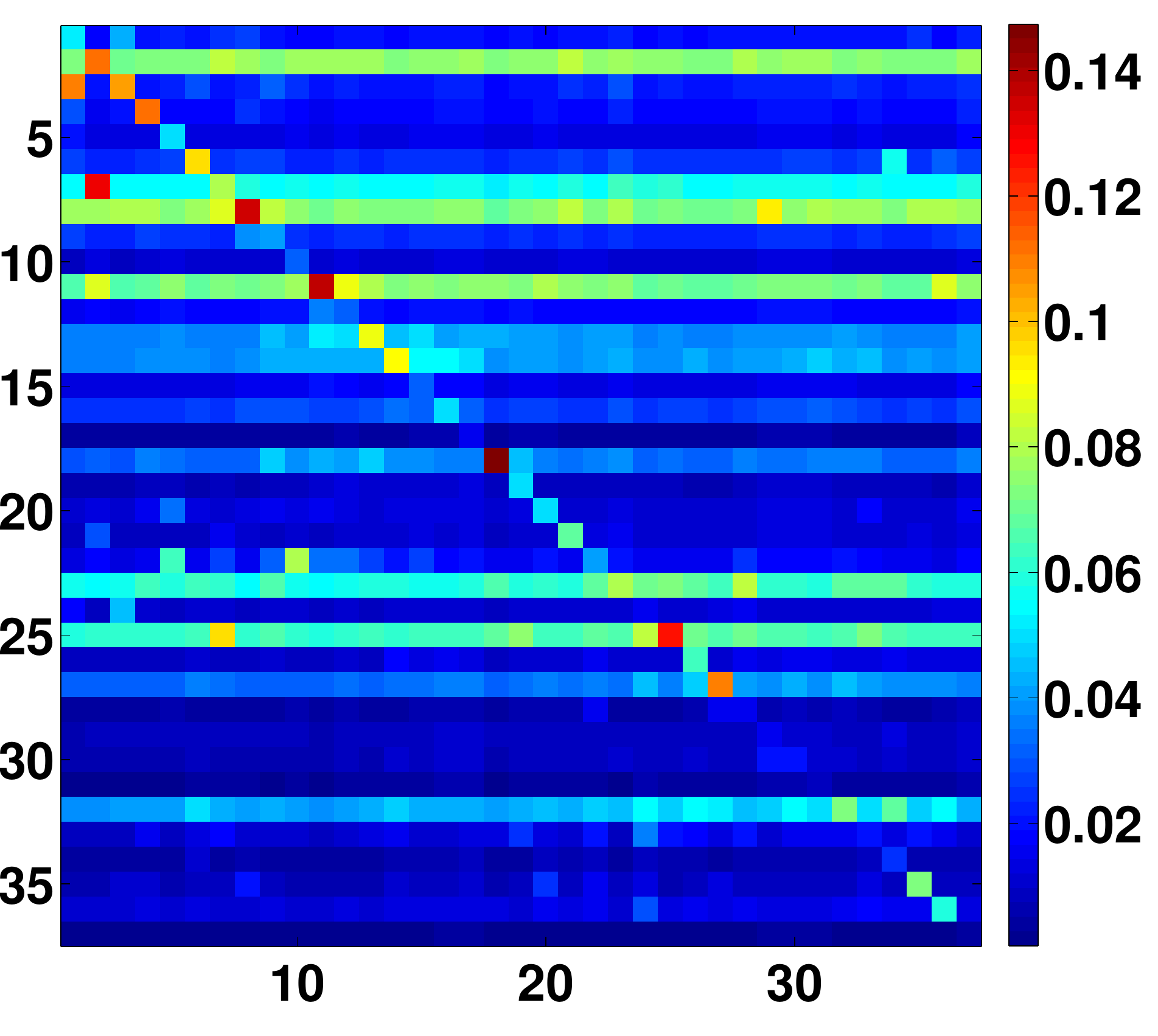}
\vglue -0.1cm
\caption {
Image of the  average reduced  transfer matrix $R_{s, s'}$ of sectors to sectors for 
for the whole world (averaged over countries) for year 2008.
Here $x$-axis represents the  initial sector $s'$ and $y$-axis 
represents the final sectors $s$ into which $s'$ is transformed. 
The sector numbering
is given in Table \ref{tab2}. 
Colors are proportional to matrix elements and $\eta=0.7$.
}
\label{fig23}
\end{center}
\end{figure}

From the global matrix $T$ of size $N$ we obtain the reduced matrix $R_{s s'}(c)$ 
of size $N_s$ describing
the transformation for activity sectors for a country $c$. We have
$R_{s s'}(c')= \sum_c T_{s, s', c, c'}$ where $c'$ is a target 
country we are interested in. The matrices  $R_{s s'}(c')$ giving 
the transformation of sector $s'$ to all other sectors $s$
for $c'$ of China, USA, Germany are given in \cite{ourwebpage}. 
The reduced transformation matrix for the whole world
is obtained by averaging over countries
with $R_{s s'} = \sum_{c'} R_{s s'}(c')/N_{c'}$ (see Fig.~\ref{fig23}).
The results of Fig.~\ref{fig23} show a few characteristic features:
the reduced transfer matrix has a strong diagonal element
(this is because each product is strong projection on itself),
there are characteristic horizontal lines corresponding to
important sectors (e.g. $s=2, 7, 11, 25$). 

\begin{figure}[!ht] 
\begin{center} 
\includegraphics[width=1\columnwidth,clip=true,trim=0 0 0 0cm]{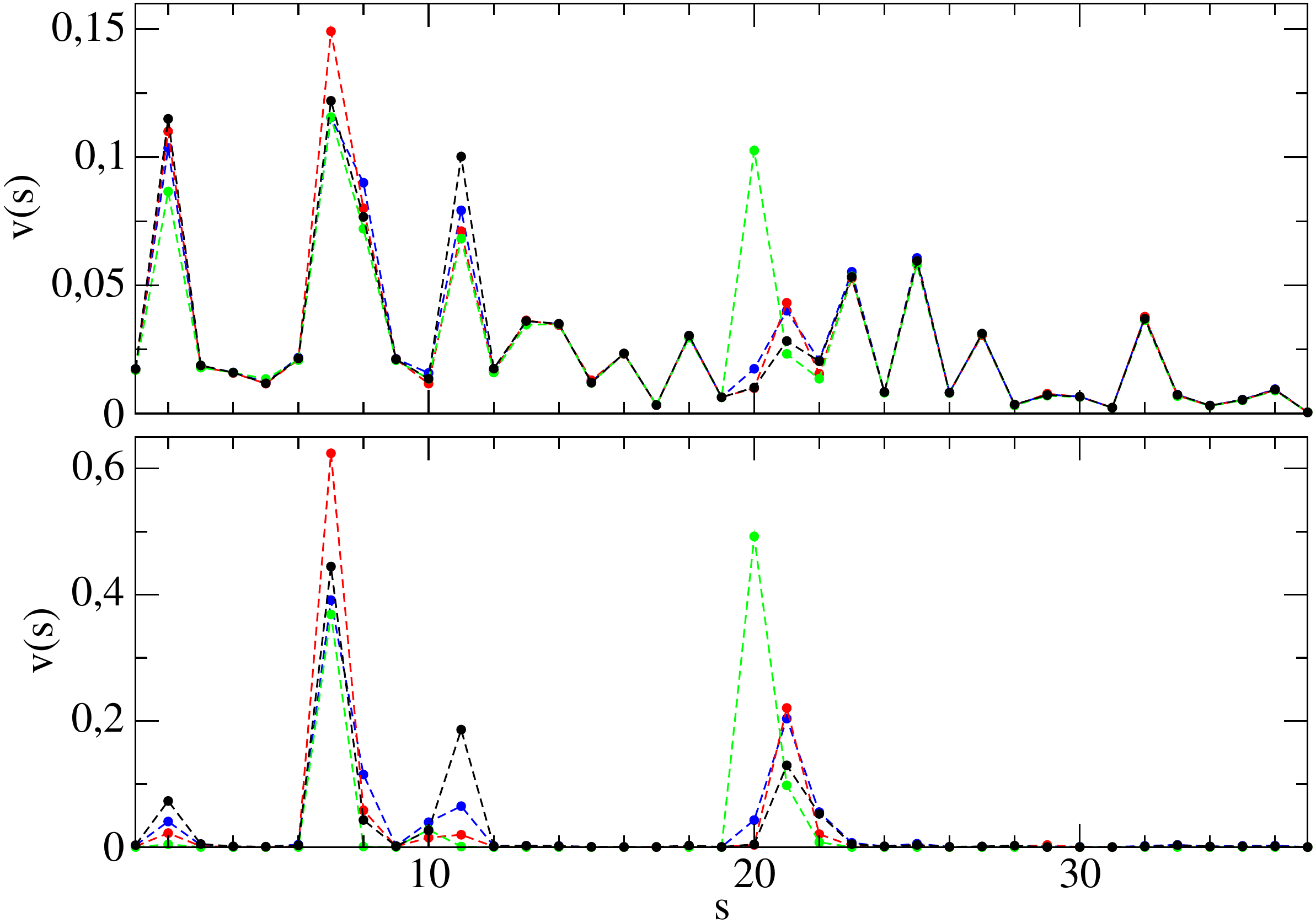}
\vglue -0.1cm
\caption {
\emph{Top panel:} Examples of profile $v(S)$ for transformation vector
from the reduced transfer matrix 
for several countries in 2008. Here the initial sector is $s=2$ (mining)
while the transformed vector $v(s)$ is formed by the matrix defined in  Fig.~\ref{fig23}; 
the countries are France (blue), Germany (red), Switzerland (green) and USA (black).
\emph{Bottom panel:} For comparison, we show here the same as top panel 
but instead of $T, R$ matrices we use the input/output matrix $M$ with normalized columns 
(dangling nodes are not replaced here, transitions inside one country
are taken to be zero); a column $s'$ of such 
a matrix for country $c'$ is given by $\sum_c M_{s s', c c'}$; here the 
same countries are shown by same colors as in top panel..
}
\label{fig24}
\end{center}
\end{figure}

By considering a transformation of a given sector 
to all other sectors for a given country. For $s'=2$  (mining) we
present the resulting transformed vector $v(s)$ in Fig.~\ref{fig24}
for France, Germany, Switzerland and USA. 
The global profiles are similar but there are significant
enhancement for Germany at sector $s=7$ (petroleum)
and for Switzerland at sector $s=20$ (manufacturing and recycling).
For comparison we show the results of transformation of input/output matrix $M$
of (\ref{eq1}). The comparison shows a drastic difference between
two approaches which we attribute to the fact that $M$ does not take into
account the multiple network transitions.

\begin{figure}[!ht] 
\begin{center} 
\includegraphics[width=1\columnwidth,clip=true,trim=0 0 0 0cm]{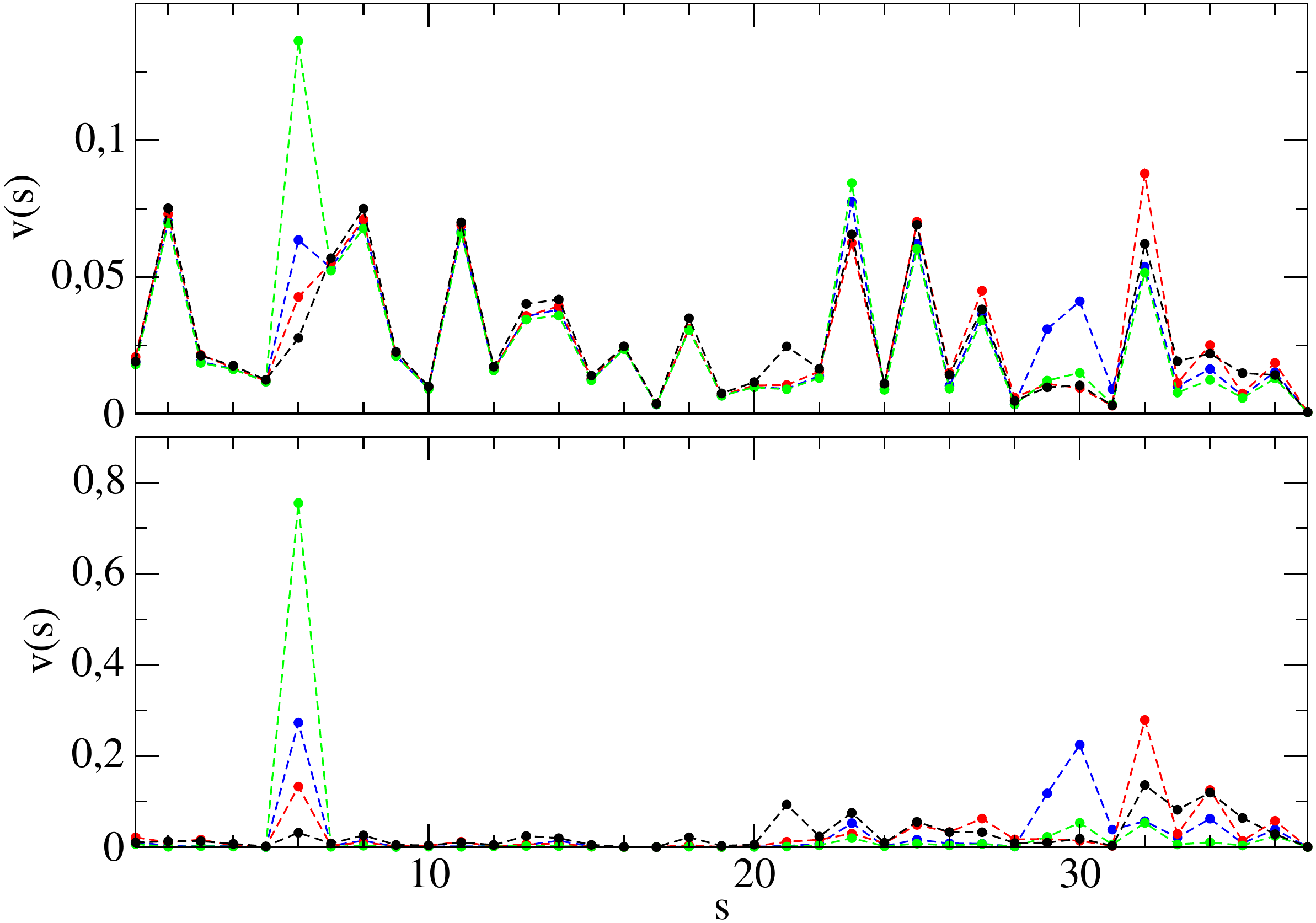}
\vglue -0.1cm
\caption {
Same as in Fig.~\ref{fig24} for the initial sector $s'=34$ (education).
The results are shown for Cyprus (blue), Singapore (red), Luxembourg (green) and Malta (black).
}
\label{fig25}
\end{center}
\end{figure}

The transformation for the sector $s'=34$ are shown in Fig.~\ref{fig25}
for Cyprus (blue), Singapore (red), Luxembourg (green) and Malta (black).
We see that for Luxembourg there is a strong transformation
of $s'=20$ to $s=6$ (publishing). At the same time the global profile, being
different from the case of Fig.~\ref{fig24} with $s'=2$,
has similar features for different countries.
The comparison with the transformation results from value exchange
matrix $ M_{s s', c c'}$ are again very different as in the case of Fig.~\ref{fig24}.

The obtained results for the activity sector transformation by the WNEA
open new possibilities for analysis of interactions between 
the world economic activities. The Google matrix approach provides
new type of results being very different from usual 
Input/Output matrix approach. This is related to the fact that 
the transformation matrix (\ref{eq14}) takes into account summation over various cycles 
over the network.

\section{Discussion}

In this work we have developed the Google matrix analysis of the 
world network of economic activities from the OECD-WTO TiVA database.
The PageRank and CheiRank probabilities allowed to obtain ranking of world countries
independently of their richness being mainly determined by the efficiency of their
economic relations.
The developed approach demonstrated the asymmetry in the economic activity sectors
some of which are export oriented and others are import oriented.
We also showed that the eigenstates of the WNEA Google matrix 
select specific quasi-isolated communities oriented to specific activity sectors.
The CheiRank-PageRank balance $B_c$ allows to determine
economically rising countries with robust network of economic relations. 
The sensitivity of this $B_c$ to price variations and labor cost in various countries
determines the hidden relations between world economies being not visible via 
usual Export-Import exchange analysis. The Google matrix analysis determines also
the transformation features of world activity sectors.

The comparison  with the multiproduct world trade network from UN COMTRADE
shows certain similarities between the two networks of WNEA and WTN.
At the same time the WNEA data provides new elements
for interactions of activity sectors while there are no direct 
interactions of products in COMTRADE database. From this viewpoint the OECD-WTO data
captures the economic reality on a deeper level. But at the same time the OECD-WTO
network is less developed compared to COMTRADE (less countries, years, sectors).
Thus it is highly desirable to extend the OECD-WTO database.

We think that the Google matrix analysis developed here and in \cite{wtngoogle,wtnproducts}
captures better the new reality of multifunctional directed tensor interactions
and that the universal features of this approach can be 
also extended to multifunctional financial network flows
which now attract an active interest of researchers \cite{craig,garratt}.
Unfortunately, the data on financial flows have much less 
accessibility compared to the networks discussed here.

We point that recently some of the matrix methods, developed in physics community,
started to find active application for economy systems 
(see e.g. \cite{bouchaud,guhr}). However, usually for physicists
these matrices have been from the 
unitary or Hermitian ensembles, where the Random Matrix Theory
allowed to obtained certain universal results. Here, we show that the
directed networks and tensors appearing in the interacting economy systems
are described by the matrices of Perron-Frobenius operators
which had not been studied much in physics. Thus the new field of research
is now opened for physicists, mathematicians and computer scientists
with application to complex interacting economy systems.

\section{Acknowledgments}
We thank the representatives of OECD \cite{oecd2014}
and WTO \cite{wto2014} for providing us with 
the friendly access to the data sets investigated in this work.
One of us (VK) thanks 
the  Economic Research and Statistics Division, WTO Gen\`eve
for hospitality during his intership there.
We thank L.Ermann for useful discussions
and advices on preparation of figures.
This research is supported in part by the EC FET Open project
``New tools and algorithms for directed network analysis''
(NADINE $No$ 288956).

\onecolumn

\clearpage

\begin{table}
\resizebox{\columnwidth}{!}{
\begin{tabular}{|c|c|c|c||c|c|c|c|} 
\hline 
 & country name & country code & country flag & & country name & country code & country flag \\ 
\hline 
\hline 
1 & Australia & AUS & \includegraphics[scale=0.4]{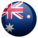} & 30 & Sweden & SWE & \includegraphics[scale=0.4]{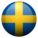}  \\ 
2 & Austria & AUT & \includegraphics[scale=0.4]{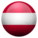} & 31 & Switzerland & CHE & \includegraphics[scale=0.4]{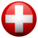}  \\ 
3 & Belgium & BEL & \includegraphics[scale=0.4]{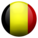} & 32 & Turkey & TUR & \includegraphics[scale=0.4]{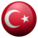}  \\ 
4 & Canada & CAN & \includegraphics[scale=0.4]{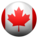} & 33 & United Kingdom & GBR & \includegraphics[scale=0.4]{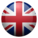}  \\ 
5 & Chile & CHL & \includegraphics[scale=0.4]{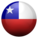} & 34 & United States & USA & \includegraphics[scale=0.4]{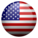}  \\ 
6 & Czech Republic & CZE & \includegraphics[scale=0.4]{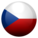} & 35 & Argentina & ARG & \includegraphics[scale=0.4]{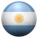}  \\ 
7 & Denmark & DNK & \includegraphics[scale=0.4]{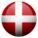} & 36 & Brazil & BRA & \includegraphics[scale=0.4]{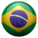}  \\ 
8 & Estonia & EST & \includegraphics[scale=0.4]{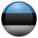} & 37 & China & CHN & \includegraphics[scale=0.4]{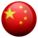}  \\ 
9 & Finland & FIN & \includegraphics[scale=0.4]{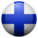} & 38 & Chinese Taipei & TWN & \includegraphics[scale=0.4]{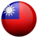}  \\ 
10 & France & FRA & \includegraphics[scale=0.4]{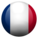} & 39 & India & IND & \includegraphics[scale=0.4]{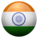}  \\ 
11 & Germany & DEU & \includegraphics[scale=0.4]{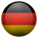} & 40 & Indonesia & IDN & \includegraphics[scale=0.4]{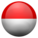}  \\ 
12 & Greece & GRC & \includegraphics[scale=0.4]{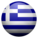} & 41 & Russia & RUS & \includegraphics[scale=0.4]{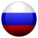}  \\ 
13 & Hungary & HUN & \includegraphics[scale=0.4]{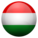} & 42 & Singapore & SGP & \includegraphics[scale=0.4]{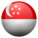}  \\ 
14 & Iceland & ISL & \includegraphics[scale=0.4]{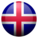} & 43 & South Africa & ZAF & \includegraphics[scale=0.4]{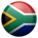}  \\ 
15 & Ireland & IRL & \includegraphics[scale=0.4]{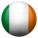} & 44 & Hong Kong & HKG & \includegraphics[scale=0.4]{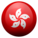}  \\ 
16 & Israel & ISR & \includegraphics[scale=0.4]{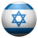} & 45 & Malaysia & MYS & \includegraphics[scale=0.4]{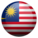}  \\ 
17 & Italy & ITA & \includegraphics[scale=0.4]{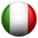} & 46 & Phillippines & PHL & \includegraphics[scale=0.4]{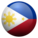}  \\ 
18 & Japan & JPN & \includegraphics[scale=0.4]{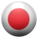} & 47 & Thailand & THA & \includegraphics[scale=0.4]{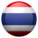}  \\ 
19 & Korea & KOR & \includegraphics[scale=0.4]{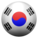} & 48 & Romania & ROU & \includegraphics[scale=0.4]{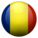}  \\ 
20 & Luxembourg & LUX & \includegraphics[scale=0.4]{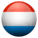} & 49 & Vietnam & VNM & \includegraphics[scale=0.4]{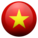}  \\ 
21 & Mexico & MEX & \includegraphics[scale=0.4]{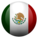} & 50 & Saudi Arabia & SAU & \includegraphics[scale=0.4]{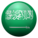}  \\ 
22 & Netherlands & NLD & \includegraphics[scale=0.4]{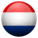} & 51 & Brunei Darussalam & BRN & \includegraphics[scale=0.4]{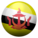}  \\ 
23 & New Zealand & NZL & \includegraphics[scale=0.4]{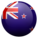} & 52 & Bulgaria & BGR & \includegraphics[scale=0.4]{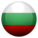}  \\ 
24 & Norway & NOR & \includegraphics[scale=0.4]{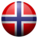} & 53 & Cyprus & CYP & \includegraphics[scale=0.4]{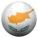}  \\ 
25 & Poland & POL & \includegraphics[scale=0.4]{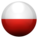} & 54 & Latvia & LVA & \includegraphics[scale=0.4]{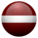}  \\ 
26 & Portugal & PRT & \includegraphics[scale=0.4]{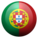} & 55 & Lithuania & LTU & \includegraphics[scale=0.4]{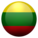}  \\ 
27 & Slovak Republic & SVK & \includegraphics[scale=0.4]{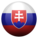} & 56 & Malta & MLT & \includegraphics[scale=0.4]{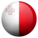}  \\ 
28 & Slovenia & SVN & \includegraphics[scale=0.4]{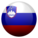} & 57 & Cambodia & KHM & \includegraphics[scale=0.4]{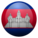}  \\ 
29 & Spain & ESP & \includegraphics[scale=0.4]{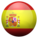} & 58 & Rest of the World & ROW & \includegraphics[scale=0.4]{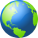}  \\ 
\hline 
\end{tabular} 
}
\caption{List of $N_c=58$ countries (with rest of the world ROW) with country name, code and flag.}
\label{tab1}
\end{table}

\clearpage

\begin{table}
\resizebox{\columnwidth}{!}{
\begin{tabular}{|c|c|l|} 
\hline 
 & OECD ICIO Category & ISIC Rev. 3 correspondence \\ 
\hline 
\hline 

1 &  C01T05 AGR	& \shortstack[l]{ 01 - Agriculture, hunting and related service activities \\
		  02 - Forestry, logging and related service activities \\
		  05 - Fishing, operation of fish hatcheries and fish farms; service activities incidental to fishing} \\ \hline
2 &  C10T14 MIN	& \shortstack[l]{ 10 - Mining of coal and lignite; extraction of peat \\
		  11 - Extraction of crude petroleum and natural gas; service activities incidental to oil and gas extraction excluding surveying \\
		  12 - Mining of uranium and thorium ores \\
		  13 - Mining of metal ores \\
		  14 - Other mining and quarrying} \\ \hline
3 &  C15T16 FOD & \shortstack[l]{ 15 - Manufacture of food products and beverages \\
		  16 - Manufacture of tobacco products} \\ \hline
4 &  C17T19 TEX	& \shortstack[l]{ 17 - Manufacture of textiles \\
		  18 - Manufacture of wearing apparel; dressing and dyeing of fur \\
		  19 - Tanning and dressing of leather; manufacture of luggage, handbags, saddlery, harness and footwear} \\ \hline
5 &  C20 WOD	& \shortstack[l]{ 20 - Manufacture of wood and of products of wood and cork, except furniture; \\
                       Manufacture of articles of straw and plaiting materials} \\ \hline
6 &  C21T22 PAP	& \shortstack[l]{ 21 - Manufacture of paper and paper products \\
		  22 - Publishing, printing and reproduction of recorded media} \\ \hline
7 &  C23 PET	& 23 - Manufacture of coke, refined petroleum products and nuclear fuel \\ \hline
8 &  C24 CHM	& 24 - Manufacture of chemicals and chemical products \\ \hline
9 &  C25 RBP	& 25 - Manufacture of rubber and plastics products \\ \hline
10 & C26 NMM	& 26 - Manufacture of other non-metallic mineral products \\ \hline
11 & C27 MET	& 27 - Manufacture of basic metals \\ \hline
12 & C28 FBM	& 28 - Manufacture of fabricated metal products, except machinery and equipment \\ \hline
13 & C29 MEQ	& 29 - Manufacture of machinery and equipment n.e.c. \\ \hline
14 & C30 ITQ	& 30 - Manufacture of office, accounting and computing machinery \\ \hline
15 & C31 ELQ	& 31 - Manufacture of electrical machinery and apparatus n.e.c. \\ \hline
16 & C32 CMQ	& 32 - Manufacture of radio, television and communication equipment and apparatus \\ \hline
17 & C33 SCQ	& 33 - Manufacture of medical, precision and optical instruments, watches and clocks \\ \hline
18 & C34 MTR	& 34 - Manufacture of motor vehicles, trailers and semi-trailers \\ \hline
19 & C35 TRQ	& 35 - Manufacture of other transport equipment \\ \hline
20 & C36T37 OTM	& \shortstack[l]{ 36 - Manufacture of furniture; manufacturing n.e.c. \\
		  37 - Recycling} \\ \hline
21 & C40T41 EGW	& \shortstack[l]{ 40 - Electricity, gas, steam and hot water supply \\
		  41 - Collection, purification and distribution of water} \\ \hline
22 & C45 CON	& 45 - Construction \\ \hline
23 & C50T52 WRT	& \shortstack[l]{ 50 - Sale, maintenance and repair of motor vehicles and motorcycles; retail sale of automotive fuel \\
		  51 - Wholesale trade and commission trade, except of motor vehicles and motorcycles \\
		  52 - Retail trade, except of motor vehicles and motorcycles; repair of personal and household goods} \\ \hline
24 & C55 HTR	& 55 - Hotels and restaurants \\ \hline
25 & C60T63 TRN	& \shortstack[l]{ 60 - Land transport; transport via pipelines \\
		  61 - Water transport \\
		  62 - Air transport \\
		  63 - Supporting and auxiliary transport activities; activities of travel agencies} \\ \hline
26 & C64 PTL	& 64 - Post and telecommunications \\ \hline
27 & C65T67 FIN	& \shortstack[l]{ 65 - Financial intermediation, except insurance and pension funding \\
		  66 - Insurance and pension funding, except compulsory social security \\
		  67 - Activities auxiliary to financial intermediation} \\ \hline
28 & C70 REA	& 70 - Real estate activities \\ \hline
29 & C71 RMQ	& 71 - Renting of machinery and equipment without operator and of personal and household goods \\ \hline
30 & C72 ITS	& 72 - Computer and related activities \\ \hline
31 & C73 RDS	& 73 - Research and development \\ \hline
32 & C74 BZS	& 74 - Other business activities \\ \hline
33 & C75 GOV	& 75 - Public administration and defense; compulsory social security \\ \hline
34 & C80 EDU	& 80 - Education \\ \hline
35 & C85 HTH	& 85 - Health and social work \\ \hline
36 & C90T93 OTS	& \shortstack[l]{ 90 - Sewage and refuse disposal, sanitation and similar activities \\
		  91 - Activities of membership organizations n.e.c. \\
		  92 - Recreational, cultural and sporting activities \\
		  93 - Other service activities} \\ \hline
37 & C95 PVH	& 95 - Private households with employed persons \\ \hline
\end{tabular}
} 
\caption{List of sectors considered by Input/Output matrices from OECD database, their correspondence to the ISIC classification is also given.}
\label{tab2}
\end{table}

\clearpage

\begin{table}
\resizebox{\columnwidth}{!}{
\begin{tabular}{|c||c|c|c|c||c|c|c|c|} 
\hline 
Sector & $\hat{K}$ (1995) & \% vol (1995) & $\hat{K}^*$ (1995) & \% vol (1995) & $\hat{K}$ (2008) & \% vol (2008) & $\hat{K}^*$ (2008) & \% vol (2008) \\ 
\hline 
\hline 
1 & 19 & 2.2979 & 16 & 2.9763 & 20 & 1.9532 & 16 &2.0902 \\ 
2 & 27 & 1.2993 & 2 & 8.6183 & 24 & 1.5245 & 1 &15.8784 \\ 
3 & 3 & 6.0117 & 12 & 3.3271 & 11 & 3.9327 & 17 &1.9835 \\ 
4 & 10 & 3.9579 & 14 & 3.0831 & 17 & 2.0934 & 19 &1.8634 \\ 
5 & 30 & 1.108 & 20 & 1.9037 & 33 & 0.60075 & 22 &1.3001 \\ 
6 & 11 & 3.5687 & 6 & 4.2128 & 18 & 2.0608 & 14 &2.3736 \\ 
7 & 4 & 5.9126 & 19 & 2.2783 & 1 & 11.589 & 4 &6.34 \\ 
8 & 2 & 6.251 & 1 & 10.6954 & 3 & 6.0558 & 2 &9.1103 \\ 
9 & 17 & 2.4035 & 15 & 3.0546 & 19 & 1.9785 & 13 &2.5549 \\ 
10 & 28 & 1.2 & 21 & 1.8337 & 29 & 1.0389 & 21 &1.3177 \\ 
11 & 8 & 4.4393 & 3 & 8.0658 & 4 & 5.4907 & 3 &8.3184 \\ 
12 & 20 & 2.2646 & 17 & 2.7194 & 23 & 1.6212 & 15 &2.2182 \\ 
13 & 9 & 4.0642 & 8 & 4.0365 & 9 & 4.0117 & 9 &4.0597 \\ 
14 & 12 & 3.3353 & 13 & 3.158 & 8 & 4.0642 & 6 &5.0066 \\ 
15 & 18 & 2.3789 & 9 & 4.0148 & 25 & 1.456 & 18 &1.8673 \\ 
16 & 15 & 2.7053 & 10 & 3.8054 & 14 & 2.7844 & 11 &3.6339 \\ 
17 & 31 & 1.0034 & 23 & 1.1434 & 34 & 0.31041 & 29 &0.40161 \\ 
18 & 7 & 5.2722 & 7 & 4.1643 & 6 & 5.1478 & 10 &3.9907 \\ 
19 & 26 & 1.3665 & 22 & 1.7813 & 26 & 1.3028 & 23 &1.2752 \\ 
20 & 24 & 1.6331 & 27 & 0.67546 & 22 & 1.6652 & 20 &1.3858 \\ 
21 & 21 & 2.1673 & 30 & 0.34377 & 10 & 3.946 & 30 &0.39969 \\ 
22 & 1 & 6.538 & 32 & 0.22022 & 2 & 6.8692 & 32 &0.15209 \\ 
23 & 5 & 5.8472 & 4 & 7.9296 & 7 & 4.6893 & 8 &4.6745 \\ 
24 & 25 & 1.5283 & 29 & 0.37682 & 27 & 1.2377 & 27 &0.62202 \\ 
25 & 6 & 5.8385 & 5 & 6.5023 & 5 & 5.2454 & 5 &5.8065 \\ 
26 & 29 & 1.1862 & 26 & 0.6839 & 28 & 1.2179 & 26 &0.62929 \\ 
27 & 13 & 2.7584 & 18 & 2.3006 & 15 & 2.5623 & 12 &3.3487 \\ 
28 & 33 & 0.70446 & 24 & 0.93849 & 31 & 0.84772 & 33 &0.105 \\ 
29 & 36 & 0.16329 & 33 & 0.18955 & 36 & 0.21276 & 24 &0.81082 \\ 
30 & 34 & 0.53799 & 28 & 0.39581 & 32 & 0.67481 & 28 &0.61668 \\ 
31 & 35 & 0.36919 & 31 & 0.33351 & 35 & 0.24684 & 31 &0.24177 \\ 
32 & 16 & 2.618 & 11 & 3.372 & 13 & 3.0455 & 7 &4.7163 \\ 
33 & 14 & 2.7071 & 34 & 0.064931 & 12 & 3.3939 & 35 &0.06377 \\ 
34 & 32 & 0.89993 & 36 & 0.0416 & 30 & 1.036 & 34 &0.09439 \\ 
35 & 22 & 1.8912 & 35 & 0.045551 & 16 & 2.2601 & 36 &0.025979 \\ 
36 & 23 & 1.7326 & 25 & 0.7136 & 21 & 1.8131 & 25 &0.72283 \\ 
37 & 37 & 0.03899 & 37 & 0 & 37 & 0.019524 & 37 &0 \\ 
\hline 
\end{tabular}
} 
\caption{First column gives the sectors from OECD database, for each of them the following columns give the ImportRank $\hat{K}$ with the sector fraction in global trade value and ExportRank $\hat{K}^*$ with sector fraction in global trade value. Data are shown for 1995 and 2008.}
\label{tab3}
\end{table}

\clearpage

\begin{table}
\resizebox{\columnwidth}{!}{
\begin{tabular}{|c|l|l|l|l|l|} 
\hline 
 & $K$ & $K^*$ & $K_2$ & $\hat{K}$ & $\hat{K}^*$ \\ 
\hline 
\hline 
1 & DEU C34 MTR & ROW C10T14 MIN & DEU C24 CHM & USA C23 PET & ROW C10T14 MIN \\ 
2 & USA C75 GOV & RUS C10T14 MIN & USA C65T67 FIN & JPN C23 PET & SAU C10T14 MIN \\ 
3 & ROW C75 GOV & SAU C10T14 MIN & DEU C29 MEQ & USA C75 GOV & RUS C10T14 MIN \\ 
4 & SAU C85 HTH & USA C24 CHM & DEU C34 MTR & ROW C45 CON & USA C24 CHM \\ 
5 & GBR C85 HTH & DEU C24 CHM & DEU C27 MET & CHN C32 CMQ & CAN C10T14 MIN \\ 
6 & USA C34 MTR & DEU C27 MET & USA C74 BZS & CHN C27 MET & DEU C24 CHM \\ 
7 & ROW C45 CON & NOR C10T14 MIN & DEU C50T52 WRT & USA C45 CON & NOR C10T14 MIN \\ 
8 & ROW C15T16 FOD & RUS C27 MET & USA C24 CHM & DEU C34 MTR & AUS C10T14 MIN \\ 
9 & USA C15T16 FOD & USA C50T52 WRT & DNK C60T63 TRN & KOR C23 PET & CHN C30 ITQ \\ 
10 & RUS C50T52 WRT & DEU C29 MEQ & GBR C74 BZS & DEU C23 PET & USA C30 ITQ \\ 
11 & USA C45 CON & USA C74 BZS & JPN C34 MTR & JPN C40T41 EGW & JPN C30 ITQ \\ 
12 & USA C85 HTH & CHN C27 MET & GBR C65T67 FIN & ROW C75 GOV & DEU C29 MEQ \\ 
13 & DEU C15T16 FOD & USA C60T63 TRN & CHN C32 CMQ & CHN C24 CHM & DEU C34 MTR \\ 
14 & ROW C60T63 TRN & GBR C65T67 FIN & CHN C24 CHM & USA C34 MTR & KOR C32 CMQ \\ 
15 & USA C65T67 FIN & USA C23 PET & DEU C60T63 TRN & USA C24 CHM & USA C23 PET \\ 
16 & GBR C50T52 WRT & GBR C74 BZS & FRA C50T52 WRT & CHN C30 ITQ & USA C74 BZS \\ 
17 & DEU C24 CHM & USA C65T67 FIN & USA C50T52 WRT & CHN C23 PET & TWN C32 CMQ \\ 
18 & DEU C29 MEQ & CHN C30 ITQ & CHN C50T52 WRT & ROW C60T63 TRN & CHN C27 MET \\ 
19 & DEU C50T52 WRT & DEU C34 MTR & CHN C29 MEQ & CHN C29 MEQ & DEU C27 MET \\ 
20 & DEU C27 MET & USA C30 ITQ & ROW C60T63 TRN & DEU C29 MEQ & GBR C74 BZS \\ 
\hline 
\end{tabular} 
}
\caption{Top 20 ranks for global PageRank $K$, CheiRank$K^*$, 2DRank $K_2$, ImportRank $K$ and ExportRank $K^*$ for the year 2008.}
\label{tab4}
\end{table}


\begin{table}
\resizebox{\columnwidth}{!}{
\begin{tabular}{|c||c|c||c|c||c|c||c|c|} 
\hline 
$K_i$ & $|\psi_i|$ & node & $|\psi_i|$ & node & $|\psi_i|$ & node & $|\psi_i|$ & node \\ 
\hline 
\hline 
1 & 0.037606 & ROW C17T19 TEX & 0.050431 & ARG C34 MTR & 0.054681 & CHN C32 CMQ & 0.052248 & RUS C10T14 MIN \\ 
2 & 0.025695 & CHN C17T19 TEX & 0.049991 & BRA C34 MTR & 0.053306 & KOR C32 CMQ & 0.03948 & SAU C10T14 MIN \\ 
3 & 0.021618 & ITA C17T19 TEX & 0.029753 & JPN C34 MTR & 0.053253 & TWN C32 CMQ & 0.026187 & ROW C10T14 MIN \\ 
4 & 0.017075 & USA C17T19 TEX & 0.026592 & DEU C34 MTR & 0.027361 & SGP C32 CMQ & 0.022125 & NOR C10T14 MIN \\ 
5 & 0.016216 & CHN C32 CMQ & 0.018372 & THA C34 MTR & 0.025189 & MYS C32 CMQ & 0.019764 & USA C71 RMQ \\ 
6 & 0.013003 & CHN C30 ITQ & 0.01531 & IDN C34 MTR & 0.018824 & USA C30 ITQ & 0.013899 & USA C50T52 WRT \\ 
7 & 0.010963 & FRA C17T19 TEX & 0.0093875 & ROW C21T22 PAP & 0.016965 & PHL C32 CMQ & 0.011638 & ROW C29 MEQ \\ 
8 & 0.010175 & TUR C17T19 TEX & 0.0093382 & DEU C15T16 FOD & 0.01534 & JPN C30 ITQ & 0.010871 & RUS C27 MET \\ 
9 & 0.010161 & USA C75 GOV & 0.0090288 & USA C15T16 FOD & 0.014664 & GBR C65T67 FIN & 0.0082943 & DEU C29 MEQ \\ 
10 & 0.0099839 & USA C65T67 FIN & 0.0086552 & USA C75 GOV & 0.013713 & CHN C30 ITQ & 0.0082905 & RUS C23 PET \\ 
\hline 
\end{tabular} 
}
\caption{Top 10 values of 4 different eigenvectors from Fig.\ref{fig9}, Fig.~\ref{fig10}. The corresponding eigenvalues from left to right are $\lambda=0.4993$ (red), $\lambda=0.3746+0.0126i$ (green), $\lambda=0.6256$ (blue) and $\lambda=-0.0001+0.1687i$ (magenta).}
\label{tab5}
\end{table}

\end{document}